\documentclass[12pt]{iopart}
\pdfoutput=1

\expandafter\let\csname equation*\endcsname\relax
\expandafter\let\csname endequation*\endcsname\relax
\usepackage{amsmath}

\usepackage{amssymb}
\usepackage{amsthm}

\usepackage{thmtools}
\usepackage{enumitem}

\usepackage[usenames,dvipsnames]{xcolor}
\usepackage{graphicx}
\usepackage{multirow}
\usepackage{float}
\usepackage{color}

\usepackage{xparse}

\usepackage[labelformat=simple]{subcaption}

\usepackage{todonotes}

\usepackage[hidelinks]{hyperref}
\hypersetup{
    colorlinks,
    linkcolor={black!100!black},
    citecolor={blue!100!black},
    urlcolor={blue!50!black}
}

\usepackage[normalem]{ulem}

\usepackage{pdflscape}

\usepackage{etoolbox}

\makeatletter
\def\@mkboth#1#2{}
\newlength\appendixwidth
\preto\appendix{\addtocontents{toc}{\protect\patchl@section}}
\newcommand{\patchl@section}{%
  \settowidth{\appendixwidth}{\textbf{Appendix }}%
  \addtolength{\appendixwidth}{1.5em}%
  \patchcmd{\l@section}{1.5em}{\appendixwidth}{}{\ddt}%
}
\makeatother

\makeatletter
\newrobustcmd{\fixappendix}{%
  \patchcmd{\l@section}{1.5em}{7em}{}{}%
  \patchcmd{\l@subsection}{2.3em}{7em}{}{}%
}
\makeatother

\newcommand \expe[2]	{\mathbb{E}_{#1}\left[#2\right]}
\newcommand \gexpe[1]	{\expe{*}{#1}}
\newcommand \lexpe[1]	{\expe{rla}{#1}}

\newcommand \expet[2]	{\expe{#1}{\gamma_{#2}}}
\newcommand \gexpet[1]	{\expet{*}{#1}}
\newcommand \lexpet[1]	{\expet{rla}{#1}}
\newcommand \sexpet[1]	{\expet{rsa}{#1}}
\newcommand \pexpet[1]	{\expet{rap}{#1}}
\newcommand \expetw[1]	{\expet{#1}{\omega}}
\newcommand \gexpetw	{\expetw{*}}
\newcommand \lexpetw	{\expetw{rla}}

\newcommand \var[2]		{\mathbb{V}_{#1}\left[#2\right]}
\newcommand \gvar[1]	{\var{*}{#1}}
\newcommand \lvar[1]	{\var{rla}{#1}}

\newcommand \probdelta[1]	{\delta_{#1}}
\newcommand \gprobdelta		{\probdelta{*}}
\newcommand \lprobdelta		{\probdelta{rla}}
\newcommand \sprobdelta		{\probdelta{rsa}}
\newcommand \pprobdelta		{\probdelta{rap}}

\newcommand \probalphast[2]		{p_{#1,\:#2}}
\newcommand \gprobalphast[1]	{\probalphast{*}{#1}}
\newcommand \lprobalphast[1]	{\probalphast{rla}{#1}}
\newcommand \sprobalphast[1]	{\probalphast{rsa}{#1}}
\newcommand \pprobalphast[1]	{\probalphast{rap}{#1}}
\newcommand \probalphastw[1]	{\probalphast{#1}{\omega}}
\newcommand \gprobalphastw		{\probalphastw{*}}
\newcommand \lprobalphastw		{\probalphastw{rla}}

\newcommand \prob[1] {\mathbb{P}r\left[#1\right]}

\newcommand \arr[1] {\pi_{#1}}
\newcommand \garr {\arr{*}}
\newcommand \linarr {\arr{la}}

\newcommand \gC {C_*}
\newcommand \Crla {C_{rla}}
\newcommand \Cla {C_{la}}
\newcommand \Clamax {C_{la,max}}

\NewDocumentCommand \cycle { O{n} }{ \mathcal{C}_{#1}}
\NewDocumentCommand \lintree { O{n} }{ \mathcal{L}_{#1}}
\NewDocumentCommand \startree { O{n} }{ \mathcal{S}_{#1}}
\NewDocumentCommand \complete { O{n} }{ \mathcal{K}_{#1}}
\NewDocumentCommand \quasistar { O{n} }{ \mathcal{Q}_{#1}}
\NewDocumentCommand \compbip { O{n_1} O{n_2} }{ \mathcal{K}_{#1,#2}}

\NewDocumentCommand \zeroreg { O{n} }{ \textbf{0}_{#1}}
\NewDocumentCommand \onereg { O{n} }{ \textbf{1}_{#1}}
\NewDocumentCommand \tworeg { O{n} }{ \textbf{2}_{#1}}
\NewDocumentCommand \maxdegreg { O{n} }{ \textbf{n-1}_{#1}}
\NewDocumentCommand \kreg { O{n} }{ \textbf{k}_{#1}}

\newcommand \nsquares { n_G(\cycle[4]) }

\newcommand \mmtdeg[1] {
	\langle k^{#1} \rangle
}

\newtheoremstyle{theoremstyle}
    {15pt} % Space above
    {} % Space below
    {\itshape} % Body font
    {} % Indent amount
    {\bfseries} % Theorem head font
    {.} % Punctuation after theorem head
    {.5em} % Space after theorem head
    {} % Theorem head spec (can be left empty, meaning `normal')

\theoremstyle{theoremstyle}

\begin{document}

\title{Edge crossings in random linear arrangements}
\author{Llu\'is Alemany-Puig$^1$ \& Ramon Ferrer-i-Cancho$^1$}
\address{
$^1$ Complexity \& Quantitative Linguistics Lab, \\
Departament de Ci\`encies de la Computaci\'o, \\
Laboratory for Relational Algorithmics, Complexity and Learning (LARCA), \\
Universitat Polit\`ecnica de Catalunya, \\
Barcelona, Catalonia, Spain.
}

\begin{abstract}
In spatial networks vertices are arranged in some space and edges may cross. When arranging vertices in a 1-dimensional lattice edges may cross when drawn above the vertex sequence as it happens in linguistic and biological networks. Here we investigate the general of problem of the distribution of edge crossings in random arrangements of the vertices. We generalize the existing formula for the expectation of this number in random linear arrangements of trees to any network and derive an expression for the variance of the number of crossings in an arbitrary layout relying on a novel characterization of the algebraic structure of that variance in an arbitrary space. We provide compact formulae for the expectation and the variance in complete graphs, complete bipartite graphs, cycle graphs, one-regular graphs and various kinds of trees (star trees, quasi-star trees and linear trees). In these networks, the scaling of expectation and variance as a function of network size is asymptotically power-law-like in random linear arrangements. Our work paves the way for further research and applications in 1-dimension or investigating the distribution of the number of crossings in lattices of higher dimension or other embeddings.
\end{abstract}



\maketitle

\tableofcontents

% Section
\section{Introduction}

\begin{figure}
	\centering
	\includegraphics{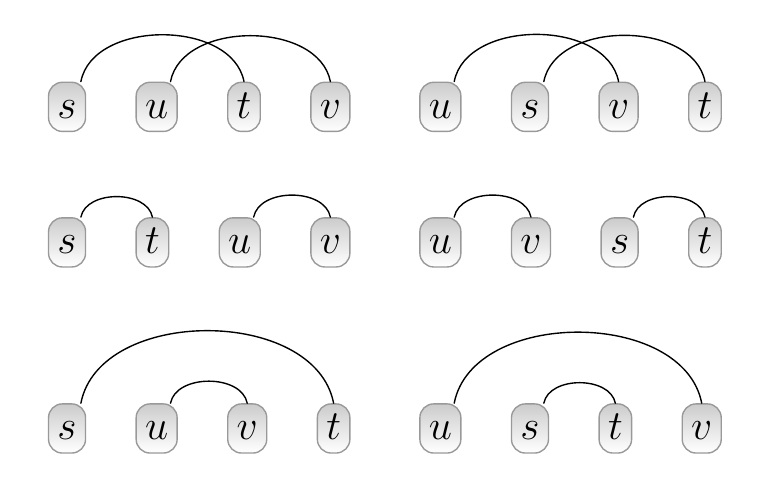}
	\caption{Linear arrangements of 4 different vertices forming two edges: $\{s,t\}$ and $\{u,v\}$.}
	\label{fig:simple_crossings}
\end{figure}

The organization of many complex systems can be described with the help of a spatial network, where nodes are embedded in some space \cite{Barthelemy2011a}. Edges may cross when vertices are arranged in some space (figure \ref{fig:simple_crossings}). A spatial graph without edge crossings is planar. In street networks, crossings are practically impossible by the construction of the network \cite{Barthelemy2011a}. In road networks, crossings typically involve bridges and tunnels \cite{Eppstein2017a}. Thus, crossings in road networks are costly and consequently scarce. Crossings are also scarce in syntactic dependency networks, networks linking the words of a sentence via syntactic dependencies \cite{Ferrer2017a}. However, whether syntactic dependency crossings are inherently costly is a matter of debate \cite{Ferrer2016a}.  

Here we study the expectation and variance of the number of edge crossings in a graph whose vertices are embedded in an arbitrary {\em space}. We study these two statistical properties in a family of layouts satisfying three requirements: (1) only independent edges can cross (edges that do not share vertices), (2) two independent edges can cross in at most one point, and (3) if several edges of the graph, say $e$ edges, cross at exactly the same point then the amount of crossings equals ${e \choose 2}=e(e-1)/2$.

In this paper, we pay special attention to the one-dimensional lattice, also known as linear arrangement, where edges may cross when drawn above the lattice, as it happens in sentences \cite{Ferrer2017a} and RNA structures \cite{Chen2009a}. In the case of RNA structures, vertices are nucleotides A, G, U, and C, while edges are Watson-Crick (A-U, G-C) and (U-G) base pairs \cite{Chen2009a}. Other examples of layouts are embeddings on the plane, where vertices represent two-dimensional points on the Euclidean $\mathbb{R}^2$ plane and edges are line segments joining their endpoints \cite{Barthelemy2018a}, and spherical arrangements, studied by Moon in \cite{Moon1965a}, where vertices are distributed on the surface of a sphere and edges joining their endpoints are the corresponding geodesic in that surface. In this paper we use the concepts {\em space}, {\em layout} and {\em arrangement} interchangeably.

The expectation and variance of the number of crossings are then denoted as $\gexpe{C}$ and $\gvar{C}$, where * is an arbitrary layout meeting the three requirements above. Then $\lexpe{C}$ and $\lvar{C}$ denote the expectation and variance of the number of crossings in uniformly random linear arrangements, henceforth {\em random linear arrangements (rla)}. For example, in trees, $\lexpe{C}$ has been shown to be $|Q|/3$, where $|Q|$ is the size of the set $Q$, the pairs of edges of a network that may potentially cross \cite{Gomez2016a}. A pair of edges belongs to $Q$ if the edges do not share vertices (or equivalently, there is at least one linear arrangement where they cross). In trees \cite{Ferrer2013b,Ferrer2013d},
\begin{equation}
\label{eq:potential_num_crossings:trees}
|Q| = \frac{n}{2}(n - 1 - \mmtdeg{2}),
\end{equation}
where $\mmtdeg{2}$ is the second moment of degree about zero and $n$ is the number of vertices. In many cases, the syntactic dependency networks of sentences are not trees \cite{Gomez2016a} and RNA secondary structures are graphs where degrees do not exceed one and are then usually disconnected \cite{Chen2009a}. 

Consider a graph of $n$ vertices and $m$ edges whose vertices are arranged with a function $\garr$ that given a vertex returns its position in the space, or layout, the graph is embedded in. Throughout this article we use letters $s$, $t$, $\cdots$, $z$ to indicate distinct vertices. It is important to bear in mind that the definition of crossing is layout-dependent. Consider the case of a one-dimensional layout. The function $\linarr$, which denotes $\garr$ for the particular case of linear arrangements, gives the position of each vertex in the sequence of length $n$. Suppose that the vertices $s$ and $t$ and the vertices $u$ and $v$ are linked. Without any loss of generality suppose that $s$ precedes $t$, i.e. $\linarr(s) < \linarr(t)$, and $u$ precedes $v$, i.e. $\linarr(u) < \linarr(v)$. Then their edges cross if and only if one of the two following conditions is met
\begin{eqnarray*}
\linarr(s) < \linarr(u) < \linarr(t) < \linarr(v) \text{, or} \\
\linarr(u) < \linarr(s) < \linarr(v) < \linarr(t).
\end{eqnarray*}
Let $\gC$ be the number of edge crossings produced by an arrangement of the vertices of a graph in a certain layout $*$. Figure \ref{fig:simple_crossings} shows a couple of linear arrangements with $\Cla = 1$ on top and linear arrangements with $\Cla=0$ below. We use simply $C$ in cases where there is no ambiguity on the layout over which $C$ is measured. For example, in $\lexpe{C}$ and $\lvar{C}$, $C=\Crla$.

In this article, we derive $\gexpe{C}$ for general graphs and investigate further aspects of the distribution of crossings under the null hypothesis of a random arrangement by means of $\gvar{C}$, the variance of $C$ in an arbitrary layout. Such knowledge of the distribution of crossings under the null hypothesis in linear arrangements has potentially many applications in biology and linguistics, e.g., it would allow one to calculate $z$-scores as in other areas of network theory research \cite{Milo2002a,Ferrer2018a}. For this reason, we establish some foundations on the general layout problem and then develop further the concrete case of a random linear arrangement.

The current article is a piece of a broader research program on the statistical properties of measures on linear arrangements. Recently, the distribution of $D$, the sum of edge lengths in random linear arrangements, has been investigated \cite{Ferrer2018a}. The present article can be seen as a continuation where $D$ is replaced by $C$, bearing in mind that the analysis of $C$ is more complex. Research on such a program can be classified according to the target: $D$ \cite{Ferrer2004b,Esteban2016a,Ferrer2018a}, $C$ \cite{Ferrer2017a} or the interplay between $D$ and $C$ \cite{Ferrer2006d,Ferrer2014c,Ferrer2015c,Gomez2016a}. In some works, the two aspects are considered simultaneously \cite{Ferrer2013b,Ferrer2014f}.

The remainder of the article is organized as follows. Section \ref{sec:crossing_theory} presents the number of crossings in a linear arrangement of complete graphs, reviews the concept of $Q$, extending it to general graphs, and investigates graphs that minimize or maximize $|Q|$. It also introduces the specific graphs for which compact formulae of $\gexpe{C}$ and $\gvar{C}$, and hence for $\lexpe{C}$ and $\lvar{C}$, are derived in subsequent sections. Section \ref{sec:exp_C} presents a general expression for $\gexpe{C}$ in general graphs as well as compact formulae for specific graphs. Section \ref{sec:var_C} analyses the mathematical structure of $\gvar{C}$ providing a general expression for it. The variance turns out to depend only on the frequency of seven distinct types of subgraphs. These seven subgraphs are particular cases of graphettes, possibly disconnected substructures within a network \cite{Hasan2017a}. Thus our work is related to research on meaningful substructures, i.e. motifs, in complex networks \cite{Milo2002a}. Section \ref{sec:theoretical_examples} provides compact formulae for specific graphs. Section \ref{sec:discussion} summarizes and discusses the findings of previous sections and suggests various possibilities for further research.

% Section
\section{A mathematical theory of crossings}
\label{sec:crossing_theory} 

Henceforth, we assume the requirements on the layouts above. Obviously, we have that
\begin{eqnarray*}
\gC \leq P,
\end{eqnarray*}
where $P$ is the number of different pairs of edges that can be formed, i.e. 
\begin{eqnarray*} 
P = {m \choose 2}.
\end{eqnarray*}  
In $\complete$, i.e. a complete graph of $n$ vertices,
\begin{eqnarray*}
%\label{eq:edges_complete_graph}
m = {n \choose 2},
\end{eqnarray*}
and then the number of pairs of edges that can be formed is
\begin{eqnarray}
\label{eq:pairs_of_edges_complete_graph}
P(\complete) & = & {\frac{n(n-1)}{2} \choose 2} \nonumber \\
             & = & \frac{1}{8}(n+1)n(n-1)(n-2).
\end{eqnarray}
As $m$ is maximum in a complete graph, we have that

\begin{eqnarray*}
\gC \leq P(\complete).
\end{eqnarray*}

We show that the actual number of crossings of a complete graph in a linear arrangement is actually 3 times smaller than $P(\complete)$.

\subsection{The number of crossings in a linear arrangement of a complete graph}

In complete graphs, the number of crossings does not depend on the linear arrangement because all vertices have maximum degree. Therefore we can refer to the number of crossings of a complete graph without specifying the linear arrangement that produces it. 

In a linear arrangement, we define the shadow of an edge as the vertices that are placed in-between the endpoints of that edge. The length of an edge $\{u,v\}$ is $d=|\linarr(u) - \linarr(v)|$, which satisfies $1 \leq d \leq n - 1$. Then, in an arbitrary graph, $f(d)$, the number of edges of length $d$, satisfies \cite{Ferrer2018a}
\begin{eqnarray}
\label{eq:maximum_number_of_edges_of_given_length}
f(d) \leq f_{max}(d) = n - d,
\end{eqnarray}
and $\Cla(d)$, the number of edges that cross an edge of length $d$ in a linear arrangement, satisfies \cite{Ferrer2013b}
\begin{eqnarray}
\Cla(d) \leq \Clamax(d) = (d-1)(n-d-1).
\label{eq:maximum_number_of_crossings_with_edge_of_given_length}
\end{eqnarray}
Notice that $d-1$ is the number of vertices of the shadow of an edge of length $d$ and $n - d - 1$ is the number of vertices excluding the vertices in the shadow and the vertices of the edge. Therefore, $\Cla(d)$ cannot exceed $(d-1)(n-d-1)$. In addition, the number of crossings satisfies 

\begin{eqnarray*}
\Cla \leq \frac{1}{2} \sum_{d=1}^{n-1} f_{max}(d) \Clamax(d).
\end{eqnarray*} 
As

\begin{eqnarray}
\label{eq:raw_number_of_crossings_of_complete_graph}
\Cla(\complete) = \frac{1}{2} \sum_{d=1}^{n-1} f_{max}(d) \Clamax(d),
\end{eqnarray}
$\Cla$ is maximized by a complete graph. 

Applying equations \ref{eq:maximum_number_of_edges_of_given_length} and \ref{eq:maximum_number_of_crossings_with_edge_of_given_length} to equation \ref{eq:raw_number_of_crossings_of_complete_graph}, one obtains
\begin{eqnarray*}
\Cla(\complete) & = & \frac{1}{2} \sum_{d=1}^{n-1} (n-d) (d-1)(n-d-1) \\ 
   & = & \frac{1}{24} n(n-1)(n-2)(n-3)
\end{eqnarray*}
for $n \geq 3$. Noting that $\gC = 0$ for $n < 4$, we get  
\begin{eqnarray}
\label{eq:number_of_crossings_of_complete_graph}
\Cla(\complete) = {n \choose 4}
\end{eqnarray}
for an arbitrary $n$. The same value of $\Cla(\complete)$ has been derived recently using a different approach \cite{Chimani2018a}. Equations \ref{eq:number_of_crossings_of_complete_graph} and  \ref{eq:pairs_of_edges_complete_graph} allow one to calculate the ratio
\begin{eqnarray*}
\frac{P(\complete)}{\Cla(\complete)} = \frac{3(n+1)}{n-3}.
\end{eqnarray*}
Notice that
\begin{eqnarray*}
\lim_{n\rightarrow\infty} \frac{P(\complete)}{\Cla(\complete)} = 3.
\end{eqnarray*}
In sum, taking into account the spatial constraints of linear arrangements, it turns out that the actual number of crossings in a complete graph is about 3 times smaller than its number of edge pairs. 

\subsection{The potential number of crossings of a graph}

A pair of edges belongs to the set $Q$ if and only if there is at least one arrangement where the two edges cross \cite{Gomez2016a}.  Obviously, 
\begin{eqnarray*}
%\label{eq:lower_and_upper_bound_of_potential_number_of_crossings}
\gC \leq |Q|.
\end{eqnarray*}
where $|Q|$ is the cardinality of $Q$. As stated in the introduction, $|Q|$ can be defined equivalently as the number of distinct pairs of independent edges of a graph \cite{Piazza1991a}. Two edges are independent, or disjoint, if and only if they do not have a common endpoint. Then $|Q|$ can be easily derived as the difference between the number of pairs of edges that can be formed, i.e. ${m \choose 2}$, and the number of dependent pairs of different edges produced by every edge. Since a vertex $s$ of degree $k_s$ produces ${k_s \choose 2}$ dependent edges, we have that \cite{Piazza1991a}
\begin{eqnarray*}
%\label{eq:Piazza_et_al_definition}
|Q| = {m \choose 2} -  \sum_{s = 1}^n {k_s \choose 2},
\end{eqnarray*}
which is equivalent to \cite{Ferrer2018a}
\begin{eqnarray}
\label{eq:potential_number_of_crossings}
|Q| = \frac{1}{2} \left[ m(m+1)- n \left< k^2 \right> \right].
\end{eqnarray}
Assuming that $m = n - 1$, e.g., a tree, equation \ref{eq:potential_number_of_crossings} gives equation \ref{eq:potential_num_crossings:trees}, which has already been derived for the particular case of trees \cite{Ferrer2013b,Ferrer2013d}.

The class of $k$-regular graphs, denoted as $\kreg$, is the class formed by all graphs of $n$ nodes where each node has degree $k$ \cite[p. 4]{Bollobas1998a} (and also \cite{Wilson1996a}). For the sake of brevity, we briefly refer to a graph in the class $\kreg$ simply as a graph $\kreg$ or $k$-regular graph. In a $k$-regular graph, $\mmtdeg{} = k$, $\mmtdeg{2}=k^2$ and $m = kn/2$. Thus equation \ref{eq:potential_number_of_crossings} becomes
\begin{eqnarray}
\label{eq:potential_number_of_crossings_d_regular_graph}
|Q(\kreg)|
	&=& \frac{1}{2}
		\left[ \frac{kn}{2}\left(\frac{kn}{2} + 1\right) - n k^2 \right] \nonumber \\
	&=& \frac{1}{8}kn(k(n-4)+2).
\end{eqnarray}
In a complete graph, $k = n - 1$ and then
\begin{eqnarray*}
|Q(\complete)| = \frac{1}{8}n(n-1)(n-2)(n-3)
\end{eqnarray*}
for $n \geq 3$, in agreement with previous work \cite{Ferrer2018a}. Noting that $|Q|=0$ for $n < 4$, we obtain
\begin{eqnarray}
\label{eq:potential_number_of_crossings_complete_graph}
|Q(\complete)| =  3 {n \choose 4}
\end{eqnarray}
for an arbitrary $n$. Recalling equation \ref{eq:number_of_crossings_of_complete_graph}, it turns out that
\begin{eqnarray}
\label{eq:potential_number_of_crossings_versus_actual_number_of_crossings_complete_graph}
\Cla(\complete) = \frac{|Q(\complete)|}{3}.
\end{eqnarray}
Equation \ref{eq:potential_number_of_crossings_complete_graph} is actually equivalent to one derived in previous work \cite{Moon1965a}, i.e. 
\begin{eqnarray*}
|Q(\complete)| = \frac{1}{2} {n \choose 2} {n - 2 \choose 2}.
%\label{eq:potential_number_of_crossings_complete_graph_Moon}
\end{eqnarray*}

Next subsections are concerned about the maxima and the minima of $|Q|$. These are not only relevant for crossing theory {\em per se} but also because $\gexpe{C}$ is proportional to $|Q|$, as it is shown in section \ref{sec:exp_C}.

\subsection{When is $|Q|$ maximum?}

$\gC$, the number of crossings of a graph in some layout *, can be defined as
\begin{eqnarray*}
\gC = \frac{1}{2} \sum_{st\in E} \gC(s,t),
\end{eqnarray*}
where $\gC(s,t)$ is the number of edge crossings involving edge $st$.

$\gC(s,t)$ cannot exceed $q(s,t)$, the potential number of crossings of the edge formed by $s$ and $t$, namely the number of edges that do not share a vertex with the pair $(s,t)$.
It is easy to see that 
\begin{eqnarray}
\label{eq:crossings_of_1_edge_upper_bound}
q(s,t) = m - k_s - k_t + 1,
\end{eqnarray}
%where $k_t$ is the degree of vertex $t$
(see \ref{crossings_of_edge_appendix} for a detailed proof). Equation \ref{eq:crossings_of_1_edge_upper_bound} is actually a generalization of a previous result, i.e. 
\begin{eqnarray*}
q(s,t) = n - k_s - k_t
\end{eqnarray*}
for trees, where $m = n - 1$ \cite{Ferrer2013d}.

Let $A = \left\{ a_{st}\right\}$ be the adjacency matrix of a graph, i.e. $a_{st} = 1$ if vertices $s$ and $t$ are connected and $a_{st} = 0$ otherwise. Then $|Q|$ can be defined equivalently as
\begin{eqnarray}
\label{eq:preliminary_potential_number_of_crossings}
|Q| &=& \frac{1}{2} \sum_{st\in E} q(s,t) \nonumber \\
    &=& \frac{1}{4} \sum_{s =1}^n \sum_{t=1}^n a_{st}q(s,t).
\end{eqnarray}
Applying equation \ref{eq:crossings_of_1_edge_upper_bound} to equation \ref{eq:preliminary_potential_number_of_crossings}, we get
\begin{eqnarray}
\label{eq:raw_potential_number_of_crossings}
|Q| = \frac{1}{4} \sum_{s =1}^n \sum_{t=1}^n a_{st}(m + 1 - k_s - k_t).
\end{eqnarray}
The fact that $a_{st}\leq 1$ if $s \neq t$ and $a_{tt} = 0$, transforms equation \ref{eq:raw_potential_number_of_crossings} into
\begin{eqnarray*}
|Q| & \leq & \frac{1}{4} \sum_{s =1}^n \sum_{t=1, t \neq s}^n (m + 1 - k_s - k_t) \nonumber \\
    & =    & \frac{1}{4} \sum_{s =1}^n \sum_{t=1, t \neq s}^n (m + 1) - \frac{1}{2}\sum_{s =1}^n \sum_{t=1, t \neq s}^n k_t \nonumber \\
    & =    & \frac{1}{4}n(n - 1)(m  +1) - m(n - 1) \nonumber\\
    & =    & \frac{n - 1}{4}( (n - 4)m + n ).
\end{eqnarray*}
Replacing $m$ by its maximum value, namely that of a complete graph, we finally obtain
\begin{eqnarray*}
|Q| & \leq & \frac{1}{8}n(n-1)(n-2)(n-3) \\
    & = & |Q(\complete)|.
\end{eqnarray*}
Therefore, $|Q|$ is maximized by a complete graph. 

\subsection{When is $|Q|$ minimum?}
\label{sec:minima_of_potential_number_of_crossings}

\begin{figure}
	\centering
	\includegraphics[scale=0.5]{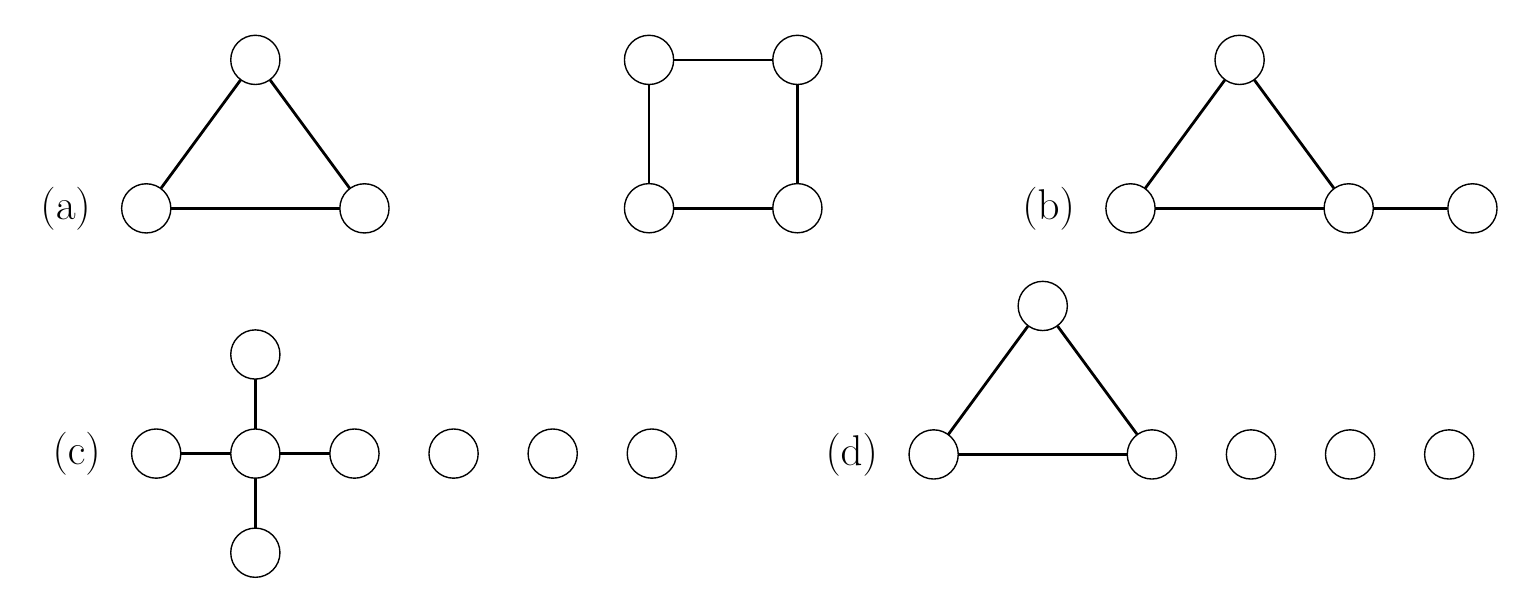}
	\caption{Drawings of simple graphs. (a) cycle graphs, (b) a paw (a triangle with a node attached), (c) a star tree with isolated vertices, and (d) a cycle with isolated vertices. }
	\label{fig:simple_graphs}
\end{figure}

In addition to $\complete$ and $\kreg$, we use specific notation to refer to other kinds of graphs: $\startree$ for a star tree of $n$ vertices, $\cycle$ for a cycle graph of $n$ vertices (figure \ref{fig:simple_graphs}(a)). These graphs are related: e.g. $\maxdegreg[n] = \complete[n]$, $\complete[3] = \cycle[3]$ and $\cycle[n]$ is a kind of $\tworeg[n]$. Let $\oplus$ denote the disjoint union of graphs, i.e. $G \oplus G'$ is the graph formed by two graphs, $G$ and $G'$, that do not share vertices \cite{Rosen2017a}. For instance, $\complete[2] \oplus \complete[2]$, a graph formed by two independent edges, is isomorphic to $\onereg[4]$. Isolated vertices, namely vertices of degree zero, are also referred as unlinked vertices. 
 
Asking when $|Q|$ is minimum is equivalent to asking when $|Q| = 0$ in equation \ref{eq:potential_number_of_crossings}, which is in turn equivalent to 
\begin{eqnarray}
\mmtdeg{2} = \frac{m(m+1)}{n}.
\label{zero_eq:potential_number_of_crossings}
\end{eqnarray}
The minima of $|Q|$ cannot have crossings ($\gC \leq |Q| = 0$). Thus these minima are trivially a subset of outerplanar graphs (a graph is outerplanar if and only if its book thickness is one \cite{Bernhart1979a}). The graphs satisfying equation \ref{zero_eq:potential_number_of_crossings} are actually sparser. Due to being outerplanar, the number of edges of the minima of $|Q|$ must satisfy \cite{Harary1969a}
\begin{eqnarray*}
m \leq 2 n - 3.
\end{eqnarray*}
% with equality if an only if the graph is a maximal outerplanar graph
% . An outerplanar graph is maximal outerplanar if it is
% not possible to add an edge such that the resulting graph is still
% outerplanar. 
Indeed, now we show that the graphs where $|Q| = 0$ are a subset of outerplanar graphs whose members satisfy
\begin{eqnarray*}
m \leq n 
\end{eqnarray*}
with equality if and only if $n = 3$.

We derive the kinds of graphs where $|Q|=0$. The derivation is based on the following principle. Let  $G'$ be a subgraph of a graph $G$ and let $Q'$ be the set of pairs of independent edges of $G'$. If $|Q'|>0$ then $|Q|>0$. Two kinds of graphs are vital for the derivation.  First, cycle graphs, where all vertices have degree 2 and then $\mmtdeg{2} = 4$ for $n\geq 3$  (for $n<3$ a regular graph with vertices of degree 2 cannot be formed) and $m = n$. Applying these two properties to equation \ref{eq:potential_number_of_crossings}, one gets 
\begin{equation}
|Q(\cycle)| = \frac{1}{2}n(n-3)  
\label{eq:size_of_Q_cycle}
\end{equation}
for $n \geq 3$. Namely, $|Q(\cycle[3])|=0$ and $|Q(\cycle)|>0$ for $n>3$. The other kind of graph is a paw, namely a triangle (i.e. a cycle graph of three vertices) with a leaf attached to it (figure \ref{fig:simple_graphs}(b)) \cite{isgci}, that has $|Q(paw)| = 1$ \footnote{A paw graph has $m= 4$ and $\mmtdeg{2}=9/2$. Applying these properties to equation \ref{eq:potential_number_of_crossings}, one gets $|Q(paw)| = 1$, as expected.}.
 
The derivation is as follows:
\begin{itemize}
\item A graph $G$ where $|Q|=0$ may have more than one connected component but must have all edges concentrated on one of the connected components. If a graph has two edges, $e_1$ and $e_2$, from different connected components, then $|Q|>0$ because $\{e_1, e_2\} \in Q$ ($e_1$ and $e_2$ cannot share vertices due to being in different components).

\item In a graph $G$ where all edges belong to just one of the connected components, whether $|Q|=0$ or not is determined by such connected component. Let $G'$ be the subgraph induced by the largest connected component of $G$. 
	\begin{itemize}
	\item Suppose that $G'$ is a tree. Then $|Q'| = 0$ if and only if $G'$ is a star tree \cite{Ferrer2013d}. 

	\item If $G'$ is not a tree then it must be a connected graph with cycles. There are only three possibilities:
		\begin{itemize}
		\item $G'$ has a cycle of 4 or more vertices and then $|Q'|>0$ borrowing the results above on cycle graphs. 

		\item $G'$ is just a cycle of three vertices and then $|Q'|=0$.

		\item $G'$ contains a cycle of three vertices and additional vertices. Then $|Q|>0$ because it contains the paw graph (figure \ref{fig:simple_graphs}(b)).
		\end{itemize}
	Notice that if $G'$ does not have any cycle then it has to be a tree because it is connected. Therefore, the only cyclic $G'$ that has $|Q| = 0$ is a $\cycle[3]$.
	
	\end{itemize}
\end{itemize}

We conclude that $|Q|=0$ is only possible in two kinds of graphs,
\begin{enumerate}
\item $\startree[\lambda] \oplus \zeroreg[n-\lambda]$, a forest consisting of a star tree of $\lambda$ vertices and a series of $n-\lambda$ unlinked vertices (figure \ref{fig:simple_graphs}(c)).

\item $\complete[3] \oplus \zeroreg[n-3]$, a graph consisting of a complete graph of three vertices (namely a cycle graph of three vertices) and $n-3$ isolated vertices (figure \ref{fig:simple_graphs}(d)).
\end{enumerate}

We check the condition in equation \ref{zero_eq:potential_number_of_crossings} is satisfied by $\startree[\lambda] \oplus \zeroreg[n-\lambda]$, 
% that has $l$ connected components and $m = n - l$ edges.
% When $l=1$, one has a star tree and when $l = n$ one has $n$ unlinked vertices. 
that has a hub vertex of degree $m=\lambda-1$, $m$ vertices of degree 1 and $n - m - 1$ isolated vertices. Therefore,
\begin{eqnarray*}
\mmtdeg{2}(\startree[\lambda] \oplus \zeroreg[n-\lambda])
	&=& \frac{1}{n} \sum_{s=1}^{n} k_s^2 \\ %\nonumber \\
	&=& \frac{1}{n} \lambda(\lambda - 1)
%\label{eq:degree_2nd_moment_of_forest}
\end{eqnarray*}
as expected by equation \ref{zero_eq:potential_number_of_crossings}. It is also easy to check that $\complete[3] \oplus \zeroreg[n-3]$ satisfies equation \ref{zero_eq:potential_number_of_crossings} because $m = 3$ and $\mmtdeg{2} = 12/n$.

Now let us derive a tight upper bound of $m$ for graphs where $|Q| = 0$. For $\startree[\lambda] \oplus \zeroreg[n-\lambda]$ we have $m \leq n - 1$ while  for $\complete[3] \oplus \zeroreg[n-3]$ we have $m = 3$. Thus we have 
\begin{eqnarray*}
m \leq \left\{ 
   \begin{array}{ll} 
      n - 1                   & \mbox{if~} n < 3 \\
      \max(n-1, 3) & \mbox{if~} n \geq 3
   \end{array} 
   \right.
\end{eqnarray*} 
and then 
\begin{eqnarray*}
m \leq n - 1,
\end{eqnarray*}
except when $n=3$, where we have $m=3$.

\subsection{Theoretical graphs}
\label{sec:theoretical_graphs}

In this article we consider a series of specific graphs. Their interest is that compact or simpler formula for $\gvar{C}$ is easy to derive for the majority of them.

First, complete graphs. They are interesting for various reasons. They maximize $m$, $\mmtdeg{2}$, $C_{rla}$ and $|Q|$; $C_{rla}$ is constant, $C_{rla} < |Q|$ (for $n > 3$) and $\lvar{C}=0$ because $C$ is constant. In addition, they are chosen in many contexts for the ease with which they allow one to obtain theoretical results, e.g. random layouts on the surface of a sphere \cite{Moon1965a}, neural networks \cite{Beren2015a} or social dynamics \cite{Catellano2009a}.

Second, complete bipartite graphs, $\compbip$, where $n_1$ and $n_2$ are the number of vertices of each partition. They are relevant for the close relationship between the present article and previous work on random layouts on the surface of a sphere \cite{Moon1965a}. It is easy to see that 
\begin{eqnarray*}
\mmtdeg{2}(\compbip)= m = n_1n_2.
\end{eqnarray*}
Applying this result to equation \ref{eq:potential_number_of_crossings}, one obtains
\begin{eqnarray*}
%\label{eq:comp-bip-graphs:Q-size}
|Q(\compbip)| = \frac{1}{2}n_1n_2(n_1 - 1)(n_2 - 1) = 2{n_1 \choose 2}{n_2 \choose 2}
\end{eqnarray*}
after some routine work.

Third, $\startree[\lambda] \oplus \zeroreg[n-\lambda]$. In these graphs $m=\lambda-1$. They are interesting because $\gC(\startree[\lambda] \oplus \zeroreg[n-\lambda])=|Q(\startree[\lambda] \oplus \zeroreg[n-\lambda])|=0$ (section \ref{sec:minima_of_potential_number_of_crossings}).

\begin{figure}
	\centering
	\includegraphics{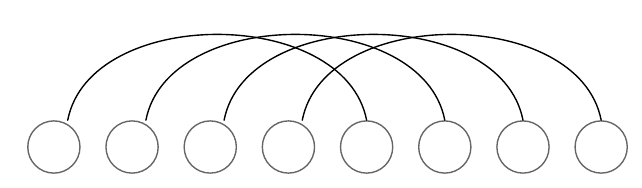}
	\caption{Linear arrangement of a 1-regular graph with $n=8$ that maximizes the number of crossings ($C_{rla} = |Q| = 6$).}
	\label{fig:1_regular_graph}
\end{figure}

Fourth, $\onereg$, one-regular graphs. By definition, $m = n/2$ and $n$ must be even. Trivially, $\mmtdeg{2} = 1$. These graphs are relevant because all edges are independent and then
\begin{eqnarray*}
|Q(\onereg)| = {m \choose 2},
\end{eqnarray*}
which in turn implies that they maximize $|Q|$ given $m$. The fact that $m = n/2$ gives 
\begin{eqnarray*}
|Q(\onereg)| = {n/2 \choose 2}.
\end{eqnarray*}
We could have reached the same conclusion applying $k=1$ to equation \ref{eq:potential_number_of_crossings_d_regular_graph}. 

Given an edge $\{u, v\}$, the initial position of the edge in a linear arrangement is $min(\linarr(u), \linarr(v))$. Recall that its length is $|\linarr(u)-\linarr(v)|$. One-regular graphs achieve maximum $\Cla$ ($\Cla(\onereg) = |Q(\onereg)|$) when the initial positions of each edge are consecutive and all edges have the same length, i.e. the length is $m=n/2$ (figure \ref{fig:1_regular_graph}). 

One-regular graphs are not the only graphs where all edges are independent. Indeed, the class of graphs where all edges are independent is formed by forests that result from the combination of a 1-regular graph (that could be empty) with an arbitrary number of unlinked vertices, i.e. $\onereg[n_1] \oplus \zeroreg[n_2]$ with $n = n_1 + n_2$. For simplicity, we restrict our analyses to pure 1-regular graphs.

Fifth, trees because they are involved in spatial networks that have received a lot of attention \cite{Barthelemy2018a,Liu2017a}. In this article we pay specific attention to kinds of trees that are relevant in crossing theory of trees \cite{Ferrer2013b,Ferrer2014f}:
\begin{itemize}
\item Star trees, $\startree$. They are a special case of $\startree[\lambda] \oplus \zeroreg[n-\lambda]$ and then $|Q|$ is minimum, i.e. $|Q(\startree)|=0$, which in turn implies $\gC(\startree)=0$ thanks to the assumption that only pairs of independent edges can cross.
In addition,  
\begin{eqnarray*}
\mmtdeg{2}(\startree) = n - 1
\end{eqnarray*}
is maximum among all trees with same $n$ \cite{Ferrer2013b}.

\item Quasi-star trees, $\quasistar$, a graph in which $n-1$ of the vertices form a star tree and the $n$-th vertex forms an edge with one of the vertices but the central \cite{Ferrer2014f}. They are interesting because among all trees with same $n$,
\begin{eqnarray*} 
|Q(\quasistar)| = n - 3
\end{eqnarray*}
is the smallest non-zero value of $|Q|$ while 
\begin{eqnarray*}
\mmtdeg{2}(\quasistar) = \frac{1}{n}(n^2-3n+6)
\end{eqnarray*}
is the second largest possible value of $\mmtdeg{2}$ \cite{Ferrer2014f}. 

\item Linear trees, $\lintree$. They are interesting because, among all trees with same $n$, 
\begin{eqnarray*}
|Q({\lintree})|={n - 2 \choose 2}
\end{eqnarray*}
is maximum, and 
\begin{eqnarray}
\label{eq:2nd_mmt_deg:linear_tree}
\mmtdeg{2}(\lintree) = 4 - \frac{6}{n}
\end{eqnarray}  
is minimum \cite{Ferrer2013b}. 
\end{itemize}
Sixth, $\cycle$, cycle graphs of $n$ vertices. They are interesting for being cyclic graphs with only one cycle, as opposed to complete graphs, where the number of cycles is maximized. $|Q(\cycle)|$ is found in equation \ref{eq:size_of_Q_cycle}.
% As cycle graphs are $\tworeg$, equation \ref{eq:potential_number_of_crossings_d_regular_graph} with $k=2$ gives
% \begin{eqnarray*}
% |Q(\cycle)| = \frac{1}{2}n(n-3).
% \end{eqnarray*}   
Notice that cycle graphs are interesting {\em a priori} for being like a linear tree but with ``periodic boundary conditions'', as in lattice field theory \cite{Smit2002a}. Indeed, we recycle calculations for cycle graphs to derive $\gvar{C}$ in linear trees in a straightforward fashion in section \ref{sec:linear_trees}.

Table \ref{table:special_graphs_summary} summarizes the properties of these special graphs above. Their values of $\mmtdeg{2}$ and $|Q|$ have been presented above (either derived in this article or borrowed from previous work). The values of $\gexpe{C}$ and $\gvar{C}$ are derived in the coming sections.  

\fulltable{\label{table:special_graphs_summary} A summary of the properties of the special graphs considered in this article as a function of $n$, the number of vertices. $\mmtdeg{2}$ is the average of squared degrees, $|Q|$ is the number of pairs of independent edges; $\lexpe{C}$ and $\lvar{C}$ are, respectively, the expectation and the variance of the number of crossings $C$ in uniformly random linear arrangements.}
\br
Graph                           & $\mmtdeg{2}$ 			& $|Q|$ 			& $\lexpe{C}$				& $\lvar{C}$ \\
\mr 
$\complete$                  	& $(n-1)^2$ 			& $3{n \choose 4}$ 	& ${n \choose 4}$ 				& 0 \\
$\compbip$        				& $n_1 n_2$				& $2{n_1 \choose 2}{n_2 \choose 2}$ 
																			& $\frac{2}{3}{n_1 \choose 2}{n_2 \choose 2}$
																											& 
% Eq. \ref{eq:var-bip-graphs} \\ 
$\frac{1}{90}{n_1 \choose 2}{n_2 \choose 2}((n_1 + n_2)^2 + n_1 + n_2)$ \\

$\onereg$						& $1$					& ${n/2 \choose 2}$	& $\frac{1}{3}{n/2 \choose 2}$ 	& $\frac{1}{360}(n - 2)n(n + 6)$ \\  
$\cycle$						& $4$					& $\frac{n(n-3)}{2}$& $\frac{n(n-3)}{6}$			& $\frac{1}{90}n(2n^2-n-30)$ \\  
$\startree[\lambda] \oplus \zeroreg[n-\lambda]$
								& $\frac{1}{n} \lambda(\lambda - 1)$
														& 0 				& 0 							& 0 \\
$\startree$						& $n -1$				& 0					& 0								& 0 \\
$\quasistar$					& $\frac{1}{n}(n^2-3n+6)$
														& $n-3$				& $\frac{n}{3}-1$				& $\frac{1}{18}n(n-3)$ \\
$\lintree$						& $4-6/n$				& ${n-2 \choose 2}$	& $\frac{1}{3}{n-2 \choose 2}$	& $\frac{1}{90}n(2n^3-5n^2-22n+60)$ \\
\br
\endfulltable

% Section
\section{The expected number of crossings of a random arrangement}
\label{sec:exp_C}

Writing any edge $\{u,v\}$ as $uv$, $\gC$, the number of crossings of a graph for a concrete arrangement $\garr$ of its vertices can be defined as 
\begin{eqnarray}
\label{eq:crossings_over_Q}
\gC = \sum_{\{st, uv\} \in Q} \alpha(st, uv)
\end{eqnarray}
where $\alpha(st,uv)$ is an indicator variable such that $\alpha(st,uv)=1$ if the edges $st$ and $uv$ cross, and $\alpha(st,uv)=0$ otherwise. The expectation of $\gC$ in a random arrangement of a graph is 

\begin{eqnarray}
\label{eq:ra:exp_C}
\gexpe{C}
	&=& \gexpe{\sum_{\{st, uv\} \in Q} \alpha(st, uv)} \nonumber\\
	&=& \sum_{\{st, uv\} \in Q} \gexpe{\alpha(st, uv) } \nonumber\\
	&=& |Q|\gprobdelta,
\end{eqnarray} 
where
\begin{eqnarray}
\label{eq:ra:prob_indep_edge_crossing}
\gprobdelta = \gexpe{\alpha(st, uv)}
\end{eqnarray}
is the probability that two independent edges cross in an arbitrary layout $*$. In the particular case of random linear arrangements, the probability $\lprobdelta$ is easy to calculate. The outline of such a calculation follows. Assume, without loss of generality, that $s$ precedes $t$ and $u$ precedes $v$ in a given linear arrangement. Then the edges formed by $s$ and $t$ and by $u$ and $v$ cross only in two relative orderings out of ${4 \choose 2}$ as illustrated in figure \ref{fig:simple_crossings}. Therefore \cite{Ferrer2013d}

\begin{eqnarray}
\label{eq:prob_delta:rla}
\lprobdelta
	= \lexpe{\alpha(st, uv)}
	= \prob{\alpha(st, uv)=1}
	= \frac{2}{{4 \choose 2}}
	= \frac{1}{3}
\end{eqnarray}

and finally
\begin{eqnarray}
\label{eq:rla:exp_C}
\lexpe{C} = \frac{|Q|}{3}.
\end{eqnarray}
Note that, in a complete graph, the number of crossings in a linear arrangement does not depend on the ordering of the vertices, therefore
\begin{eqnarray*}
\lexpe{C} = \Cla(\complete).
\end{eqnarray*}
Combining the previous equation with equation \ref{eq:potential_number_of_crossings_versus_actual_number_of_crossings_complete_graph}, we confirm that equation \ref{eq:rla:exp_C} holds in complete graphs as expected. Knowing $|Q|$ and applying equation \ref{eq:ra:exp_C}, obtaining the value of $\gexpe{C}$ for each of the special graphs considered in this article is straightforward given their already known values of $|Q|$ (table \ref{table:special_graphs_summary}), and provided $\gprobdelta$ is known.

In the one-dimensional layout, combining equation \ref{eq:potential_num_crossings:trees} and equation \ref{eq:rla:exp_C} one recovers the value of $\lexpe{C}$ that has been obtained previously for trees \cite{Ferrer2013d}, i.e. 
\begin{eqnarray*}
%\label{eq:expected_number_of_crossings_trees}
\lexpe{C} = \frac{n}{6} \left( n-1 - \mmtdeg{2} \right).
\end{eqnarray*}

\subsection{The scaling of $\lexpe{C}$ as a function of $n$ in uniformly random linear arrangements }

\begin{figure}
	\centering
	\includegraphics[scale = 0.8]{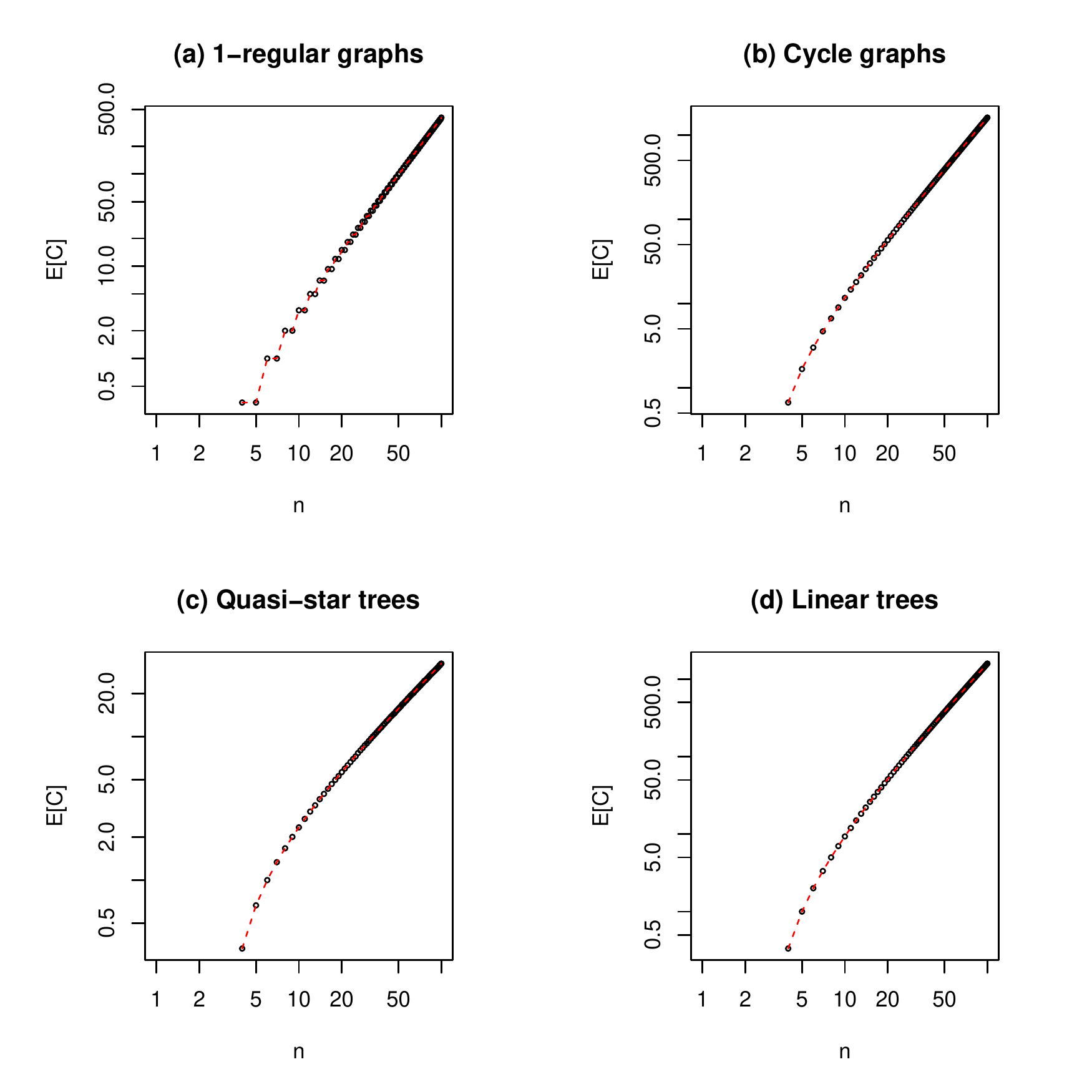}
	\caption{The mean of $C$, the number of crossings in random linear arrangements, as a function of $n$, the number of vertices of the graph. For every $n$, the mean $C$ is estimated over $T$ linear arrangements, i.e. all the $T=n!$ distinct arrangements for $n \le n^* = 10$ and $T=10^5$ uniformly random linear arrangements for $n > n^*$ (circles). Thus, the mean $C$ matches $\lexpe{C}$, the theoretical expectation, when $n \leq n^*$, and estimates it when $n > n^*$. $\lexpe{C}$, according to table \ref{table:special_graphs_summary} is also shown (dashed red line). As the scale is log-log, values with $n<4$ are not shown because $\lexpe{C}=0$.}
	\label{fig:expected_number_of_crossings}
\end{figure}

Figure \ref{fig:expected_number_of_crossings} shows $\lexpe{C}$ as a function of $n$ for the special graphs where $\lexpe{C}$ depends only on the number of vertices of the graph. Notice that $\lexpe{C}$ is constant in complete graphs and in $\startree[\lambda] \oplus \zeroreg[n-\lambda]$. In complete bipartite graphs $\compbip$, $|Q|$ depends on both $n_1$ and $n_2$. According to table \ref{table:special_graphs_summary}, $\lexpe{C}$ is expected to scale as $\sim n^{\gamma}$, with $\gamma = 1$ for quasi-star trees and $\gamma = 2$ for the remainder of graphs in figure \ref{fig:expected_number_of_crossings}. Figure \ref{fig:expected_number_of_crossings} shows the agreement between $\lexpe{C}$ and numerical estimates.

% Section
\section{The variance of the number of crossings of a random arrangement}
\label{sec:var_C}

By definition, $\gvar{C}$, the variance of $C$ in a random arrangement, is 

\begin{eqnarray*}
\gvar{C} = \gexpe{(C - \gexpe{C})^2}.
\end{eqnarray*}

We now present a derivation of $\gvar{C}$ inspired by the derivation of the variance of $C$ of the arrangement of the vertices of a complete graph on the surface of a sphere \cite{Moon1965a}. Recall that the number of crossings (equation \ref{eq:crossings_over_Q}) can be expressed as
\begin{eqnarray*}
\gC = \sum_{\{st, uv\} \in Q} \alpha(st, uv).
\end{eqnarray*}
It is easy to see that
\begin{eqnarray*}
\gC - \gexpe{C} = \sum_{\{st, uv\} \in Q} \beta(st, uv)
\end{eqnarray*}
with
\begin{eqnarray*}
\beta(st, uv) = \alpha(st, uv) - \gprobdelta,
\end{eqnarray*}
where $\gprobdelta$ is defined in equation \ref{eq:ra:prob_indep_edge_crossing}, and then

\begin{eqnarray*}
\gvar{C} = \gexpe{\left(\sum_{\{st, uv\} \in Q} \beta(st, uv)\right)^2}.
\end{eqnarray*}

Expanding the previous expression, one finds that $\gvar{C}$ can be decomposed into a sum of $|Q \times  Q| = |Q|^2$ summands of the form $\gexpe{\beta(st,uv)\beta(wx,yz)}$, i.e.

\begin{eqnarray}
\label{eq:variance_of_crossings}
\gvar{C}
	&=&
	\gexpe{\sum_{\{st,uv\} \in Q} \sum_{\{wx,yz\} \in Q} \beta(st,uv)\beta(wx,yz)} \nonumber\\
	&=&
	\sum_{\{st,uv\} \in Q} \sum_{\{wx,yz\} \in Q} \gexpe{\beta(st,uv)\beta(wx,yz)}.
\end{eqnarray}

Since $\{st, uv\}, \{wx, yz\} \in Q$, we have that $\gexpe{\alpha(st, uv)} = \gexpe{\alpha(wx, yz)} = \gprobdelta$, and therefore
\begin{eqnarray*}
\gexpe{\beta(st,uv)\beta(wx,yz)}
	= \gexpe{\alpha(st,uv)\alpha(wx,yz)}
	- 2\gprobdelta\gexpe{\alpha(st, uv)}
	+ \gprobdelta^2
\end{eqnarray*}

which leads to
\begin{eqnarray}
\label{eq:decomposition}
\gexpe{\beta(st,uv)\beta(wx,yz)} = \gexpe{\alpha(st,uv)\alpha(wx,yz)} - \delta_*^2.
\end{eqnarray}

When the product $\alpha(e_1,e_2)\alpha(e_3,e_4)$ corresponds to a type $\omega$, we can replace $\gexpe{\alpha(e_1,e_2)\alpha(e_3,e_4)}$ by its shorthand $\gprobalphastw$. Then, equation \ref{eq:decomposition} gives
\begin{eqnarray}
\label{eq:short_decomposition}
\gexpetw
	&=& \gprobalphastw - \gprobdelta^2.
\end{eqnarray}

As we show in the following section, the analysis of the products $\beta(st,uv)\beta(wx,yz)$, or $\alpha(st,uv)\alpha(wx,yz)$, allows one to classify the combinations $(\{st,uv\}, \{wx,yz\})$ in the double summation in equation \ref{eq:variance_of_crossings} into 9 types, which are always the same irrespective of the layout where the graph is embedded in.

\subsection{The types of products}

\begin{table}
\caption{The classification of the elements of $Q \times Q$ into types of products abstracting away from the order of the elements of the pair. $\omega\in\Omega$ is the code that identifies the product type; these codes have two (or three digits) that results from concatenating $\tau$ and $\phi$ (except for types 021-022, where a third digit is required). $(\{e_1,e_2\},\{e_3,e_4\})$ is an element of $Q \times Q$ where the symbols $s,t,u,v,w,x,y,z$ indicate distinct vertices, $|\upsilon|$ is the number of different vertices of the type, $\tau$ is the size of the intersection between $\{e_1, e_2\}$ and $\{e_3, e_4\}$, $\phi$ is the number of edge intersections, $\lprobalphastw$ is the probability that $\alpha(e_1,e_2)\alpha(e_3,e_4)=1$ in a uniformly random permutation of all the vertices of the graph, and $\lexpetw = \lprobalphastw - \lprobdelta^2$ (equation \ref{eq:short_decomposition}), where $\lprobdelta=1/3$ (equation \ref{eq:prob_delta:rla}).
}
\label{table:types_of_combinations}
\begin{indented}
\item[]
\begin{tabular}{@{}llllllll}
\br
$\omega\in\Omega$ & $(\{e_1,e_2\},\{e_3,e_4\})$ & $|\upsilon|$ & $\tau$ & $\phi$ & $\lprobalphastw$ & $\lexpetw$ \\ 
\mr
00 & $(\{st,uv\},\{wx,yz\})$ & 8 & 0 & 0 & 1/9  & 0 \\
24 & $(\{st,uv\},\{st,uv\})$ & 4 & 2 & 4 & 1/3  & 2/9 \\
13 & $(\{st,uv\},\{st,uw\})$ & 5 & 1 & 3 & 1/6  & 1/18 \\
12 & $(\{st,uv\},\{st,wx\})$ & 6 & 1 & 2 & 2/15 & 1/45 \\
04 & $(\{st,uv\},\{su,tv\})$ & 4 & 0 & 4 & 0    & -1/9 \\ 
03 & $(\{st,uv\},\{su,vw\})$ & 5 & 0 & 3 & 1/12 & -1/36 \\
021 & $(\{st,uv\},\{su,wx\})$ & 6 & 0 & 2 & 1/10 & -1/90 \\
022 & $(\{st,uv\},\{sw,ux\})$ & 6 & 0 & 2 & 7/60 & 1/180 \\
01 & $(\{st,uv\},\{sw,xy\})$ & 7 & 0 & 1 & 1/9  & 0 \\
\br
\end{tabular}
\end{indented}
\end{table}

Suppose that $\eta = (\{e_1,e_2\},\{e_3,e_4\})$ is an element of $Q \times Q$ of type $\omega$. By definition, $\{e_1,e_2\},\{e_3,e_4\}\in Q$. The set of vertices of $\eta$ is
\begin{eqnarray*}
\upsilon = e_1 \cup e_2 \cup e_3 \cup e_4.
\end{eqnarray*}  
One the one hand, the 4 edges of $\eta$ contribute with at most 2 different vertices each. On the other hand, $\{e_1,e_2\},\{e_3,e_4\}\in Q$ implies that $|e_1 \cap e_2| = |e_3 \cap e_4| = 0$. Then 
\begin{eqnarray*}
4 \leq |\upsilon| \leq 8.
\end{eqnarray*}
As a first approximation, we classify $\eta$ based on two parameters. The first parameter is $\tau$, the size of the intersection between the two sets of edges making an element of $Q \times Q$, i.e. 
\begin{eqnarray}
\label{eq:tau}
\tau=|\{e_1, e_2\} \cap \{e_3, e_4\}|.
\end{eqnarray}
The second parameter is $\phi$, the number of edge intersections, i.e. 
\begin{eqnarray}
\label{eq:phi}
\phi=|e_1 \cap e_3| + |e_1 \cap e_4| + |e_2 \cap e_3| + |e_2 \cap e_4|.
\end{eqnarray}
Table \ref{table:types_of_combinations} summarizes the 9 types of products. The set of all type of products is
\begin{eqnarray*}
\Omega=\{00,24,13,12,04,03,021,022,01\}.
\end{eqnarray*}
Henceforth, we use $\omega$ to denote one of the 9 types of products of $\Omega$. Type $00$ and type $24$ represent two extreme configurations: type $00$ is the case where all the vertices are actually different while type $24$ is the case where the pairs of edges are the same. Types $13$-$01$ represent intermediate possibilities. Types $00$ to $04$ are found in the pioneering analysis by Moon \cite{Moon1965a} on complete graphs (type $00$ is implicit in p. 506 and types $24$-$04$ are enumerated as the types that have non-zero contribution in p. 506). Moon omitted types 03-01 but we do not know if the exclusion of a type was due to not being aware of the existence of the type or not considering it relevant for the calculation of the variance.

\begin{figure}
	\centering
	\includegraphics[scale = 0.8]{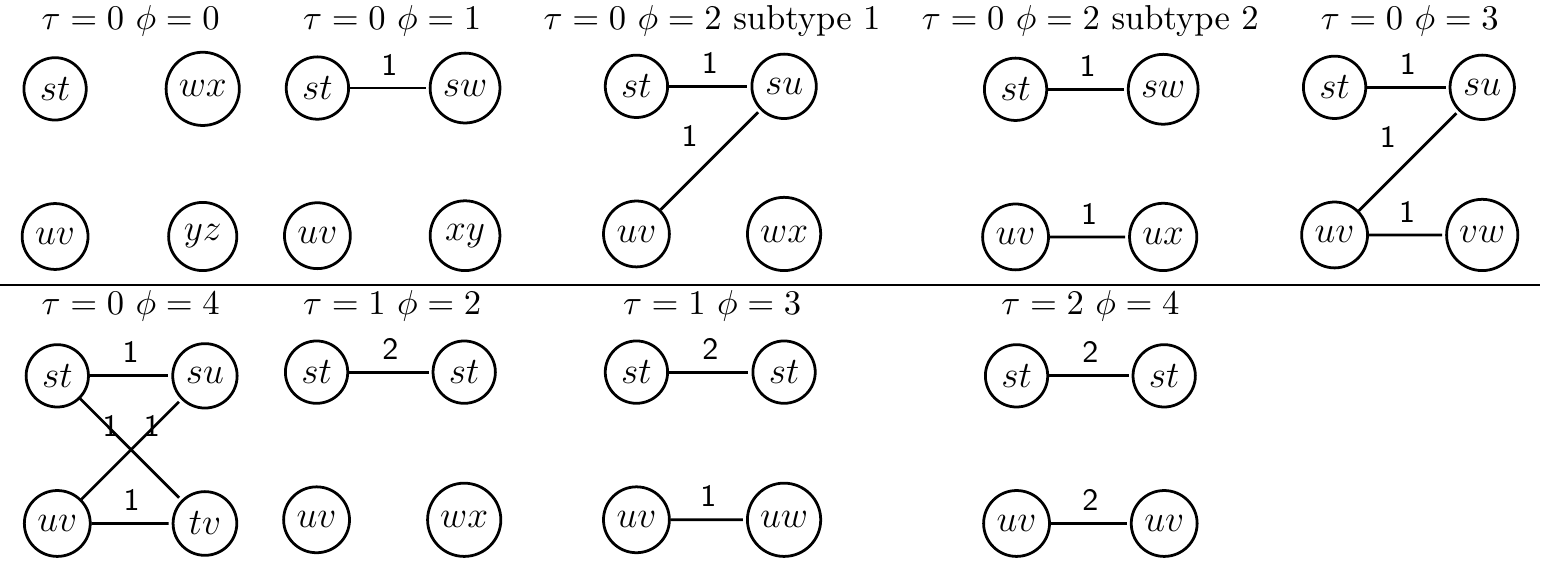}
	\caption{A summary of all the unlabeled weighted bipartite graphs produced by elements of $Q \times Q$, classified by $\tau$ and $\phi$. In these bipartite graphs, vertices are edges in the original graph and edges indicate that two edges of the original graph are not independent. Edge weights indicate the number of vertices of the original graph shared by vertices of the bipartite graph.}
	\label{fig:bipartite_graphs}
\end{figure}

Every combination of $\tau$ and $\phi$ yields a unique type of product except $\tau=0$ and $\phi=2$, that yields two types (types $021$ and $022$).  The latter follows from a further graph analysis which shows, in addition, that the classification into 9 types is exhaustive. The analysis is based on an equivalence between $\eta = (\{e_1,e_2\},\{e_3,e_4\})$ and a labeled weighted bipartite graph where 
\begin{enumerate}
\item The set of vertices is the set of edges $\{e_1, e_2, e_3, e_4\}$.
\item Two vertices $e_i$ and $e_j$ are linked if and only if $|e_i \cap e_j|>0$.
\item The weight of an edge is $w(e_i, e_j) = |e_i \cap e_j|$. Therefore $w(e_i, e_j) \in \{1,2\}$.
\end{enumerate}
The graph is bipartite: one partition is $\{e_1, e_2\}$ and the other is $\{e_3, e_4\}$ (within each partition, edges cannot be formed because $\{e_1,e_2\},\{e_3,e_4\} \in Q$). All the unlabeled weighted bipartite graphs that can be produced by all the possible values of $\tau$ and $\phi$ are summarized in figure \ref{fig:bipartite_graphs}. In that figure, vertex labels are merely used to show that all these graphs are actually possible. The ensemble of labeled bipartite graphs that can be produced is, of course, larger. A detailed analysis follows.

By definition (equations \ref{eq:tau} and \ref{eq:phi}), $0 \leq \tau\leq 2$ and $0 \leq \phi\leq 4$. First, suppose that $\tau>0$. Then, suppose without any loss of generality that $e_1 = e_3$. Then $e_1$ and $e_3$ are linked with a weight of 2. In turn, this implies that $|e_1 \cap e_4| = |e_2 \cap e_3| = 0$ because $\{e_1,e_2\},\{e_3,e_4\} \in Q$. Now consider two cases, i.e.
\begin{itemize}
\item $\tau=2$. Then $e_2 = e_4$ because $\tau=2$ and $e_2$ and $e_4$ are linked with a weight of 2. Therefore, there is only one possible unlabeled bipartite graph and $\phi=4$ as a result of plugging all the results above into the definition of $\phi$ (equation \ref{eq:phi}).

\item $\tau=1$. Then $2 \leq \phi$ and $|e_2 \cap e_4| < 2$, implying that there are only two possibilities: (1) $e_2$ and $e_4$ are  unlinked so $\phi=2$ or (2) linked with a weight of 1 and $\phi=3$. Thus there is only one unlabeled bipartite graph for $\phi=2$ and another one for $\phi=3$.
\end{itemize}
Second, we consider the case $\tau=0$. Then 
\begin{eqnarray*}
0 \leq |e_1 \cap e_3|, |e_1 \cap e_4|, |e_2 \cap e_3|, |e_2 \cap e_4| \leq 1.
\end{eqnarray*}
Recall that $0 \leq \phi\leq 4$. To obtain $\phi=0$, there is only one possibility, i.e. 
\begin{eqnarray*}
|e_1 \cap e_3| = |e_1 \cap e_4| = |e_2 \cap e_3| = |e_2 \cap e_4| = 0. 
\end{eqnarray*} 
To obtain $\phi=4$ (the complementary of the case $\phi=0$), there is also only one possibility, i.e. 
\begin{eqnarray*}
|e_1 \cap e_3| = |e_1 \cap e_4| = |e_2 \cap e_3| = |e_2 \cap e_4| = 1. 
\end{eqnarray*}
Therefore $\phi=0$ and $\phi=4$ produce only one unlabeled bipartite graph each. To obtain $\phi=1$, we should have $|e_i \cap e_j|=1$ only in one pair and $|e_i \cap e_j|=0$ in the remainder. To obtain $\phi=3$ (the complementary of the case $\phi=1$), we should have $|e_i \cap e_j|=0$ only in one pair and $|e_i \cap e_j|=1$ in the remainder.

The interesting case is $\phi=2$. Suppose without any loss of generality that $|e_1 \cap e_3| = 1$, namely the bipartite graph has an edge between $e_1$ and $e_3$ with a weight of 1. We have to link an additional pair of edges to achieve $\phi=2$. There are only three possibilities, 
\begin{enumerate}
\item $|e_1 \cap e_4| = 1$ (e.g. $(\{st,uv\}, \{sw,tx\})$).
\item $|e_2 \cap e_3| = 1$ (e.g. $(\{st,uv\}, \{su,wx\})$).
\item $|e_2 \cap e_4| = 1$ (e.g. $(\{st,uv\}, \{sw,ux\})$).
\end{enumerate}
where $s,t,u,v,w,x$ are all distinct. The 1st and the 2nd configurations are symmetric (one gives the other by swapping the contents of the two partitions), namely they represent the same unlabeled bipartite graph. The third yields a different unlabeled bipartite graph (notice that the degree sequence of the 1st and 2nd configurations differ from that of the 3rd). See figure \ref{fig:bipartite_graphs} for examples of the only two different unlabeled bipartite graphs.

As a result of the arguments above, every type can be meaningfully described by a code of two digits that results from concatenating $\tau$ and $\phi$ as shown in table \ref{table:types_of_combinations}. The only exception is $02$ that requires an additional digit to distinguish the two unlabeled bipartite graphs it can produce.

\begin{figure}
	\centering
	\includegraphics[scale=0.8]{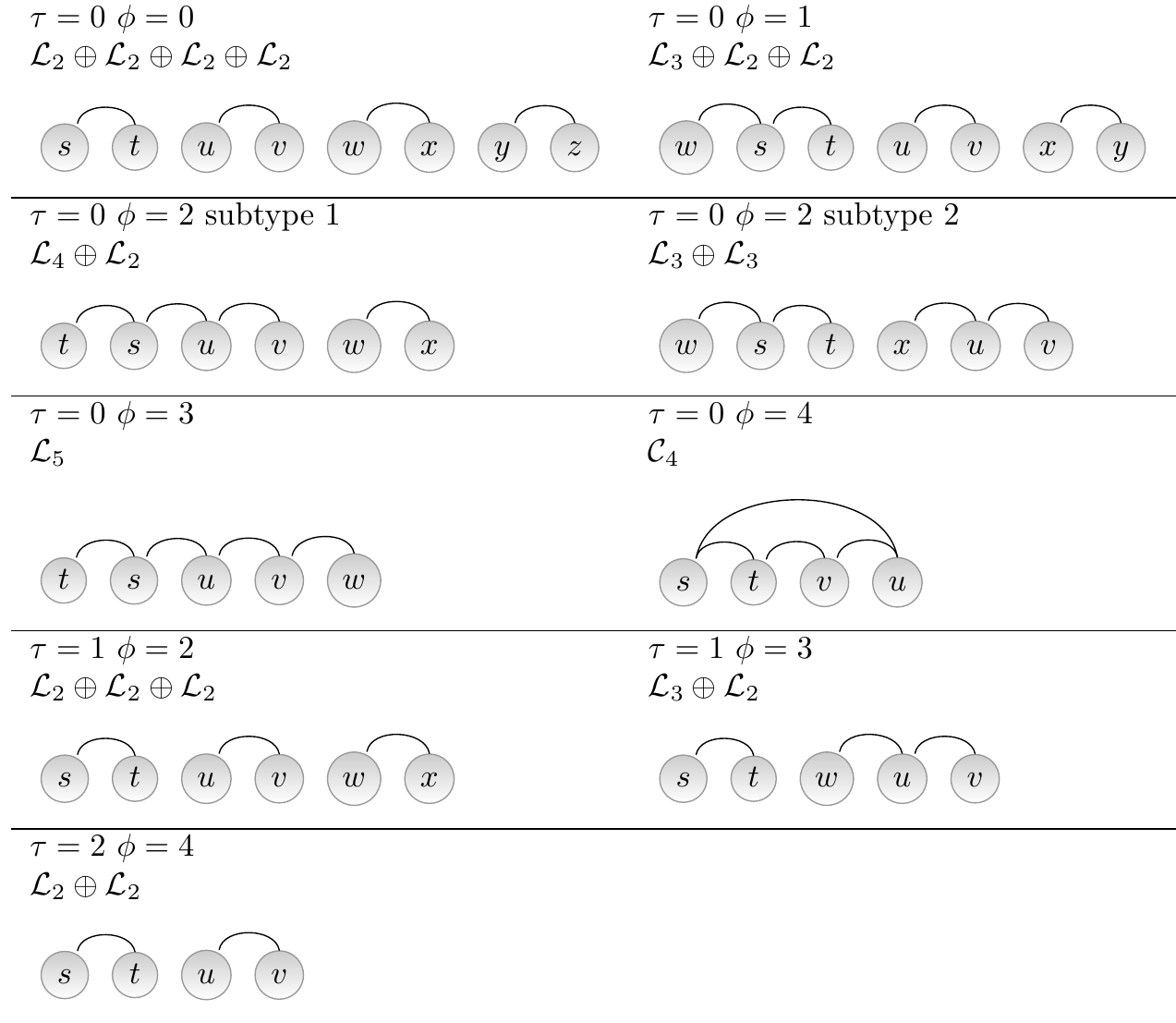}
	\caption{A summary of all the unipartite graphs generated by the bipartite graphs of figure \ref{fig:bipartite_graphs}, classified by $\tau$ and $\phi$.}
	\label{fig:unipartite_graphs}
\end{figure}

\subsection{The variance of $C$ as the function of the number of products}
\label{sec:variance-C:freqs}

Let $\omega=\mathcal{T}(\{e_1,e_2\}, \{e_3,e_4\})\in\Omega$ be the type of product of $\beta(e_1,e_2)\beta(e_3,e_4)$. If $\omega=\mathcal{T}(\{e_1,e_2\}, \{e_3,e_4\})$, then $\gamma_\omega=\beta(e_1,e_2)\beta(e_3,e_4)$, allowing one to express $\gvar{C}$ (equation \ref{eq:variance_of_crossings}) equivalently as
\begin{eqnarray}
\label{eq:variance:freq-times-exp}
\gvar{C} = \sum_{\omega\in\Omega} f_\omega \gexpetw,
\end{eqnarray}
where $f_\omega$ is the number of products of type $\omega$, defined as 
\begin{eqnarray}
\label{eq:frequency_template}
f_\omega = \sum_{q_1 \in Q} \sum_{\substack{q_2 \in Q \\ \omega=\mathcal{T}(q_1, q_2)}} 1.
\end{eqnarray} 

We can draw some initial conclusions from the analysis of the value of $|\upsilon|$ in table \ref{table:types_of_combinations}. For each type $\omega$, this value implies that
\begin{itemize}
\item	$f_{00} = 0$ if $n < |\upsilon_{00}| = 8$,
\item	$f_{24} = f_{04} = 0$ if $n < |\upsilon_{24}| = |\upsilon_{04}| = 4$,
\item	$f_{13} = f_{021} = f_{022} = 0$ if $n < |\upsilon_{13}| = |\upsilon_{021}| = |\upsilon_{022}| = 6$,
\item	$f_{12} = f_{03}$ if $n < |\upsilon_{12}| = |\upsilon_{03}| = 5$,
\item	$f_{01} = 0$ if $n < |\upsilon_{01}| = 7$.
\end{itemize}

Furthermore, given the definition of the $f_\omega$'s in equation \ref{eq:frequency_template}, it is easy to see that any pair $\{q_1,q_2\}$ such that $q_1,q_2 \in Q$ and $q_1 \neq q_2$ is counted twice. In contrast, pairs such that $q_1 = q_2$ (i.e. type $24$) are counted only once. Therefore, $f_\omega$ is even for $\omega \neq 24$. It is easy to see that $f_{24} = |Q|$, and that $f_{04} = 0$ in trees because the edges $st$, $uv$, $su$, $tv$ define some $\cycle[4]$. 

Also, we have, by definition,
\begin{eqnarray}
\label{eq:total_frequency_of_types}
\sum_{\omega\in\Omega} f_\omega = |Q|^2. 
\end{eqnarray}

Moreover, the unipartite graphs in figure \ref{fig:unipartite_graphs} generated by the bipartite graphs in figure \ref{fig:bipartite_graphs} allow one to identify necessary (sometimes sufficient) conditions for finding a given type in a graph. These unipartite graphs allow one to identify necessary conditions for finding a given type in a graph. For instance, as mentioned above, as the unipartite graph of type $04$ is a cycle graph of four vertices, a cycle is needed for $f_{04}>0$. Then $f_{04}=0$ in trees. Type $03$ needs paths of five vertices so that $f_{03}>0$. Types $022$, $13$, $01$ need paths of three vertices so that $f_{022},f_{13},f_{01}>0$. Types $021$, $01$, $00$, and $13$ need paths of two vertices (edges) so that $f_{021},f_{01},f_{00},f_{13}>0$.

In any a generic layout, some of the values $\gprobalphastw$ are easy to calculate. Firstly, let $\{e_1,e_2\},\{e_3,e_4\}\in Q$ such that $(\{e_1,e_2\},\{e_3,e_4\})$ is of type 00. Recalling that $\prob{\alpha(e_1,e_2)=1}=\gprobdelta$ (equation \ref{eq:ra:prob_indep_edge_crossing}), we have that
\begin{eqnarray*}
\gprobalphast{00}
	&=& \prob{\alpha(e_1,e_2)\alpha(e_3,e_4)=1} \\
	&=& \prob{\alpha(e_1,e_2)=1 }\prob{\alpha(e_3,e_4)=1} \\
	&=& \gprobdelta^2.
\end{eqnarray*}
since no vertices are shared among edges, and then 
\begin{eqnarray*}
\gexpet{00}=0.
\end{eqnarray*}
Secondly, let $\{e_1,e_2\},\{e_3,e_4\}\in Q$ such that $(\{e_1,e_2\},\{e_3,e_4\})$ is of type 01. In this case, we also have that
\begin{eqnarray*}
\gprobalphast{01} = \gprobdelta^2.
\end{eqnarray*}
Without loss of generality, let $e_1=st$, $e_2=uv$, $e_3=sw$, $e_4=xy$. The only interaction between between $\alpha(e_1,e_2)$ (edges $e_1$ and $e_2$ crossing) and $\alpha(e_3,e_4)$ (edges $e_3$ and $e_4$ crossing) is only through a vertex $s$ and thus $\alpha(e_1,e_2)$ and $\alpha(e_3,e_4)$ are mutually independent, hence
\begin{eqnarray*}
\gexpet{01}=0.
\end{eqnarray*}
Thus types 00 and 01 do not contribute to the variance. For the remainder of the types, independence is not guaranteed {\em a priori} because either more than two vertices are shared ($\phi>1$) or the interaction is strong via sharing edges ($\tau>0$). Thirdly,
\begin{eqnarray*}
\lprobalphast{04} = \pprobalphast{04} = \sprobalphast{04} = 0,
\end{eqnarray*}
where \textit{rap} denotes a random arrangement on the plane where edges between vertices are line segments \cite{Barthelemy2018a}, and \textit{rsa} denotes Moon's spherical arrangement \cite{Moon1965a}. In all those layouts, if one of the pairs of edges of the type $04$ crosses, the other pair cannot possibly cross. Therefore
\begin{equation*}
\lexpet{04} = -\lprobdelta^2, \qquad
\pexpet{04} = -\pprobdelta^2, \qquad
\sexpet{04} = -\sprobdelta^2.
\end{equation*}
Whether $\gprobalphast{04}=0$ requires future investigation. And, finally,
\begin{eqnarray*}
\gprobalphast{24}
	= \prob{\alpha(st,uv)\alpha(st,uv)=1}
	= \prob{\alpha(st,uv)=1}
	= \gprobdelta,
\end{eqnarray*}
where $\{st,uv\}\in Q$, which leads to
\begin{eqnarray*}
\gexpet{24}=\gprobdelta(1 - \gprobdelta).
\end{eqnarray*}

In case of linear arrangements, the values $\gprobalphastw$ can be calculated exactly and easily by means of a computational procedure. In such layouts, $\lprobalphastw$ is the proportion of permutations of the vertices of $\upsilon$ where $\alpha(\cdots,\cdots)\alpha(\cdots,\cdots)=1$. The different values of $\lprobalphastw$ and $\lexpetw$ are summarized in table \ref{table:types_of_combinations}. Applying the values of $\lexpetw$ in table \ref{table:types_of_combinations} to equation \ref{eq:variance:freq-times-exp}, allows one to express $\lvar{C}$ as a function of the amount of times every product appears, namely
\begin{eqnarray}
\lvar{C}
	&=& \sum_{\omega\in\Omega} f_\omega \lexpetw \nonumber \\
	&=& \frac{1}{9}\left[ 2|Q| + \frac{1}{20}f_{022} + \frac{1}{5}f_{12} + \frac{1}{2}f_{13} - \left(\frac{1}{10}f_{021} + f_{04} + \frac{1}{4}f_{03} \right) \right].
\label{eq:variance_of_number_of_crossings}
\end{eqnarray} 
In a tree, $f_{04} = 0$ and then 
\begin{equation}
\label{eq:variance_of_number_of_crossings_trees}
\lvar{C}
	= 	\frac{1}{9}
		\left[
			2|Q|
			+ \frac{1}{20}f_{022}
			+ \frac{1}{5}f_{12}
			+ \frac{1}{2}f_{13}
			- \left(
				\frac{1}{10}f_{021}
				+ \frac{1}{4}f_{03}
			  \right)
		\right]
\end{equation}
with
\begin{equation*}
f_{00} + f_{01} + f_{021} + f_{022} + f_{03} + f_{04} + f_{12} + f_{13} = |Q|(|Q|-1)
\end{equation*}
in both equations \ref{eq:variance_of_number_of_crossings} and \ref{eq:variance_of_number_of_crossings_trees}.

In the coming sections, we first derive general expressions for the $f_\omega$'s in simple graphs. Then we derive specific formulae for particular kinds of graphs. These expressions allow one to obtain simple arithmetic expressions for $\gvar{C}$ via equation \ref{eq:variance_of_number_of_crossings} in these kinds of graphs. Before we proceed, we give a chance to the reader to check detailed counts of the $f_\omega$'s and the calculation of $\lvar{C}$ for small graphs (\ref{toy_examples_appendix}). They can help gain intuitions for the mathematical calculations to follow. In addition, these examples are a component of the protocol (\ref{testing_protocol_appendix}) to validate the formulae that are derived.

% 	subsection
\subsection{Preliminaries}
\label{sec:preliminaries}

Before moving on to formalizing each type $f_\omega$ and deriving general expressions for them, we first define the notation used. 

\begin{figure}
	\centering
	\includegraphics[scale=0.75]{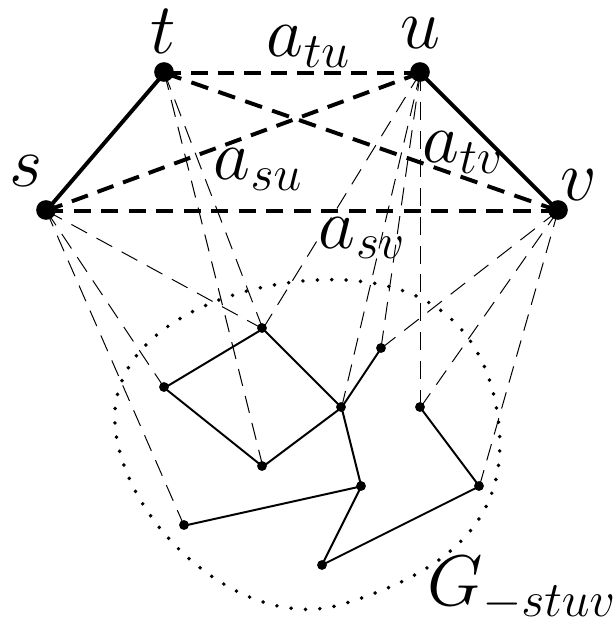}
	\caption{The effect of removing vertices $s,t,u,v$ such that $\{st,uv\} \in Q$ from a graph $G$ to produce the graph $G_{-stuv}$. Solid thick lines indicate the edges formed by these vertices in $G$, solid thin lines indicate edges in $G_{-stuv}$ and dashed lines indicate potential edges, namely edges that may not exist between these vertices and vertices in $G_{-stuv}$.}
	\label{fig:graph-without-stuv}
\end{figure}

For any undirected simple graph $G=(V,E)$, we define $G_{-L}$ as the induced graph resulting from the removal of the vertices in $L \subseteq V$. Figure \ref{fig:graph-without-stuv} shows an example of $G_{-\{s,t,u,v\}}$. The format of this figure is used in similar figures involved in the derivation of the $f_\omega$'s. Unless stated otherwise, we use $Q=Q(G)$, to refer to the set of pairs of independent edges of $G$.

Since it is used extensively, we denote $G_{-\{s,t,u,v\}}$ as simply $G_{-stuv}$. We use $\setminus$ to indicate the set difference operator. Then $G_{-stuv\setminus k}$ is used to mean $G_{-(\{s,t,u,v\}\setminus\{k\})}$. We use $\Gamma(s,-L) = \Gamma(s)\setminus L$ to indicate the set of neighbors of $s \in L \subseteq V$ in $V(G_{-L})$. Notice that its size is
\begin{eqnarray*}
%\label{eq:amount-neighbours-from-G-minus}
|\Gamma(s,-L)| = k_{s} - |\{ w \in L \;|\; \{w,s\} \in E \}|
= k_{s} - \sum_{w \in L} a_{sw}.
\end{eqnarray*}
Since $\Gamma(k,-\{s,t,u,v\})$, with $k\in\{s,t,u,v\}$, is used extensively in this work, we use $\Gamma(k,-stuv)$ instead. We use the shorthand ``$n$-path'' to refer to paths of $n$ vertices. Finally, we use $n_G(F)$ to denote the number of subgraphs isomorphic to $F$ in $G$.

% The coming generalisation of the types $f_{00}, f_{01}, ..., f_{24}$ 
The calculations of $f_{00}, f_{01}, ..., f_{24}$ to come, require a clear notation that states the vertices shared between each pair of elements of $Q$ for an arbitrary graph $G$. Throughout this article, we need to use summations of the form
\begin{eqnarray*}
\sum_{
	\substack{ s,t,u,v \in V \;: \\ \{st,uv\} \in Q(G) }
}
\sum_{
	\substack{ w,x,y,z \in V \;: \\ \{wx,yz\} \in Q(G_{-\{s,t,u,v\}}) }
} \square,
\end{eqnarray*}
where, below each summation operand, we specify the scope before ``:'' followed below by the condition. The ``$\square$'' represents any term. For the sake of brevity, we contract them as 
\begin{eqnarray*}
\sum_{ \{st,uv\} \in Q } \sum_{ \{wx,yz\} \in Q(G_{-stuv}) } \square.
\end{eqnarray*}
Notice that the scope is omitted in the new notation. This detail is crucial for the countings performed with the help of these compact summations. Likewise, if we want to denote when two elements of $Q$ from each of the summations share one or more vertices, we use
\begin{eqnarray*}
\sum_{ \{st,uv\} \in Q } \sum_{ \{sx,tz\} \in Q(G_{-uv}) } \square =
\sum_{
	\substack{ s,t,u,v \in V \;: \\ \{st,uv\} \in Q(G) }
}
\sum_{
	\substack{ x,z \in V \;: \\ \{sx,tz\} \in Q(G_{-\{u,v\}})}
} \square.
\end{eqnarray*}
This expression indicates the summation over the pairs of elements of $Q$ in which the second one shares two vertices with the first one. Again, the expression to the left is a shorthand for the one to the right. For the sake of comprehensiveness, we also present two more definitions
\begin{eqnarray*}
\sum_{ \{st,uv\} \in Q } \sum_{ \{st,yz\} \in Q(G_{-uv}) } \square =
\sum_{
	\substack{ s,t,u,v \in V \;: \\ \{st,uv\} \in Q(G) }
}
\sum_{ 
	\substack{ y,z \in V \;: \\ \{st, yz\} \in Q(G_{-\{u,v\} }) }
} \square,
\end{eqnarray*}
\begin{eqnarray*}
\sum_{ \{st,uv\} \in Q } \sum_{ \{sv,yz\} \in Q(G_{-tu}) } \square =
\sum_{
	\substack{ s,t,u,v \in V \;: \\ \{st,uv\} \in Q(G) }
}
\sum_{ 
	\substack{ y,z \in V \;: \\ \{sv,yz\} \in Q(G_{-tu}) }
} \square.
\end{eqnarray*}

% 	subsection
\subsection{Theoretical formulae}
\label{sec:general_formulas}

In the following subsections, we formalize the number of products of each type for any simple graph providing general expressions that are to be considered a first approach. The general formulae for the $f_\omega$'s presented in this section are designed based on two principles: compactness and connection with network theory, in particular, the problem of counting the number of subgraphs of a certain kind \cite{Milo2002a,Przulj2007a}. The link to graph theory is established showing that
\begin{eqnarray}
\label{eq:counting_subgraphs}
f_\omega = a_\omega n_G(F_\omega),
\end{eqnarray}
where $a_\omega$ is a constant (an even natural number except $a_{24} = 1$) that depends on $\omega$ and $F_\omega$ is some subgraph that also depends on $\omega$. $F_\omega$ is either an elementary graph (a linear tree or a cycle graph) or a combination of them with the operator $\oplus$. These elementary subgraphs are graphlets, i.e. connected subgraphs \cite{Przulj2007a}, while their combinations define graphettes, a generalization of graphlets to disconnected structures \cite{Hasan2017a}. The subgraph for each type of product is found in figure \ref{fig:unipartite_graphs}. An overview of the types of expressions that are derived for the $f_\omega$'s, is shown in table \ref{table:summary_frequencies}.

\begin{table}
\caption{\label{table:summary_frequencies} $f_\omega$, the number of products of type $\omega$ as a function of the $n_G(F_\omega)$, the number of subgraphs of a certain kind, where $F_\omega$ is a graph that depends on $\omega$.}
\begin{indented}
\item[]
\begin{tabular}{@{}lll}
\br
$\omega\in\Omega$
		& 	$f_\omega = a_\omega n_G(F_\omega)$
		&	Equation \\
\mr
$00$	&	$6n_G(\lintree[2]\oplus\lintree[2]\oplus\lintree[2]\oplus\lintree[2])$
		&	\ref{eq:general:00:geometric} \\
$24$	&	$\phantom{1}n_G(\lintree[2]\oplus\lintree[2])$
		&	\ref{eq:general:24:geometric} \\
$13$	&	$2n_G(\lintree[3]\oplus\lintree[2])$
		&	\ref{eq:general:13:geometric} \\
$12$	&	$6n_G(\lintree[2]\oplus\lintree[2]\oplus\lintree[2])$
		&	\ref{eq:general:12:geometric} \\
$04$	&	$2n_G(\cycle[4])$
		&	\ref{eq:general:04:geometric} \\
$03$	&	$2n_G(\lintree[5])$
		&	\ref{eq:general:03:geometric} \\
$021$	&	$2n_G(\lintree[4]\oplus\lintree[2])$
		&	\ref{eq:general:021:geometric} \\
$022$	&	$4n_G(\lintree[3]\oplus\lintree[3])$
		&	\ref{eq:general:022:geometric} \\
$01$	&	$4n_G(\lintree[3]\oplus\lintree[2]\oplus\lintree[2])$
		&	\ref{eq:general:01:geometric} \\
\br
\end{tabular}
\end{indented}
\end{table}

We use the same approach to obtain all the expressions of the $f_\omega$'s shown in table \ref{table:summary_frequencies}. We first instantiate equation \ref{eq:frequency_template} with the help of table \ref{table:types_of_combinations} and figure \ref{fig:bipartite_graphs}. Then we produce an equation of the form of equation \ref{eq:counting_subgraphs} by means of figure \ref{fig:unipartite_graphs}. All the initial definitions of the $f_\omega$'s that follow stem from equation \ref{eq:frequency_template} although it is only mentioned for the first types so as to avoid repetition. Although we have proved that $\gexpet{00}=\gexpet{01}=0$, we also include the analysis of these types.

\subsubsection{$\tau=2$, $\phi=4$}
\label{sec:general_formulas:24}

Thanks to equation \ref{eq:frequency_template}, 
\begin{eqnarray*}
f_{24} =
	\sum_{\{st,uv\} \in Q}
	\sum_{ \{st,uv\}  \in Q} 1.
\end{eqnarray*}
As $\{st,uv\}$ in the inner summation is constant (determined by the outer summation),
\begin{eqnarray}
\label{eq:general:24:final}
f_{24} = \sum_{\{st,uv\} \in Q} 1 = |Q|.
\end{eqnarray}

Then $f_{24}$ can be calculated easily thanks to the definition of $|Q|$ in equation \ref{eq:potential_number_of_crossings}. By definition of $Q$,
\begin{eqnarray}
\label{eq:general:24:geometric}
f_{24} = n_G(\lintree[2] \oplus \lintree[2]).
\end{eqnarray}
\subsubsection{$\tau=0$, $\phi=0$}
\label{sec:general_formulas:00}

Thanks to equation \ref{eq:frequency_template},
\begin{eqnarray*}
f_{00} = \sum_{\{st,uv\} \in Q} \sum_{ \{wx,yz\} \in Q(G_{-stuv}) } 1.
%\label{eq:general:00:raw}
\end{eqnarray*}
Noting that the inner summation defines the size of the set of pairs of independent edges of $G_{-stuv}$,  the expression above can be simplified, and then we obtain
\begin{eqnarray}
\label{eq:general:00}
f_{00} = \sum_{\{st,uv\} \in Q} |Q(G_{-stuv})|.
\end{eqnarray}
The last result comes to say that for any element $\{st,uv\}\in Q$ we only need to  calculate the size of $Q(G_{-stuv})$, the set of pairs of independent edges of $G_{-stuv}$. 

The summation in equation \ref{eq:general:00} counts over combinations of a pair of edges $\{st, uv\}$ from $Q$ with any two other independent edges $\{wx, yz\}$, defining a set $H=\{st,uv,wx, yz\}$ of independent edges. Each of these sets defines some subgraph $\lintree[2] \oplus \lintree[2] \oplus \lintree[2] \oplus \lintree[2]$. Every distinct set $H$ is produced by
\begin{eqnarray*}
{|H| \choose 2} = 6 
\end{eqnarray*}
elements of $Q$. Therefore, equation \ref{eq:general:00} gives
\begin{eqnarray}
\label{eq:general:00:geometric}
f_{00} = 6n_G(\lintree[2] \oplus \lintree[2] \oplus \lintree[2] \oplus \lintree[2]).
\end{eqnarray}

\subsubsection{$\tau=1$, $\phi=3$}
\label{sec:general_formulas:13}

\begin{figure}
	\centering
	\includegraphics[scale=0.75]{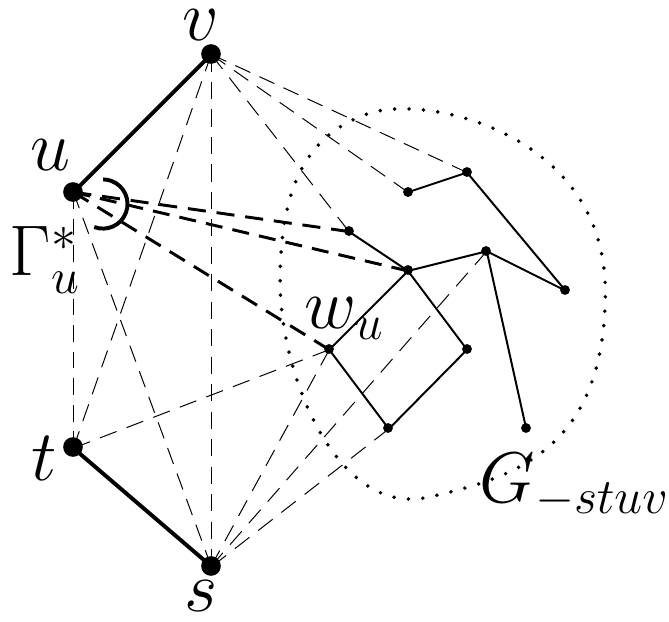}
	\caption{Illustration of the first inner summation in equation
	\ref{eq:general:13:formalisation:intermediate}. In the figure,
	$\{st,uv\}\in Q$, and $w_u \in \Gamma_u^* = \Gamma(u, -stuv)$. }
	\label{fig:general:13}
\end{figure}

This type deals with the pairs of edges sharing exactly one edge ($\tau=1$) and that have three vertices in common ($\phi=3$). Via equation \ref{eq:frequency_template}, a possible formalization of $f_{13}$ is
\begin{equation}
\label{eq:general:13:formalisation}
f_{13} = \sum_{\{st,uv\} \in Q}
\left(
	\sum_{\{st,uw\} \in Q(G_{-v})} 1 +
	\sum_{\{st,vw\} \in Q(G_{-u})} 1 +
	\sum_{\{uv,sw\} \in Q(G_{-t})} 1 +
	\sum_{\{uv,tw\} \in Q(G_{-s})} 1
\right).
\end{equation}
The first inner summation in the previous equation denotes the amount of vertices neighbors of $u$ in $G$ that are not $s,t,v$, the second the amount of vertices neighbors of $v$ in $G$ that are not $s,t,u$, and so on (figure \ref{fig:general:13}).

A formal definition for the first inner summation is, given a fixed $\{st,uv\}\in Q$,
\begin{eqnarray*}
\sum_{\{st,uw\} \in Q(G_{-v})} 1 = |\Gamma(u, -stuv)|.
\end{eqnarray*}
Likewise for the other inner summations. Then, equation \ref{eq:general:13:formalisation} becomes
%\begin{eqnarray*}
%\sum_{\{st,uv\} \in Q} \sum_{\{st,vw\} \in Q(G_{-u})} 1 & = &
%\sum_{\{st,uv\} \in Q} k_v - (a_{vs} + a_{vt} + a_{vu}), \\
%\sum_{\{st,uv\} \in Q} \sum_{\{uv,sw\} \in Q(G_{-t})} 1 & = &
%\sum_{\{st,uv\} \in Q} k_s - (a_{st} + a_{su} + a_{sv}), \\
%\sum_{\{st,uv\} \in Q} \sum_{\{uv,tw\} \in Q(G_{-s})} 1 & = &
%\sum_{\{st,uv\} \in Q} k_t - (a_{ts} + a_{tu} + a_{tv}),
%\end{eqnarray*}
\begin{eqnarray}
\label{eq:general:13:formalisation:intermediate}
f_{13}
&=& 	\overbrace{\sum_{\{st,uv\} \in Q} |\Gamma(u, -stuv)|}^{(1)} +
		\overbrace{\sum_{\{st,uv\} \in Q} |\Gamma(v, -stuv)|}^{(2)} \nonumber \\
&+& 	\overbrace{\sum_{\{st,uv\} \in Q} |\Gamma(s, -stuv)|}^{(3)} +
		\overbrace{\sum_{\{st,uv\} \in Q} |\Gamma(t, -stuv)|}^{(4)}.
\end{eqnarray}

$f_{13}$ counts over subgraphs $\lintree[3] \oplus \lintree[2]$ because, in equation \ref{eq:general:13:formalisation:intermediate},
\begin{enumerate}
\item[(1)] Counts the number of combinations of a 2-path $(s,t)$ with a 3-path $(w,u,v)$.
\item[(2)] Counts the number of combinations of a 2-path $(s,t)$ with a 3-path $(u,v,w)$.
\item[(3)] Counts the number of combinations of a 2-path $(u,v)$ with a 3-path $(w,s,t)$.
\item[(4)] Counts the number of combinations of a 2-path $(u,v)$ with a 3-path $(s,t,w)$.
\end{enumerate}
Figure \ref{fig:general:13} shows an example of summation (1): the edge $st$ is independent of all $\lintree[3]$ of the form $(v,u,w_u)$, where $w_u \in V(G_{-stuv})$. This can also be seen in the corresponding unipartite graph in figure \ref{fig:unipartite_graphs}. Then, since summands (1) and (2) count the same subgraphs as summands (3) and (4) (summands (1) and (2)  give summands (3) and (4) exchanging $s$ and $t$ by $u$ and $v$), we have that
\begin{eqnarray}
\label{eq:general:13:geometric}
f_{13} = 2n_G(\lintree[3] \oplus \lintree[2]).
\end{eqnarray}

We outline an alternative argument that leads to the same conclusion. Assume $\{(s,t), (u,v,w)\}$, a $\lintree[3]\oplus\lintree[2]$, is a subgraph of $G$. Then, the only elements of $Q\times Q$ classified as type $13$ with these vertices are $(\{st,uv\},\{st,vw\})$ and its reverse. It is easy to see that there are other elements of $Q\times Q$ of type 13 with the same vertices but they do not correspond to $\{(s,t), (u,v,w)\}$. Therefore, equation \ref{eq:frequency_template} counts two elements of $Q\times Q$ for a single $\lintree[3]\oplus\lintree[2]$. This argumentation is also used in some of the types to follow.

\subsubsection{$\tau=1$, $\phi=2$}
\label{sec:general_formulas:12}

\begin{figure}
	\centering
	\includegraphics[scale=0.75]{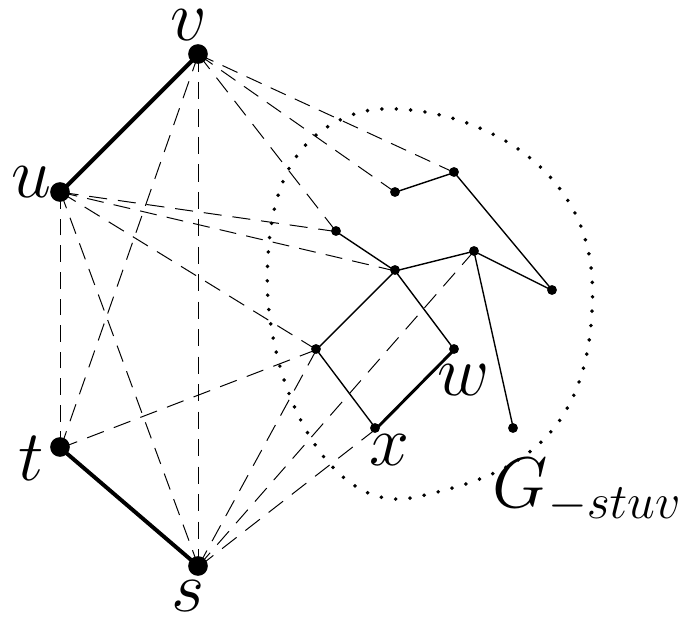}
	\caption{Illustration of the inner summations of equation \ref{eq:general:12:formalisation}, where both summations represent the edges in $G_{-stuv}$.}
	\label{fig:general:12}
\end{figure}

In this type, as in type $13$, one edge is shared, but this time only two vertices are equal. Therefore, we can formalize $f_{12}$ as
\begin{eqnarray}
\label{eq:general:12:formalisation}
f_{12} = \sum_{\{st,uv\} \in Q}
\left(
	\overbrace{\sum_{\{st,wx\} \in Q(G_{-uv})} 1}^{(1)} +
	\overbrace{\sum_{\{uv,wx\} \in Q(G_{-st})} 1}^{(2)}
\right).
\end{eqnarray}
The first summation is illustrated in figure \ref{fig:general:12}.

As $st$ in summation (1) is constant (determined by the outer summation), summation (1) inside the expression above counts the amount of edges $wx$ where $w,x \neq s,t,u,v$. Likewise for summation (2). Then equation \ref{eq:general:12:formalisation} can be simplified as
\begin{eqnarray}
\label{eq:general:12}
f_{12} = 2\sum_{\{st,uv\} \in Q} |E(G_{-stuv})|.
\end{eqnarray}

In equation \ref{eq:general:12}, the summation is counting over configurations that are produced combining two edges from $Q$ with a third independent edge, giving three independent edges from a set $H= \{(s,t), (u,v), (w,x)\}$, which defines some subgraph $\lintree[2] \oplus \lintree[2] \oplus \lintree[2]$. The summation in equation \ref{eq:general:12} visits the same set $H$   
\begin{eqnarray*}
{|H| \choose 2} = 3  
\end{eqnarray*}
times. Therefore, the summation in equation \ref{eq:general:12} matches $3n_G(\lintree[2] \oplus \lintree[2] \oplus \lintree[2])$ and then 
\begin{eqnarray}
\label{eq:general:12:geometric}
f_{12} = 6n_G(\lintree[2] \oplus \lintree[2] \oplus \lintree[2]).
\end{eqnarray}

\subsubsection{$\tau=0$, $\phi=4$}
\label{sec:general_formulas:04}

All pairs of elements of $Q$ classified in this type share no edges. However, they have four vertices in common. This allows one to obtain a simple formalization for $f_{04}$,
\begin{eqnarray}
\label{eq:general:04}
f_{04}
= \sum_{\{st,uv\} \in Q}
\left(
	\sum_{\{su,tv\} \in Q} 1 + \sum_{\{sv,tu\} \in Q} 1
\right)
= \sum_{\{st,uv\} \in Q} (a_{su}a_{tv} + a_{sv}a_{tu}).
\end{eqnarray}
In the previous summation, for each element $\{st,uv\} \in Q$, two distinct $\cycle[4]$ are counted, i.e.
\begin{itemize}
\item $b_1 = (s,t,v,u,s)$ if $a_{su}a_{tv} = 1$\footnote{Notice that the cycles $(t,v,u,s,t)$, $(v,u,s,t,v)$, $(u,s,t,v,u)$ are the same as cycle $(s,t,v,u,s)$.}.

\item $b_2 = (s,t,u,v,s)$ if $a_{tu}a_{sv} = 1$\footnote{Notice that the cycles $(t,u,v,s,t)$, $(u,v,s,t,u)$, $(v,s,t,u,v)$ are the same as cycle $(s,t,u,v,s)$.}.
\end{itemize}
Notice that for every element $\{st,uv\} \in Q$ that forms a $\cycle[4]$, say $b_1$, we have that $a_{su}a_{tv} = 1$ and thus there exists another element $\{su,tv\} \in Q$. For this other element $\{su,tv\} \in Q$ we can make the cycle $(s,u,v,t,s)$ because $st, uv \in E$ (as $\{st,uv\} \in Q$), which is isomorphic to $b_1$. The same reasoning can be applied to the second potential $\cycle[4]$ we can make. Thus, each $\cycle[4]$ formed by one element of $Q$ is counted twice in equation \ref{eq:general:04}. Trivially, all $\cycle[4]$ are counted in the summation above because $|Q(\cycle[4])|>0$. For these reasons, equation \ref{eq:general:04} becomes
\begin{eqnarray}
\label{eq:general:04:geometric}
f_{04} & = & 2 \nsquares.
\end{eqnarray}

\subsubsection{$\tau=0$, $\phi=3$}
\label{sec:general_formulas:03}

This type denotes those pairs in $Q \times Q$ that do not share an edge completely but that have three vertices in common. Given an element $\{st,uv\}\in Q$, the other possible elements of $Q$ that make the pair follow this type's characterization are
\begin{flalign*}
\{su,tw\} \qquad \{su,vw\} \qquad &\{sv,tw\} \qquad \{sv,uw\}\\
\{tu,sw\} \qquad \{tu,vw\} \qquad &\{tv,sw\} \qquad \{tv,uw\} \qquad w \neq s,t,u,v.
\end{flalign*}

\begin{figure}
	\centering
	\includegraphics[scale=0.65]{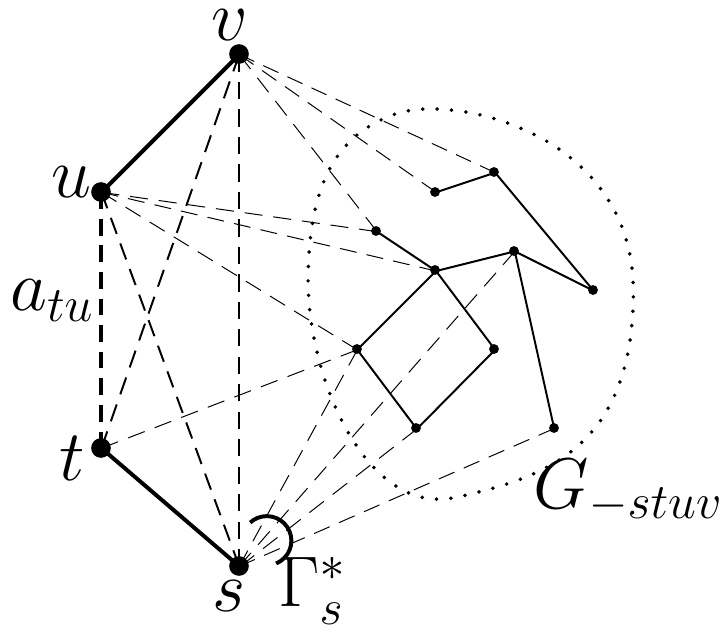}
	\caption{Illustration of $\varphi_{tus}$, that is exactly $|\Gamma_s^*| = |\Gamma(s,-stuv)|$ (the amount of neighbors of $s$ in $G_{-stuv}$), provided that the edge $tu$ exists (i.e. $a_{tu} = 1$).}
	\label{fig:general:03}
\end{figure}

Therefore, to calculate the value of $f_{03}$, we have to count how many elements of the previous list there are in the graph. Formally,
\begin{equation*}
%\label{eq:general:03}
f_{03} =
	\sum_{\{st,uv\} \in Q}
	(
		\varphi_{sut} + \varphi_{svt} + \varphi_{tus} + \varphi_{tvs} +
		\varphi_{svu} + \varphi_{tvu} + \varphi_{tuv} + \varphi_{suv}
	),
\end{equation*}
where $\varphi_{sut}$, $\varphi_{svt}$, $\cdots$ are functions with implicit parameter $\{st,uv\}\in Q$ and explicit parameters are three of the four vertices $s$,$t$,$u$ or $v$. These $\varphi_{\cdots}$ are defined as
\begin{eqnarray}
\label{eq:general:f03:varphis}
\varphi_{xyz} = a_{xy} |\Gamma(z, -stuv)|,
\end{eqnarray}
where $x,y,z \in \{s,t,u,v\}$, with $s,t,u,v$ the vertices of the implicit parameter. Then
\begin{align}
\label{eq:general:03:expanded}
f_{03}
& = \sum_{\{st,uv\} \in Q} a_{su} |\Gamma(t, -stuv)| + 
    \sum_{\{st,uv\} \in Q} a_{sv} |\Gamma(t, -stuv)| + 
    \sum_{\{st,uv\} \in Q} a_{tu} |\Gamma(s, -stuv)|  \nonumber \\ 
&   +
    \sum_{\{st,uv\} \in Q} a_{tv} |\Gamma(s, -stuv)| +
    \sum_{\{st,uv\} \in Q} a_{sv} |\Gamma(u, -stuv)| + 
    \sum_{\{st,uv\} \in Q} a_{tv} |\Gamma(u, -stuv)| \nonumber \\
&   +
    \sum_{\{st,uv\} \in Q} a_{tu} |\Gamma(v, -stuv)| +
    \sum_{\{st,uv\} \in Q} a_{su} |\Gamma(v, -stuv)|.
\end{align}
Looking at each $\varphi_{sut}, \varphi_{tus}, \cdots, \varphi_{suv}$ separately we see that, given $\{st,uv\} \in Q$, $\varphi_{sut}$ counts the amount of neighbors of $t$ in $G_{-stuv}$ if $su \in E$, $\varphi_{tus}$ counts the amount of neighbors of $s$ in $G_{-stuv}$ if $tu \in E$, and so on (figure \ref{fig:general:03}).

We can express this type as the amount of a certain type of subgraph with the help of figure \ref{fig:general:03} and the corresponding unipartite graph in figure \ref{fig:unipartite_graphs}. Figure \ref{fig:general:03} shows the interpretation of the value $\varphi_{tus}$ (formalized in equation \ref{eq:general:f03:varphis}) which, given an element $\{st,uv\} \in Q$ and the existence of edge $a_{tu}$, counts the amount of neighbors of $s$ in $G_{-stuv}$, namely $|\Gamma_s^*| = |\Gamma(s,-stuv)|$. Notice that if $\{st,uv\} \in Q$, and assuming that $a_{tu}=1$, then we have a $4$-path: $(v,u,t,s)$, and that by appending any vertex $w \in \Gamma_s^*$ to it we can make a $5$-path $(v,u,t,s,w)$. The same reasoning applies to the other $\varphi_{...}$. Here ... is used to indicate ``anything'' for each of the three explicit parameters of $\varphi$. The aforementioned unipartite graph supports that $f_{03}$ counts $5$-paths.

\begin{table}[H]
	\caption{The pattern of the 5-paths counted by each summand in equation \ref{eq:general:03:expanded} for a given $\{st,uv\} \in Q$. A dot is used to indicate an arbitrary neighbor of the nearest vertex in the path different from $s,t,u,v$.}
	\label{table:03:5-paths}

	\begin{indented}
	\item[]
	\begin{tabular}{@{}cc}
		$(.,s,t,u,v)$ & $(s,t,u,v,.)$ \\ 
		$(.,s,t,v,u)$ & $(s,t,v,u,.)$ \\
		$(.,t,s,v,u)$ & $(t,s,v,u,.)$ \\ 
		$(.,t,s,u,v)$ & $(t,s,u,v,.)$ 
	\end{tabular}
	\end{indented}
\end{table}

Each summation in equation \ref{eq:general:03:expanded} can be analyzed at two levels. At the local level, each summand counts over distinct 5-paths. At a global level, the 5-paths counted are also different. The distinct patterns of the 5-paths counted by each summation of equation \ref{eq:general:03:expanded} are shown in table \ref{table:03:5-paths}.  Notice that the paths in the right column of table \ref{table:03:5-paths} are obtained by shifting the vertices of the paths in the left column. Crucially, the edges of $Q$ yielding the path are consecutive in the path (i.e. they correspond to four consecutive vertices in the path). Therefore any 5-path of the graph is counted exactly by two different summations in equation \ref{eq:general:03:expanded}, i.e.
\begin{eqnarray}
\label{eq:general:03:geometric}
f_{03} = 2n_G(\lintree[5]).
\end{eqnarray}
\subsubsection{$\tau=0$, $\phi=2$, Subtype 1}
\label{sec:general_formulas:021}

\begin{figure}
	\centering
	\includegraphics[scale=0.8]{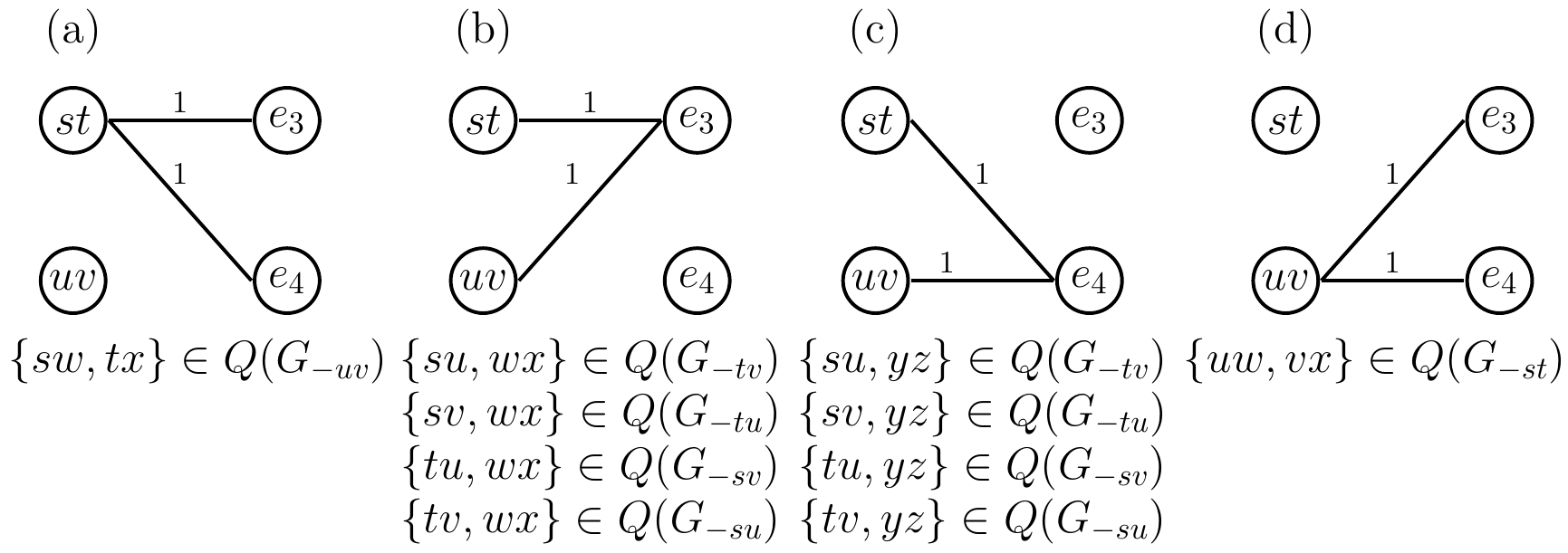}
	\caption{Elements of $Q$ such that when paired with element $\{st,uv\}\in Q$, the pair is classified as type $021$. Elements in (b) and (c) are symmetric.}
	\label{fig:combinations-02:1}
\end{figure}

Recall the definition of type 021 as an unlabeled bipartite graph (figure \ref{fig:bipartite_graphs}). Figure \ref{fig:combinations-02:1} shows the 10 possible forms that elements of $Q$ such that when paired with $\{st,uv\}\in Q$ yield a pair of $Q\times Q$ classified as $021$ can take by labeling the left partition with ${st,uv}$ and considering all the possible labelings of the right partition of that type. However, by symmetry between the forms in figure \ref{fig:combinations-02:1}(b) and those of figure \ref{fig:combinations-02:1}(c), there are only 6 unique forms.

Then we can formalize $f_{021}$ as
\begin{eqnarray}
\label{eq:general:021}
f_{021}
	&=& \sum_{\{st,uv\} \in Q} \varphi_{st} +
		\sum_{\{st,uv\} \in Q} \varphi_{uv} \nonumber \\
	&+& \sum_{\{st,uv\} \in Q} \varepsilon_{su} +
	\sum_{\{st,uv\} \in Q} \varepsilon_{sv} +
	\sum_{\{st,uv\} \in Q} \varepsilon_{tu} +
	\sum_{\{st,uv\} \in Q} \varepsilon_{tv},
\end{eqnarray}
where $\varphi_{xy}$ and $\varepsilon_{xy}$ are auxiliary functions with implicit parameter $\{st,uv\}\in Q$ and explicit parameters $xy$, that are defined as
\begin{eqnarray}
\varphi_{xy} & = & \sum_{w \in \Gamma(x, -stuv)} \sum_{w' \in \Gamma(y, -stuvw)} 1 \label{eq:general:f021:varphis-varepsilons} \\
             & = & \sum_{w \in \Gamma(x, -stuv)} |\Gamma(y, -stuvw)|, \nonumber \\
\varepsilon_{xy} & = & a_{xy} |E(G_{-stuv})|, \qquad x,y\in\{s,t,u,v\}. \nonumber
\end{eqnarray}

\begin{figure}
	\centering
	\includegraphics[width=10cm]{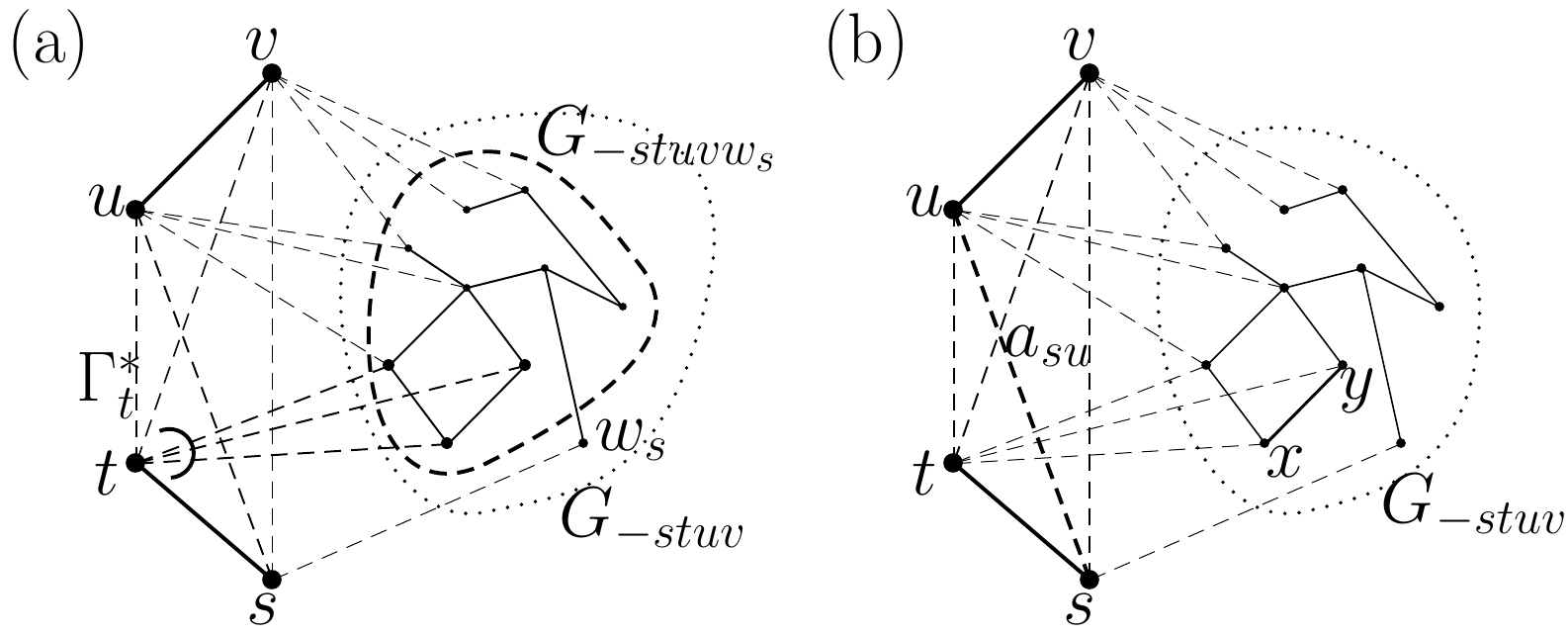}
	\caption{Illustration of (a) $\varphi_{st}$, and (b) $\varepsilon_{su}$. In (a), $w_s$ represents the only neighbor of $s$ different from $t,u,v$. Therefore, in this case, $\varphi_{st}$ is exactly the amount of vertices in $G_{-stuvw_s}$ neighbors of $t$, indicated with $\Gamma_t^* = \Gamma(t, -stuvw_s)$. $\varepsilon_{su}$ requires the existence of an edge between $s$ and $u$, indicated with $a_{su}$, and is equal to the amount of edges in $G_{-stuv}$.}
	\label{fig:general:021}
\end{figure}

The functions $\varphi_{st}$ and $\varphi_{uv}$ count the elements of the form of those illustrated in figures \ref{fig:combinations-02:1}(a) and \ref{fig:combinations-02:1}(d) ($\varphi_{st}$ is depicted in figure \ref{fig:general:021}(a)). The first function counts, for each neighbor of $s$, $w_s \neq t,u,v$, the number of neighbors of $t$, $w_t \neq s,t,u,v,w_s$. Likewise for the second function. On the other hand, the values $\varepsilon_{su}$, $\varepsilon_{sv}$, $\varepsilon_{tu}$, $\varepsilon_{tv}$ count the edges $xy\in E$, $x,y\neq s,t,u,v$ such that when paired with $su$, $sv$, $tu$, $tv$ form an element of $Q$ whose form is that of those elements illustrated in figures \ref{fig:combinations-02:1}(b) and \ref{fig:combinations-02:1}(c) ($\varepsilon_{su}$ is depicted in figure \ref{fig:general:021}(b)). These amounts are counted only if such edges exist in the graph, hence the $a_{su}$ for $\varepsilon_{su}$, and likewise for the other $\varepsilon_{..}$. Here .. is used to indicate ``anything'' for each of the two explicit parameters of $\varepsilon$.

We can take common factor in equation \ref{eq:general:f021:varphis-varepsilons} and simplify it. We obtain
\begin{eqnarray}
\label{eq:general:021:algorithmic}
f_{021} =
\sum_{\{st,uv\} \in Q}
	(\varphi_{st} + \varphi_{uv} +
	(a_{su} + a_{sv} + a_{tu} + a_{tv})|E(G_{-stuv})|).
\end{eqnarray}

We split the r.h.s. of equation \ref{eq:general:021:algorithmic} into two halves: the $\varphi$'s and the $\varepsilon$'s.  On the one hand,  
\begin{eqnarray}
\sum_{\{st,uv\} \in Q} \varphi_{st} + \sum_{\{st,uv\} \in Q} \varphi_{uv} = n_G(\lintree[4] \oplus \lintree[2]) 
\label{eq:021:sum1}
\end{eqnarray}
because the 1st summation is counting all the $\lintree[4] \oplus \lintree[2]$ such that $\lintree[2]$ is the edge $uv$ and $\lintree[4]$ is the path $(w,s,t,w')$ and the 2nd summation is counting all the $\lintree[4] \oplus \lintree[2]$ such that $\lintree[2]$ is the edge $st$ and $\lintree[4]$ is the path $(w,u,v,w')$. On the other hand,  
\begin{eqnarray}
	\sum_{\{st,uv\} \in Q} \varepsilon_{su} + \sum_{\{st,uv\} \in Q} \varepsilon_{sv} + \sum_{\{st,uv\} \in Q} \varepsilon_{tu} + \sum_{\{st,uv\} \in Q} \varepsilon_{tv} = n_G(\lintree[4] \oplus \lintree[2])
\label{eq:021:sum2}
\end{eqnarray}
because each summation is counting all the $\lintree[4] \oplus \lintree[2]$ such that the $\lintree[4]$ is build on the vertices $s,t,u,v$ and $\lintree[2]$ is any edge that is not formed by these vertices. Every summation is in charge of one of the four different ways in which a distinct $\lintree[4]$ can be produced linking one vertex of the edge $st$ with a vertex of the edge $uv$. Therefore, 
\begin{eqnarray}
\label{eq:general:021:geometric}
f_{021} = 2n_G(\lintree[4] \oplus \lintree[2]).
\end{eqnarray}
A detailed proof of equation \ref{eq:general:021:geometric} with a technique similar to the one applied to types 022 and 01 can be found in \ref{alternative_proof_f021_appendix}.

\subsubsection{$\tau=0$, $\phi=2$, Subtype 2}
\label{sec:general_formulas:022}

\begin{figure}
	\centering
	\includegraphics[scale=0.8]{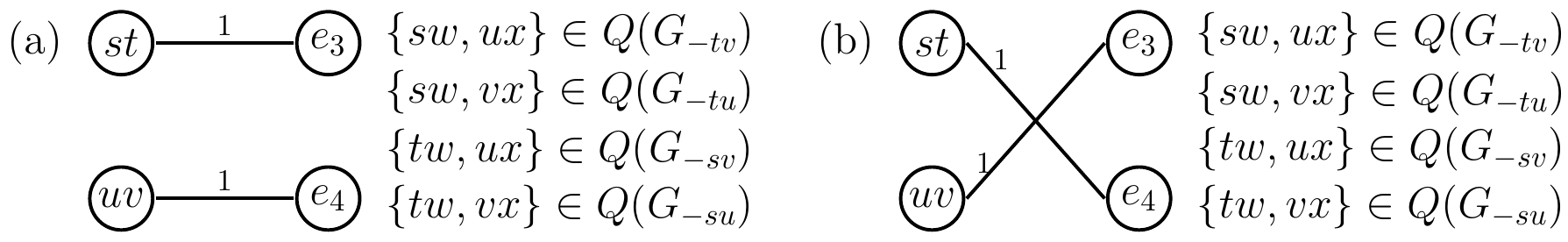}
	\caption{\label{fig:combinations-02:2} Elements of $Q$ such that when paired with element $\{st,uv\}\in Q$, the pair is classified as type $022$. (a) One of the bipartite graphs of type 022. (b) The other bipartite graph of type 022. The elements in (a) are symmetric to those of (b). }
\end{figure}

We follow the same approach as the one applied in type 021 (this second subtype is simpler to formalize). Figure \ref{fig:combinations-02:2} shows all the elements of $Q$ such that when paired with $\{st, uv \} \in Q$ yield a pair of $Q \times Q$ classified as type 022 for each of the two labeled bipartite graphs of that type. This gives eight configurations that are constructed by making two new independent edges ($e_3$ and $e_4$), one edge linking a new vertex, say $w$, to one of the vertices of $st$ and another edge linking another vertex, say $x$,  to edge $uv$, ($w,x \neq s,t,u,v$), so that the pair of new edges belongs to $Q$. However, only four configurations are distinct by symmetry: $x$ and $w$ are interchangeable. Therefore, the elements of $Q \times Q$ defined in figure \ref{fig:combinations-02:2}(a) are the same as those of \ref{fig:combinations-02:2} (b). As a result of this analysis, $f_{022}$ can be defined as
\begin{eqnarray}
\label{eq:general:022}
f_{022} =
	\sum_{\{st,uv\} \in Q} \varphi_{su} +
	\sum_{\{st,uv\} \in Q} \varphi_{sv} +
	\sum_{\{st,uv\} \in Q} \varphi_{tu} +
	\sum_{\{st,uv\} \in Q} \varphi_{tv},
\end{eqnarray}
where $\varphi_{xy}$ is an auxiliary function defined as in equation \ref{eq:general:f021:varphis-varepsilons}. 
$\varphi_{su}$ can be understood from the case of $\varphi_{st}$ in figure \ref{fig:general:021}(a).

Now, notice that this type counts pairs of $\lintree[3]\oplus\lintree[3]$: given a fixed $\{st,uv\}\in Q$, the value $\varphi_{su}$, for example, counts the neighbors of $s$ in $G_{-stuv}$ and the neighbors of $u$ in $G_{-stuvw}$, where $w \in \Gamma(s,-stuv)$. Similarly for the other $\varphi_{..}$. Therefore, the form of the subgraphs counted by each summation of equation \ref{eq:general:022} for a fixed $\{st,uv\}\in Q$ are  
\begin{align*}
\varphi_{su}&: \{(z_s,s,t),(z_u,u,v)\}, &\quad \varphi_{sv}&: \{(z_s,s,t),(u,v,z_v)\}, \\
\varphi_{tu}&: \{(s,t,z_t),(z_u,u,v)\}, &\quad \varphi_{tv}&: \{(s,t,z_t),(u,v,z_v)\},
\end{align*}
where $z_k$ indicates a neighbor of vertex $k$. It is easy to see that a fixed $\lintree[3]\oplus\lintree[3]$ of the form $H=\{(s,t,u),(v,w,x)\}$ is counted four times in equation \ref{eq:general:022}. First, notice that $Q(H)$ has exactly for elements, i.e.
\begin{equation*}
q_1=\{st,vw\}, \; q_2=\{st,wx\}, \; q_3=\{tu,vw\}, \; q_4=\{tu,wx\}.
\end{equation*}
Given one of them, say $q_1$, our graph $H$ is counted only by $\varphi_{tw}$ because for $q_1$ the subgraphs counted in equation \ref{eq:general:022} are of the form
\begin{align*}
\varphi_{sv}&: \{(z_s,s,t),(z_v,v,w)\}, &\quad \varphi_{sw}&: \{(z_s,s,t),(v,w,z_w)\}, \\
\varphi_{tv}&: \{(s,t,z_t),(z_v,v,w)\}, &\quad \varphi_{tw}&: \{(s,t,z_t),(v,w,z_w)\}.
\end{align*}
Similarly, for each of the other $q_i$'s, there is a unique $\varphi_{..}$ where $H$ is counted. Then, when calculating $f_{022}$ with equation \ref{eq:general:022}, every distinct $H$ is counted four times, i.e.
\begin{eqnarray}
\label{eq:general:022:geometric}
f_{022} = 4n_G(\lintree[3] \oplus \lintree[3]).
\end{eqnarray}

\subsubsection{$\tau=0$, $\phi=1$}
\label{sec:general_formulas:01}

\begin{figure}
	\centering
	\includegraphics[scale=0.75]{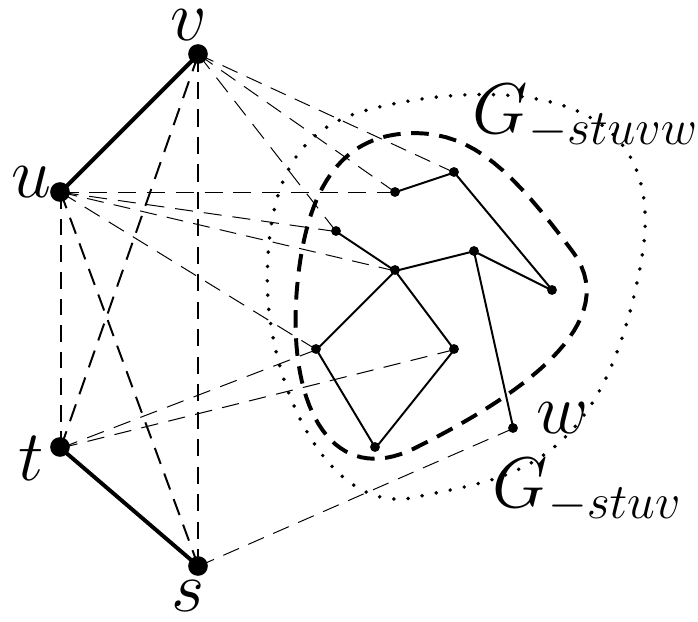}
	\caption{Illustration of $\varphi_s$. Here $w$ represents the only neighbor of $s$ different from $s,t,u,v$. Thus $\varphi_s$ is exactly the amount of edges in $G_{-stuvw}$.}
	\label{fig:general:01}
\end{figure}

Finally, $f_{01}$ can be formalized as
\begin{eqnarray}
\label{eq:general:01}
f_{01} =
	\sum_{\{st,uv\} \in Q} \varphi_s +
	\sum_{\{st,uv\} \in Q} \varphi_t +
	\sum_{\{st,uv\} \in Q} \varphi_u +
	\sum_{\{st,uv\} \in Q} \varphi_v,
\end{eqnarray}
where $\varphi_k$ is a function with an implicit parameter $\{st,uv\} \in Q$ and an explicit parameter $k \in \{s,t,u,v\}$, that is defined as
\begin{eqnarray*}
\varphi_k
	&=& \sum_{\{kw,xy\} \in Q(G_{-stuv\setminus k })} 1 \\ %\label{eq:general:01:varphis} \\
	&=& \sum_{w \in \Gamma(k, -stuv)} \sum_{xy \in E(G_{-stuvw})} 1 \\
	&=& \sum_{w \in \Gamma(k, -stuv)} |E(G_{-stuvw})|.
\end{eqnarray*}
The particular case of $\varphi_s$ is illustrated in figure \ref{fig:general:01}.

The fact that equation \ref{eq:general:01} counts over $\lintree[3]\oplus\lintree[2]\oplus\lintree[2]$ is readily seen by noting that, for a fixed $\{st,uv\}\in Q$, the summations of the $\varphi_{.}$ count subgraphs of the form
\begin{align*}
\varphi_s&: \{(z_s,s,t),(u,v),(z_1,z_2)\}, &\varphi_t&: \{(s,t,z_t),(u,v),(z_1,z_2)\} \\
\varphi_u&: \{(s,t),(z_u,u,v),(z_1,z_2)\}, &\varphi_v&: \{(s,t),(u,v,z_v),(z_1,z_2)\},
\end{align*}
where $z_k\in V$ denotes a neighbor of $k\in V$, and $z_1,z_2\in V$ vertices different from $z_k,s,t,u,v$. Now, consider some $\lintree[3]\oplus\lintree[2]\oplus\lintree[2]$ of the form $H = \{(s,t,u),(v,w),(x,y)\}$.  Notice that $Q(H)$ has exactly four elements, i.e.
\begin{equation*}
q_1=\{st,vw\}, \; q_2=\{st,xy\}, \; q_3=\{tu,vw\}, \; q_4=\{tu,xy\}.
\end{equation*}
Given one of them, say $q_1$, our graph $H$ is counted in only one of the $\varphi_.$ (by $\varphi_t$ in the example) because for $q_1$ the subgraphs counted in equation \ref{eq:general:01} are of the form
\begin{align*}
\varphi_s&: \{(z_s,s,t),(v,w),(z_1,z_2)\}, &\varphi_t&: \{(s,t,z_t),(v,w),(z_1,z_2)\} \\
\varphi_v&: \{(s,t),(z_v,v,w),(z_1,z_2)\}, &\varphi_w&: \{(s,t),(v,w,z_w),(z_1,z_2)\}.
\end{align*}
Similarly, for each of the other $q_i$'s, there is a unique $\varphi_.$ where $H$ is counted. Then, when calculating $f_{01}$ with equation \ref{eq:general:01}, every distinct $H$ is counted 4 times, i.e. 
\begin{eqnarray}
\label{eq:general:01:geometric}
f_{01} = 4n_G(\lintree[3] \oplus \lintree[2] \oplus \lintree[2]).
\end{eqnarray}
%\input{4-9-bounds-variance}

% section
\section{Theoretical examples}
\label{sec:theoretical_examples}

In the coming sections we derive compact formulae for different types of graphs, which can be found in table \ref{table:special_graphs_summary}.

%   subsubsection
\subsection{One-regular graphs}

The general characterization of the $f_\omega$'s in table \ref{table:summary_frequencies} based on equation \ref{eq:counting_subgraphs} allows one to see that
\begin{eqnarray*}
f_{01}(\onereg) = f_{021}(\onereg) = f_{022}(\onereg) = f_{03}(\onereg)
                = f_{04}(\onereg) = f_{13}(\onereg) = 0
\end{eqnarray*}
and that

\begin{eqnarray*}
%\label{eq:one-regular:00-24-12}
f_{00}(\onereg) = 6{m \choose 4}, \quad
f_{24}(\onereg) = {m \choose 2}, \quad
f_{12}(\onereg) = 6 {m \choose 3}.
\end{eqnarray*}

The application of the results above to equation \ref{eq:variance:freq-times-exp} with $m = n/2$ yields, after some algebra,
\begin{eqnarray*}
\gvar{C(\onereg)} = \frac{1}{8}n(n - 2)((n - 4)\gexpet{12} + \gexpet{24}),
\end{eqnarray*}
which leads to
\begin{eqnarray*}
%\label{eq:variance:one-regular}
\lvar{C(\onereg)} = \frac{1}{360}(n-2)n(n+6)
\end{eqnarray*} 
by applying the values $\lexpet{\omega}$ in table \ref{table:types_of_combinations}.

% 	subsubsection
\subsection{Quasi-star trees}
\label{sec:quasistar_trees}

The general characterization of the $f_\omega$'s in table \ref{table:summary_frequencies} based on equation \ref{eq:counting_subgraphs} allows one to see that

\begin{equation*}
%\label{eq:quasistar-trees:00-12-04-03-022-021-01}
f_{00}(\quasistar) =
f_{12}(\quasistar) =
f_{04}(\quasistar) = f_{03}(\quasistar) =
f_{022}(\quasistar) = f_{021}(\quasistar) =
f_{01}(\quasistar) = 0.
\end{equation*}

Recalling table \ref{table:special_graphs_summary},
\begin{eqnarray*}
%\label{eq:quasistar-trees:24}
f_{24}(\quasistar) = |Q(\quasistar)| = n - 3.
\end{eqnarray*}
Since all $f_\omega$'s have to add up to $|Q(\quasistar)|^2$ (equation \ref{eq:total_frequency_of_types}), one has
\begin{eqnarray*}
%\label{eq:quasistar-trees:13}
f_{13}(\quasistar) = |Q(\quasistar)|^2 - f_{24}(\quasistar) = (n - 3)(n - 4).
\end{eqnarray*}
Therefore, thanks to equation \ref{eq:variance:freq-times-exp}
\begin{eqnarray*}
\gvar{C(\quasistar)} = (n - 3)( (n - 4)\gexpet{13} + \gexpet{24} )
\end{eqnarray*}
and finally, using the values of $\lexpetw$ in Table \ref{table:types_of_combinations},
\begin{eqnarray*}
%\label{eq:quasi-star-trees-variance}
\lvar{C(\quasistar)} = \frac{1}{18}n(n - 3)
\end{eqnarray*}
for $n \geq 3$ and $\lvar{C(\quasistar)} = 0$ otherwise. 

%	subsubsection
\subsection{Complete graphs}
\label{sec:complete_graphs}

Although $\lvar{C(\complete)}=0$, complete graphs are important as a test of the soundness of the theory and the calculations. Furthermore, there are layouts where $\gvar{C(\complete)}>0$, e.g. random spherical arrangements \cite{Moon1965a}. Then, knowing the $f_\omega$'s of a complete graph is useful for future developments beyond linear arrangements.

The derivation of many of the $f_\omega$'s requires calculating $n_{\complete}(\lintree[n'])$, the number of $n'$-paths in $\complete$, for some $1 < n' \le n$. Some $\lintree[n']$ is obtained with a path starting from a vertex and visiting new vertices $n'-1$ times. Then
\begin{eqnarray}
\label{eq:complete-graph:paths}
n_{\complete}(\lintree[n']) = \frac{1}{2}\prod_{i=0}^{n'-1} (n - i) = \frac{1}{2} \frac{n!}{(n-n')!}
\end{eqnarray}
for $1 < n' \le n$. The $1/2$ factor comes from the fact that the same $\lintree$ is obtained with a walk from the initial to the end vertex but also backwards. $n_{\complete}(\lintree[2]) = |E(\complete)|$ and $n_{\complete}(\lintree) = n!/2$ as expected. Furthermore,
\begin{eqnarray}
\label{eq:complete-graph:two-disj-paths-diff}
  n_{\complete}(\lintree[n_1] \oplus \lintree[n_2])
= n_{\complete}(\lintree[n_1]) n_{\complete[n - n_1]}(\lintree[n_2])
= \frac{1}{4}\frac{n!}{(n - n_1 - n_2)!}
\end{eqnarray}
for $1 < n_1,n_2 \leq n$ with $n_1 + n_2 \leq n$ and $n_1 \neq n_2$. In case $n_1 = n_2$ we have
\begin{eqnarray}
\label{eq:complete-graph:two-disj-paths-equal}
  n_{\complete}(\lintree[n_1] \oplus \lintree[n_2])
= \frac{1}{2}n_{\complete}(\lintree[n_1]) n_{\complete[n - n_1]}(\lintree[n_2])
= \frac{1}{8}\frac{n!}{(n - n_1 - n_2)!}.
\end{eqnarray}

\subsubsection{$\tau=0$, $\phi=0$}
\label{sec:complete_graphs:00}

$f_{00}(\complete)$ is easy to calculate via equation \ref{eq:general:00}. For any $\complete$ it is easy to see that, for any $s \in V(G)$, $G_{-s}$ is also a complete graph. This can also be generalized to $G_{-L}$ with $L \subseteq V(G)$. Adapting the formula for $Q(\complete)$ in table \ref{table:special_graphs_summary}, we obtain $|Q((\complete)_{-stuv})| = 3{n - 4 \choose 4}$. Therefore,
\begin{equation*}
%\label{eq:complete-graphs:00}
f_{00}(\complete)
	= \sum_{\{st,uv\} \in Q(\complete)} |Q((\complete)_{-stuv})|
	= \sum_{\{st,uv\} \in Q(\complete)} 3{n - 4 \choose 4}
	= 630 {n \choose 8}.
\end{equation*}

\subsubsection{$\tau=2$, $\phi=4$}
\label{sec:complete_graphs:24}

Combining equation \ref{eq:general:24:final} and table
\ref{table:special_graphs_summary}, we have that
\begin{eqnarray*}
%\label{eq:complete-graphs:24}
f_{24}(\complete) = |Q(\complete)| = 3{n \choose 4}.
\end{eqnarray*}

\subsubsection{$\tau=1$, $\phi=3$}
\label{sec:complete_graphs:13}

Combining equation \ref{eq:general:13:geometric}, and \ref{eq:complete-graph:two-disj-paths-diff}, one obtains
\begin{eqnarray*}
%\label{eq:complete-graphs:13}
f_{13}(\complete)
	= 2n_{\complete}(\lintree[3]\oplus\lintree[2])
	= 2\frac{1}{4}\frac{n!}{(n - 5)!}
	= 60{n \choose 5}.
\end{eqnarray*}

\subsubsection{$\tau=1$, $\phi=2$}
\label{sec:complete_graphs:12}

We can easily obtain an expression for $f_{12}(\complete)$ via equation \ref{eq:general:12:geometric}, i.e., $6n_{\complete}(\lintree[2]\oplus\lintree[2]\oplus\lintree[2])$. To do so we can first remove an edge of the graph and then count the amount of $\lintree[2]\oplus\lintree[2]$, or the other way around. We use the first way, so
\begin{eqnarray*}
%\label{eq:complete-graphs:12}
f_{04}(\complete)
	&= 6n_{\complete}(\lintree[2]\oplus\lintree[2]\oplus\lintree[2])
	 = \frac{1}{3}6n_{\complete}(\lintree[2])n_{\complete[n-2]}(\lintree[2]\oplus\lintree[2]) \nonumber\\
	&= 2{n \choose 2}3{n-2 \choose 4}
	 = 90{n \choose 6}.
\end{eqnarray*}

\subsubsection{$\tau=0$, $\phi=4$}
\label{sec:complete_graphs:04}

Via equation \ref{eq:general:04}, one obtains
\begin{eqnarray*}
%\label{eq:complete-graphs:04}
f_{04}(\complete)
	&=& \sum_{\{st,uv\} \in Q(\complete)} (a_{su}a_{tv} + a_{sv}a_{tu})
	 = \sum_{\{st,uv\} \in Q(\complete)} 2 \nonumber \\
	&=& 2|Q(\complete)| = 6{n \choose 4}.
\end{eqnarray*}

\subsubsection{$\tau=0$, $\phi=3$}
\label{sec:complete_graphs:03}

Applying equation \ref{eq:complete-graph:paths} to equation \ref{eq:general:03:geometric}, one obtains
\begin{eqnarray*}
f_{03}(\complete)
	&=& n(n-1)(n-2)(n-3)(n-4) = 120{n \choose 5}.
%\label{eq:complete-graphs:03}
\end{eqnarray*}

\subsubsection{$\tau=0$, $\phi=2$, subtype $1$}
\label{sec:complete_graphs:021}

Applying equation \ref{eq:complete-graph:two-disj-paths-diff} to equation \ref{eq:general:021:geometric}, one obtains
\begin{eqnarray*}
%\label{eq:complete-graphs:021}
f_{021}(\complete)
	= 2\frac{1}{4}\frac{n!}{(n - 6)!}
	= 360{n \choose 6}.
\end{eqnarray*}

\subsubsection{$\tau=0$, $\phi=2$, subtype $2$}
\label{sec:complete_graphs:022}

Likewise, by applying \ref{eq:complete-graph:two-disj-paths-equal} to equation \ref{eq:general:022:geometric}, we obtain immediately
\begin{eqnarray*}
%\label{eq:complete-graphs:022}
f_{022}(\complete)
	= 4\frac{1}{8}\frac{n!}{(n - 6)!}
	= 360{n \choose 6}.
\end{eqnarray*}

Notice that $f_{021}(\complete) = f_{022}(\complete)$.

\subsubsection{$\tau=0$, $\phi=1$}
\label{sec:complete_graphs:01}

We rely on equation \ref{eq:general:01:geometric} to calculate $f_{01}(\complete)$. First, equation \ref{eq:complete-graph:two-disj-paths-diff} indicates that 
\begin{eqnarray*}
n_{\complete}(\lintree[3] \oplus \lintree[2])
= \frac{1}{2}{n \choose 2}3{n-2 \choose 3},
\end{eqnarray*}
which allows one to calculate
\begin{eqnarray*}
n_{\complete}(\lintree[3] \oplus \lintree[2] \oplus \lintree[2])
	= {n \choose 2} n_{\complete[n-2]}(\lintree[3] \oplus \lintree[2])
	= \frac{1}{16} \prod_{i=0}^6 (n-i) 
	= 315{n \choose 7}.
\end{eqnarray*}
Finally,
\begin{eqnarray*}
%\label{eq:complete-graphs:01}
f_{01}(\complete)
	= 4n_{\complete}(\lintree[3]\oplus\lintree[2]\oplus\lintree[2])
	= 1260{n \choose 7}.
\end{eqnarray*}

\subsubsection{Variance}
\label{sec:complete_graphs:variance}

Inserting the results above into equation \ref{eq:variance:freq-times-exp} one obtains
\begin{eqnarray*}
\gvar{C(\complete)}
	&=& 3{n \choose 4} ((n - 4)(n - 5)(\gexpet{12} + 4(\gexpet{021} + \gexpet{022})) \\
	&&\phantom{3n4}	+ 4(n - 4)(\gexpet{13} + 2\gexpet{03}) \\
	&&\phantom{3n4}	+ 2\gexpet{04} + \gexpet{24})
\end{eqnarray*}
With the help of Table \ref{table:types_of_combinations}, we verify that
\begin{eqnarray*}
\lvar{C(\complete)} = 0
\end{eqnarray*}
as expected.

%	subsubsection
\subsection{Complete bipartite graphs}
\label{sec:comp_bip_graphs}

We derive the $f_\omega$'s for $\compbip$ by mere counting of subgraphs as in the previous section but with the support of a new figure (figure \ref{fig:complete-graphs:counting}).

\subsubsection{$\tau=0$, $\phi=0$}
\label{sec:comp_bip_graphs:00}

We follow the same technique used for complete graphs, i.e. 
\begin{eqnarray*}
f_{00}(\compbip)
	&=& \sum_{\{st,uv\} \in Q(\compbip)} |Q((\compbip)_{-stuv})| \\
	&=& \sum_{\{st,uv\} \in Q(\compbip)} 2{n_1 - 2 \choose 2}{n_2 - 2 \choose 2} \\
	&=& |Q(\compbip)|\cdot|Q(\compbip[n_1 - 2,n_2 - 2])| \\
	&=& 2{n_1 - 2 \choose 2}{n_2 - 2 \choose 2} 2{n_1 \choose 2}{n_2 \choose 2} \\
	&=& 144{n_1 \choose 4} {n_2 \choose 4}.
\end{eqnarray*}

\subsubsection{$\tau=2$, $\phi=4$}
\label{sec:comp_bip_graphs:24}

Equation \ref{eq:general:24:final} gives
\begin{eqnarray*}
%\label{eq:comp-bip-graphs:24}
f_{24}(\compbip) = |Q(\compbip)| = 2{n_1 \choose 2}{n_2 \choose 2}.
\end{eqnarray*}

\subsubsection{$\tau=1$, $\phi=3$}
\label{sec:comp_bip_graphs:13}

Via equation \ref{eq:general:13:geometric}
\begin{eqnarray*}
f_{13}(\compbip)
	& = & 2\sum_{st\in E(\compbip)} n_{\compbip[n_1-1,n_2-1]}(\lintree[3])
	  =   2|E(\compbip)|n_{\compbip[n_1-1,n_2-1]}(\lintree[3]).
\end{eqnarray*}
Through figure \ref{fig:comp-bip-graphs:counting:13} we can easily calculate the amount of $\lintree[3]$
\begin{eqnarray*}
%\label{eq:comp-bip-graphs:13}
f_{13}(\compbip)
	& = & 2n_1n_2\frac{1}{2}((n_1 - 1)(n_2 - 1)(n_1 - 2) + (n_2 - 1)(n_1 - 1)(n_2 - 2)) \nonumber\\
	& = & 12{n_1 \choose 3}{n_2 \choose 2} + 12{n_1 \choose 2}{n_2 \choose 3}.
\end{eqnarray*}

\subsubsection{$\tau=1$, $\phi=2$}
\label{sec:comp_bip_graphs:12}

Equation \ref{eq:general:12:geometric} gives
\begin{eqnarray*}
f_{12}(\compbip)
	&=& 6 n_{\compbip}(\lintree[2] \oplus \lintree[2] \oplus \lintree[2]) \\
	&=& 6 \frac{1}{3} |E(\compbip)| n_{\compbip[n_1-1,n_2-1]}(\lintree[2] \oplus \lintree[2]) \\
	&=& 2 |E(\compbip)| |Q(\compbip[n_1-1,n_2-1])| \\
	&=& 36{n_1 \choose 3}{n_2 \choose 3}.
\end{eqnarray*}

\subsubsection{$\tau=0$, $\phi=4$}
\label{sec:comp_bip_graphs:04}

Recall equation \ref{eq:general:04:geometric}. Checking the way of building a path of four vertices in figure \ref{fig:comp-bip-graphs:counting:04}, and noting that  a $\cycle[4]$ can be generated in four different ways depending on the initial vertex, we obtain 
\begin{eqnarray*}
n_{\compbip}(\cycle[4])
&= \frac{1}{4} n_1n_2(n_1 - 1)(n_2 - 1) =
   {n_1 \choose 2}{n_2 \choose 2},
\end{eqnarray*}
hence 
\begin{eqnarray*}
%\label{eq:comp-bip-graphs:04}
f_{04}(\compbip) = 2{n_1 \choose 2}{n_2 \choose 2}.
\end{eqnarray*}

\subsubsection{$\tau=0$, $\phi=3$}
\label{sec:comp_bip_graphs:03}

Recall equation \ref{eq:general:03:geometric}. The procedure to count all subgraphs isomorphic to $\lintree[5]$ resembles the procedure to count all the $\lintree[3] \oplus \lintree[2]$ for type 13 (figure \ref{fig:comp-bip-graphs:counting:13}), though we need to continue with two more vertices (figure \ref{fig:comp-bip-graphs:counting:03}). Then
\begin{align*}
n_{\compbip}(\lintree[5])
	&= \frac{1}{2} ( n_1n_2(n_1 - 1)(n_2 - 1)(n_1 - 2) + n_2n_1(n_2 - 1)(n_1 - 1)(n_2 - 2) )\\
	&= n_{\compbip}(\lintree[3] \oplus \lintree[2]),
\end{align*}
which implies
\begin{eqnarray*}
%\label{eq:comp-bip-graphs:03}
f_{03}(\compbip) = f_{13}(\compbip).
\end{eqnarray*}

\subsubsection{$\tau=0$, $\phi=2$ subtype 1}
% \label{sec:comp_bip_graphs:02}

Figure \ref{fig:comp-bip-graphs:counting:021} illustrates how to build pairs $\lintree[4] \oplus \lintree[2]$. Therefore,
\begin{eqnarray*}
n_{\compbip}(\lintree[4] \oplus \lintree[2])
	&=& n_{\compbip}(\lintree[6])
	 =  n_1n_2(n_1 - 1)(n_2 - 1)(n_1 - 2)(n_2 - 2)\\
	&=& 36{n_1 \choose 3}{n_2 \choose 3},
\end{eqnarray*}
hence equation \ref{eq:general:021:geometric} becomes
\begin{eqnarray*}
%\label{eq:comp-bip-graphs:021}
f_{021}(\compbip) = 72{n_1 \choose 3}{n_2 \choose 3}.
\end{eqnarray*}

\subsubsection{$\tau=0$, $\phi=2$ subtype 2}

Equation \ref{eq:general:022:geometric} indicates that $f_{021}$ counts over $\lintree[3] \oplus \lintree[3]$. Figure \ref{fig:comp-bip-graphs:counting:022} shows the procedure followed to obtain these subgraphs when starting at the left partition of vertices. The procedure when starting at the right partition is completely symmetric. Therefore, 
\begin{flalign*}
n_{\compbip}(\lintree[3] \oplus \lintree[3])
& =  \frac{1}{2} n_1n_2(n_1 - 1)\left( \frac{1}{2}(n_1 - 2)(n_2 - 1)(n_1 - 3)
  +  \frac{1}{2}(n_2 - 1)(n_1 - 2)(n_2 - 2) \right) \\
& +  \frac{1}{2} n_2n_1(n_2 - 1)\left( \frac{1}{2}(n_2 - 2)(n_1 - 1)(n_2 - 3)
  +  \frac{1}{2}(n_1 - 1)(n_2 - 2)(n_1 - 2) \right) \\
& =  12{n_1 \choose 2}{n_2 \choose 4} + 12{n_1 \choose 4}{n_2 \choose 2} + 18{n_1 \choose 3}{n_2 \choose 3}
\end{flalign*}
and then 
\begin{eqnarray*}
f_{022}(\compbip)
	&=& 4n_{\compbip}(\lintree[3] \oplus \lintree[3]) \\
	&=& 24{n_1 \choose 2}{n_2 \choose 4} + 24{n_1 \choose 4}{n_2 \choose 2} + 36{n_1 \choose 3}{n_2 \choose 3}.
%\label{eq:comp-bip-graphs:022}
\end{eqnarray*}

\subsubsection{$\tau=0$, $\phi=1$}
\label{sec:comp_bip_graphs:01}

The paths depicted in figure \ref{fig:comp-bip-graphs:counting:01} allow one to see that after removing some $\lintree[3]$ we only have to count all the $\lintree[2]\oplus\lintree[2]$, i.e. the size of $Q$ of the resulting graph, yielding
\begin{eqnarray*}
n_{\compbip}(\lintree[3] \oplus \lintree[2] \oplus \lintree[2])
	&=& \frac{1}{2}n_1n_2(n_1 - 1)n_{\compbip[n_1-2,n_2-1]}(\lintree[2] \oplus \lintree[2]) \\
	&+& \frac{1}{2}n_2n_1(n_2 - 1)n_{\compbip[n_1-1,n_2-2]}(\lintree[2] \oplus \lintree[2]) \\
	&=& \frac{1}{2}n_1n_2(n_1 - 1)|Q(\compbip[n_1-2,n_2-1])| \\
	&+& \frac{1}{2}n_2n_1(n_2 - 1)|Q(\compbip[n_1-1,n_2-2])| \\
	&=& 36{n_1 \choose 4}{n_2 \choose 3} + 36{n_1 \choose 3}{n_2 \choose 4}.
\end{eqnarray*}
Then, equation \ref{eq:general:01:geometric} becomes
\begin{eqnarray*}
%\label{eq:comp-bip-graphs:01}
f_{01}(\compbip) = 144{n_1 \choose 4}{n_2 \choose 3} + 144{n_1 \choose 3}{n_2 \choose 4}.
\end{eqnarray*}

\subsubsection{Variance}
\label{sec:comp_bip_graphs:variance}

Using the results above, we obtain
\begin{eqnarray}
\label{eq:var-bip-graphs}
\gvar{C(\compbip)}
&=\phantom{+}& 	 2(\gexpet{24} + \gexpet{04}) {n_1 \choose 2}{n_2 \choose 2} \nonumber\\
&\phantom{=}&  + 12(\gexpet{03} + \gexpet{13})\left[ {n_1 \choose 3}{n_2 \choose 2} + {n_1 \choose 2}{n_2 \choose 3} \right] \nonumber\\
&\phantom{=}&  + 36(\gexpet{12} + \gexpet{022} + 2\gexpet{021}) {n_1 \choose 3}{n_2 \choose 3} \nonumber\\
&\phantom{=}&  + 24\gexpet{022}\left[{n_1 \choose 2}{n_2 \choose 4} + {n_1 \choose 4}{n_2 \choose 2} \right].
\end{eqnarray}
Finally,
\begin{eqnarray*}
\lvar{C(\compbip)}
	&=& \frac{1}{90}{n_1 \choose 2}{n_2 \choose 2}((n_1 + n_2)^2 + n_1 + n_2)
\end{eqnarray*}
after several algebraic manipulations. Given that $\startree = \compbip[1,n-1]$, equation \ref{eq:var-bip-graphs} gives $\gvar{C(\startree)} = 0$ as expected by $\gC(\startree)=0$. 

\begin{figure}
	\centering
	\begin{subfigure}[t]{0.3\textwidth}
		\centering
		\includegraphics[scale=0.6]{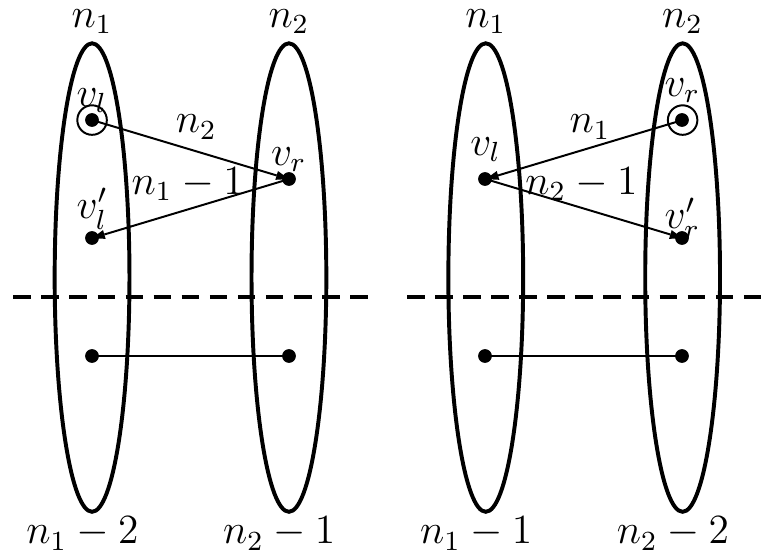}
		\caption{Pairs of independent $3$-paths and $2$-paths.}
		\label{fig:comp-bip-graphs:counting:13}
	\end{subfigure}
	\begin{subfigure}[t]{0.3\textwidth}
		\centering
		\includegraphics[scale=0.6]{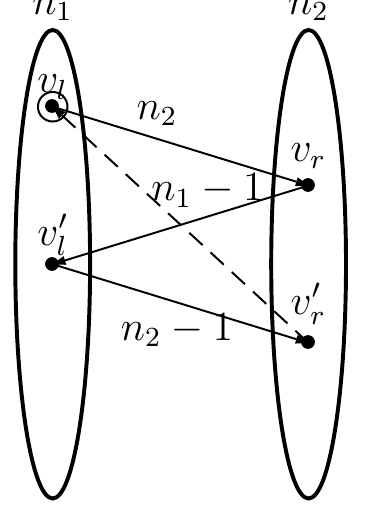}
		\caption{Cycles of four vertices.}
		\label{fig:comp-bip-graphs:counting:04}
	\end{subfigure}
	\begin{subfigure}[t]{0.3\textwidth}
		\centering
		\includegraphics[scale=0.6]{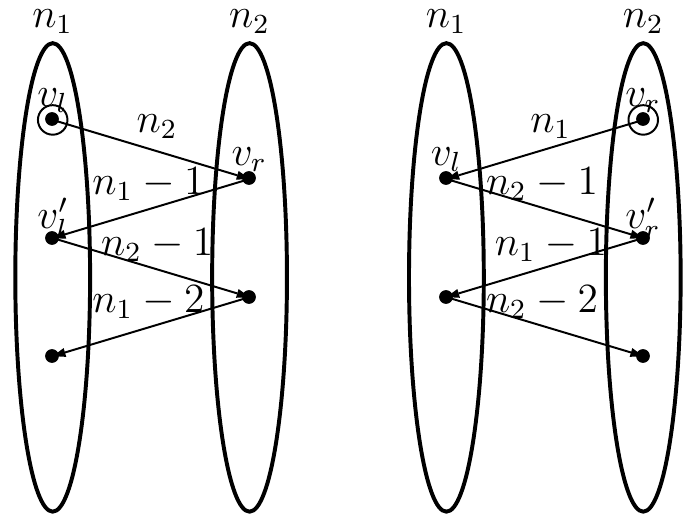}
		\caption{Paths of five vertices.}
		\label{fig:comp-bip-graphs:counting:03}
	\end{subfigure}
\end{figure}
\begin{figure}
	\ContinuedFloat
	\centering
	\begin{subfigure}[t]{0.3\textwidth}
		\centering
		\includegraphics[scale=0.6]{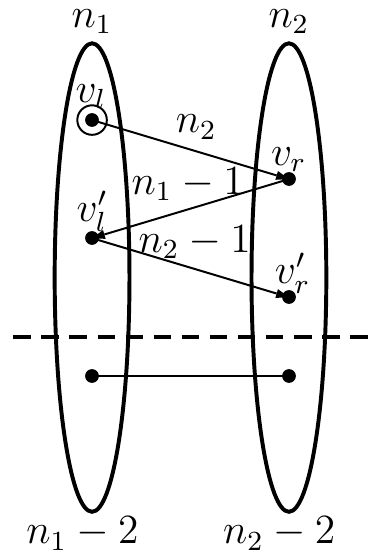}
		\caption{Pairs of independent $4$-paths and $2$-paths.}
		\label{fig:comp-bip-graphs:counting:021}
	\end{subfigure}
	\begin{subfigure}[t]{0.3\textwidth}
		\centering
		\includegraphics[scale=0.6]{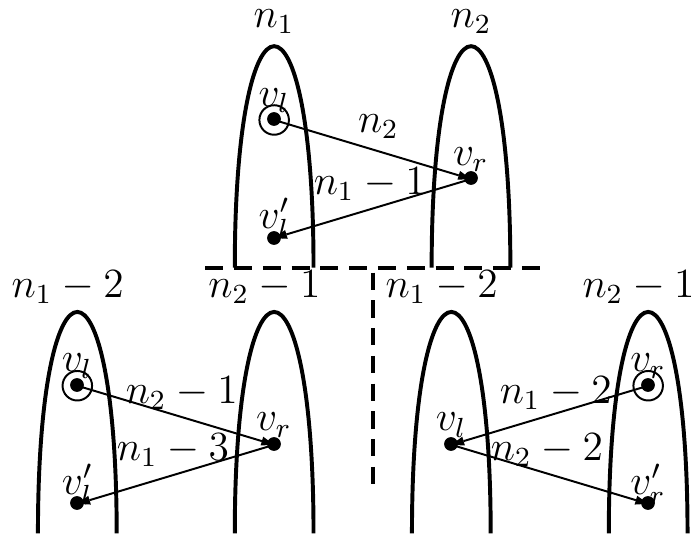}
		\caption{Half of the pairs of independent $3$-paths.}
		\label{fig:comp-bip-graphs:counting:022}
	\end{subfigure}
	\begin{subfigure}[t]{0.3\textwidth}
		\centering
		\includegraphics[scale=0.6]{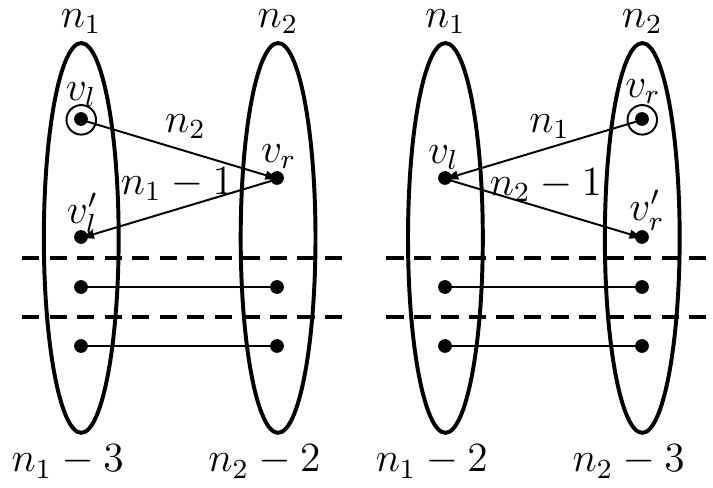}
		\caption{Triples of independent $3$-, $2$-, $2$-paths.}
		\label{fig:comp-bip-graphs:counting:01}
	\end{subfigure}
	\caption{Paths in $\compbip$. $v_l$, $v_l'$, $v_r$ and $v_r'$ are
	vertices, drawn using dots. Circled dots are used to indicate the vertex that is the start of a path.}
	\label{fig:complete-graphs:counting}
\end{figure}

%	subsubsection
\subsection{Cycle graphs}
\label{sec:cycle_graphs}

In this section we assume a labeling of the vertices of $\cycle$ from $1$ to $n$ and that its set of edges is then formed by consecutively labeled vertices
\begin{eqnarray*}
E(\cycle) = \{ \{s,t\} \;|\; 1 \le s,t \le n,\; s - t \equiv 1 \mod n \}.
\end{eqnarray*}
In cycle graphs $|E(\cycle)| = |V(\cycle)| = n$. The vertex labels define a circular arrangement of the vertices, namely the placement of the vertices on a circle \cite{Deutsch2002a}.

In this section, let $e_1,e_2\in \mathbb{N}$ be edge indices, $e_1 \neq e_2$, and $\{e_1,e_2\}\in Q$ be formed by the $e_1$-th and $e_2$-th edges.

Recall the value of $|Q(\cycle)|$ in table \ref{table:special_graphs_summary}. Before we start calculating the $f_\omega$'s, we note some useful general properties. Firstly, 
\begin{eqnarray}
\label{what_is_the_size_of_Q}
|Q| = n_G(\lintree[2]\oplus\lintree[2]).
\end{eqnarray}
Secondly,
\begin{eqnarray}
\label{number_of_linear_trees_within_cycle}
n_{\cycle[n_1]}(\lintree[n_2]) = n_1
\end{eqnarray}
for $n_2 \leq n_1$. Thirdly,
\begin{eqnarray}
\label{number_of_linear_trees_within_linear_tree}
n_{\lintree[n_1]}(\lintree[n_2]) = n_1 - n_2 + 1
\end{eqnarray}
for $n_2 \leq n_1$. Finally, we define $\cycle[n_1]\setminus\lintree[n_2]$ as the outcome of removing some $\lintree[n_2]$ from $\cycle[n_1]$ with $n_2 \leq n_1$, namely
\begin{equation}
\cycle[n_1]\setminus\lintree[n_2] = \lintree[n_1 - n_2].
\label{removal_of_linear_tree_from_cycle}
\end{equation} 
 
\subsubsection{$\tau=0$, $\phi=0$}
\label{sec:cycle_graphs:00}

\begin{figure}
\begin{center}
	\begin{subfigure}{0.32\textwidth}
		\includegraphics[scale=0.6]{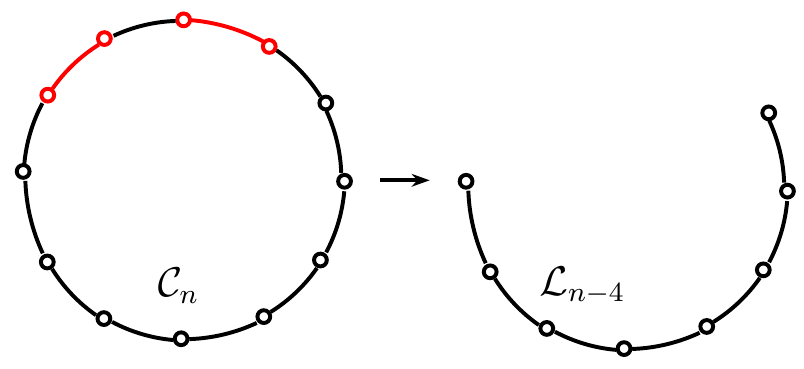}
		\caption{Case (\ref{item:cycle-graphs:00:1}).}
		\label{fig:cycle-remove-edges:00:case-1}
	\end{subfigure}
	\begin{subfigure}{0.32\textwidth}
		\includegraphics[scale=0.6]{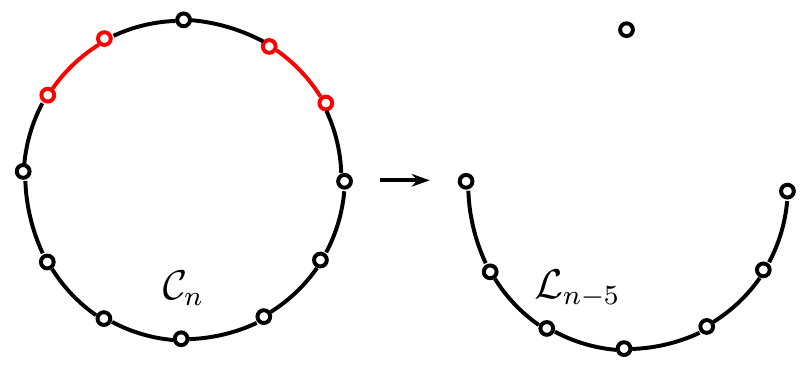}
		\caption{Case (\ref{item:cycle-graphs:00:2}).}
		\label{fig:cycle-remove-edges:00:case-2}
	\end{subfigure}
	\begin{subfigure}{0.32\textwidth}
		\includegraphics[scale=0.6]{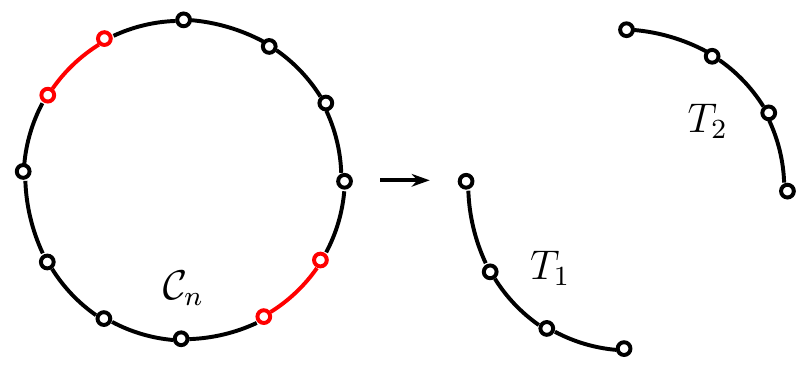}
		\caption{Case (\ref{item:cycle-graphs:00:3}).}
		\label{fig:cycle-remove-edges:00:case-3}
	\end{subfigure}
	\caption{Different graphs produced by removing pairs of red-colored edges from $\cycle$.}
	\label{fig:cycle-remove-edges:00}
\end{center}
\end{figure}

The rationale behind equation \ref{what_is_the_size_of_Q} allows one to express equation \ref{eq:general:00:geometric} as
\begin{eqnarray}
f_{00}(\cycle) & = & 6 n_{\cycle}(\lintree[2]\oplus\lintree[2]\oplus\lintree[2]\oplus\lintree[2]) \nonumber \\
               & = & 6 \sum_{\{st, uv\} \in Q} n_{(\cycle)_{-stuv}}(\lintree[2]\oplus\lintree[2]) \nonumber \\
               & = & 6 \sum_{\{st, uv\} \in Q} |Q((\cycle)_{-stuv})|. \label{eq:cycle-graphs:00:early}
\end{eqnarray}
Crucially, $(\cycle)_{-stuv}$ can only be of three mutually exclusive types depending on the ``distance'' between the vertices $s$, $t$, $u$, and $v$ in the circular arrangement, i.e.
\begin{enumerate}
	\item \label{item:cycle-graphs:00:1}
	$\lintree[n-4]$, when the edges $st$ and $uv$ are separated by just one edge in the circular arrangement  (figure \ref{fig:cycle-remove-edges:00:case-1}). This type needs $n \geq 5$. There are $n$ pairs of independent edges that are separated by one edge in the circular arrangement. These are the pairs $\{e_1,e_2\}$ such that $1 \le e_1, e_2 \le n$, $e_2 - e_1 \equiv 2 \mod n$, i.e.
	\begin{equation*}
		\{1,3\}, \{2,4\}, \{3,5\}, \cdots, \{n-3, n-1\}, \{n-2, n\}, \{n-1, 1\}, \{n, 2\}.
	\end{equation*}
	We have  
	\begin{eqnarray*}
		|Q((\cycle)_{-stuv})| = |Q(\lintree[n-4])| = {n - 6 \choose 2}.
	\end{eqnarray*}
	
	\item \label{item:cycle-graphs:00:2}
	$\zeroreg[1] \oplus \lintree[n-5]$, when the edges $st$ and $uv$ are separated by two edges in the circular arrangement (figure \ref{fig:cycle-remove-edges:00:case-2}). This type needs $n \geq 6$. There are $n$ pairs of independent edges that are separated by two edges in the circular arrangement. These pairs are the $\{e_1,e_2\}$ such that $1 \le e_1, e_2 \le n$, $e_2 - e_1 \equiv 3 \mod n$, i.e.
	\begin{equation*}
		\{1,4\}, \{2,5\}, \{3,6\}, \cdots, \{n-4, n-1\}, \{n-3, n\}, \{n-2, 1\}, \{n-1, 2\}, \{n, 3\}.
	\end{equation*}   
	We have 
	\begin{eqnarray*}
		|Q((\cycle)_{-stuv})| = |Q(\zeroreg[1] \oplus \lintree[n-5])| = |Q(\lintree[n-5])| = {n - 7 \choose 2}.
	\end{eqnarray*}
	
	\item \label{item:cycle-graphs:00:3}
	$T_1 \oplus T_2$, where $T_1$ and $T_2$ are a couple of linear trees, $\lintree[n_1]$ and $\lintree[n_2]$, such that $1 \leq n_1,n_2$ and $n_1 + n_2 = n -4$ (figure \ref{fig:cycle-remove-edges:00:case-3}). This type needs $n \geq 7$.  The amount of pairs of edges leading to this forest of two trees is just all the elements of $Q(\cycle)$ except for those leading to the previous two scenarios, i.e.
	\begin{eqnarray*}
		|Q(\cycle)| - 2n = \frac{n(n-5)}{2}.
	\end{eqnarray*}
	The size of $|Q((\cycle)_{-stuv})|$ is the same for all pairs of linear trees. As $T_1$ and $T_2$ do not share edges, 
	\begin{eqnarray*}
	|Q(T_1 \oplus T_2)|
		& = & |Q(T_1)| + |Q(T_2)| + |E(T_1)| \cdot |E(T_2)| \\
		& = & |Q(\lintree[n_1])| + |Q(\lintree[n_2])| + |E(\lintree[n_1])| \cdot |E(\lintree[n_2])| \\ 
		& = & {n_1 - 2 \choose 2} + {n_2 - 2 \choose 2} + (n_1 - 1)(n_2 - 1).
	\end{eqnarray*} 
	The substitution $n_2 = n - 4 - n_1$, leads to
	\begin{eqnarray*} 
		|Q(T_1 \oplus T_2)| = \frac{1}{2}(n-15)n + 29
	\end{eqnarray*}
	after some algebra.
\end{enumerate}
These types are illustrated in figure \ref{fig:cycle-remove-edges:00}.

As a result of the case analysis above, equation \ref{eq:cycle-graphs:00:early} can be written as 
\begin{eqnarray*}
f_{00}(\cycle[n])
	= 6\left[
		n|Q(\lintree[n-4])| + n|Q(\lintree[n-5])| + \frac{n(n-5)}{2}|Q(T_1 \oplus T_2)|
	\right],
\end{eqnarray*}
which gives
\begin{eqnarray*}
%\label{eq:cycle-graphs:00}
f_{00}(\cycle) = \frac{3}{2}n{n - 5 \choose 3}.
\end{eqnarray*}
after some algebra.

\subsubsection{$\tau=2$, $\phi=4$}
\label{sec:cycle_graphs:24}

This type is trivial. By recalling equation \ref{eq:general:24:final}
\begin{eqnarray*}
%\label{eq:cycle-graphs:24}
f_{24}(\cycle) = |Q(\cycle)| = \frac{1}{2}n(n - 3).
\end{eqnarray*}

\subsubsection{$\tau=1$, $\phi=3$}
\label{sec:cycle_graphs:13}

Thanks to equations \ref{removal_of_linear_tree_from_cycle} and \ref{number_of_linear_trees_within_linear_tree},
\begin{eqnarray}
\label{eq:cycle-minus-lintree}
n_{\cycle[n_1]\setminus \lintree[n_2]}(\lintree[n_3]) =
n_{\lintree[n_1-n_2]}(\lintree[n_3]) = n_1 - n_2 - n_3 + 1.
\end{eqnarray}
Then, by applying equations \ref{eq:cycle-minus-lintree} and \ref{number_of_linear_trees_within_cycle} to equation \ref{eq:general:13:geometric}, we get
\begin{eqnarray*}
f_{13}(\cycle)
	&=& 2n_{\cycle}(\lintree[3]\oplus\lintree[2])
	 =  2n_{\cycle}(\lintree[3])n_{\cycle[n]\setminus \lintree[3]}(\lintree[2]) \\
	&=& 2n_{\cycle}(\lintree[3])n_{\lintree[n-3]}(\lintree[2])
	 =  2n(n - 3 - 2 + 1) = 2n(n - 4).
\end{eqnarray*}

\subsubsection{$\tau=1$, $\phi=2$}
\label{sec:cycle_graphs:12}

Thanks to equation \ref{eq:general:12:geometric}, we have that 
\begin{eqnarray*}
f_{12}(\cycle)
	&=& 6n_{\cycle}(\lintree[2]\oplus\lintree[2]\oplus\lintree[2]) \\
    &=& \frac{6}{3}\sum_{st \in E} n_{(\cycle)_{-st}}(\lintree[2]\oplus\lintree[2]).
%\label{eq:cycle-graphs:12:early}
\end{eqnarray*}
As removing an edge from $\cycle$ produces $\lintree[n-2]$, the rationale behind equation \ref{what_is_the_size_of_Q} leads to
\begin{eqnarray*}
f_{12}(\cycle)
	&=& 2 n \cdot n_{\lintree[n-2]}(\lintree[2]\oplus\lintree[2]) \\
	&=& 2 n |Q(\lintree[n-2])| \\
	&=& n(n - 4)(n - 5)
%\label{eq:cycle-graphs:12} 
\end{eqnarray*}
after some algebra. 

\subsubsection{$\tau=0$, $\phi=4$}
\label{sec:cycle_graphs:04}

Applying the fact that $n_{\cycle}(\cycle[4])$ is 1 if $n=4$ and 0 otherwise to equation \ref{eq:general:04:geometric}, produces
\begin{eqnarray*}
%\label{eq:cycle-graphs:04}
f_{04}(\cycle) = 
\begin{cases}
2 & \text{ if } n = 4\\
0 & \text{ otherwise.}
\end{cases}
\end{eqnarray*}

\subsubsection{$\tau=0$, $\phi=3$}
\label{sec:cycle_graphs:03}

Applying equation \ref{number_of_linear_trees_within_cycle} with $n_2=5$ to equation \ref{eq:general:03:geometric}, we obtain
\begin{eqnarray*}
%\label{eq:cycle-graphs:03}
f_{03}(\cycle) = 2n.
\end{eqnarray*}

\subsubsection{$\tau=0$, $\phi=2$, subtype 1}

Thanks to equation \ref{eq:general:021:geometric},
\begin{eqnarray*}
f_{021}(\cycle)
	= 2n_{\cycle}(\lintree[4]\oplus\lintree[2])
	= 2n_{\cycle}(\lintree[4])n_{\lintree[n-4]}(\lintree[2]).
\end{eqnarray*}
As $n_{\cycle}(\lintree[4])=n$ (equation \ref{number_of_linear_trees_within_cycle}) and $n_{\lintree[n-4]}(\lintree[2]) = n-5$ (equation \ref{number_of_linear_trees_within_linear_tree}), we finally obtain
\begin{eqnarray*}
%\label{eq:cycle-graphs:021}
f_{021}(\cycle) = 2n(n-5).
\end{eqnarray*}

\subsubsection{$\tau=0$, $\phi=2$, subtype 2}

Following the procedure for type 021, equation \ref{eq:general:022:geometric} leads to 
\begin{eqnarray*}
%\label{eq:cycle-graphs:022}
f_{022}(\cycle)
	&=& 4n_{\cycle}(\lintree[3] \oplus \lintree[3]) 
	 =  4 \cdot \frac{1}{2}
	 	n_{\cycle}(\lintree[3]) n_{\lintree[n-3]}(\lintree[3])
	 =  2n(n-5).
\end{eqnarray*}
Notice that $f_{021}(\cycle) = f_{022}(\cycle)$.

\subsubsection{$\tau=0$, $\phi=1$}
\label{sec:cycle_graphs:01}

Equation \ref{eq:general:01:geometric}, leads to
\begin{eqnarray*}
f_{01}(\cycle)
	&=& 4 n_{\cycle}(\lintree[3] \oplus \lintree[2] \oplus \lintree[2])
     =  4 \sum_{\lintree[3]} n_{\cycle \setminus \lintree[3]}(\lintree[2] \oplus \lintree[2]) \\
	&=& 4 n_{\cycle}(\lintree[3]) |Q(\lintree[n-3])|
	 =  4 n {n - 5\choose 2} \\
   	&=& 2n(n-5)(n-6).
\end{eqnarray*}

\subsubsection{Variance}
\label{sec:cycle_graphs:variance}

Using equation \ref{eq:variance:freq-times-exp}, the results above give
\begin{eqnarray*}
\gvar{C(\cycle)}
	&=& \frac{1}{2}n (
		4(n-4)\gexpet{13} + (n-3)\gexpet{24} + 4\gexpet{03} \\
	&& + (n-5)[2(n-4)\gexpet{12} + 4(\gexpet{021} + \gexpet{022})] )
\end{eqnarray*}
which leads to (thanks to table \ref{table:types_of_combinations})
\begin{eqnarray*}
\lvar{C(\cycle)}
	= \frac{1}{45}n^3 + \frac{1}{90}n^2 - \frac{1}{3}n.
\end{eqnarray*}

\subsubsection{Notes on the occurrences of edges}
\label{sec:cycle:edge-frequency}

In this section we study the amount of times an edge $e$ is involved in the elements of $Q(G)\times Q(G)$ of a certain type $\omega$. With this information, we obtain the $f_\omega(\lintree)$ in a much more systematic way in section \ref{sec:linear_trees}, with the help of the fact that a linear tree is obtained by removing one edge from a cycle graph. First, for a simple graph $G$ we define the set of elements of $Q(G)\times Q(G)$ of a certain type $\omega$, $\rho_\omega(G)\subseteq Q(G)\times Q(G)$, and the set of elements of $Q(G)\times Q(G)$ of a certain type $\omega$ that contain a certain edge $e\in E$, $\rho_\omega(G, e)\subseteq Q(G)\times Q(G)$. Then
\begin{eqnarray*}
\rho_\omega(G)
	= \{ (q_1, q_2) \in Q(G)\times Q(G) \;|\;
		\mathcal{T}(q_1,q_2)=\omega\}, \\
\rho_\omega(G, e)
	= \{ (q_1, q_2) \in \rho_\omega(G) \;|\;
		e \in q_1 \cup q_2
	\}.
%\label{eq:rho-i-of-edge}
\end{eqnarray*}
Notice
\begin{eqnarray*}
|\rho_\omega(G, e)| =
	\sum_{q_1\in Q}
	\sum_{
		\substack{
			q_2\in Q \;: \\
			e \in q_1\cup q_2 \\
			\mathcal{T}(q_1,q_2)=\omega
		}
	} 1.
\end{eqnarray*}
Interestingly,
\begin{eqnarray}
\label{eq:sum-cardinality}
\sum_{e\in E} |\rho_\omega(G,e)| = c_\omega f_\omega(G),
\end{eqnarray}

where, given a pair $(q_1,q_2)\in Q\times Q$ of type $\omega$,
\begin{eqnarray*}
c_\omega = c_\omega(q_1,q_2) =
	\sum_{
		\substack{
			e\in E \;: \\
			e \in q_1\cup q_2
		}
	} 1
\end{eqnarray*}
is the number of distinct edges of any element of $Q\times Q$ of type $\omega$. Therefore, $c_\omega=c_\omega(q_1,q_2)=|q_1\cup q_2|=4-\tau$. It is easy to see that
\begin{eqnarray}
\label{eq:values-of-ci}
c_{00} = c_{01} = c_{021} = c_{022} = c_{03} = c_{04} = 4, \quad
c_{12} = c_{13} = 3, \quad
c_{24} = 2.
\end{eqnarray}
Now we show that equation \ref{eq:sum-cardinality} is true. Recall the definition of the $f_\omega$ in equation \ref{eq:frequency_template}. Let us change it slightly as
\begin{eqnarray*}
f_\omega =
	\sum_{q_1\in Q}
	\sum_{\substack{q_2\in Q \;: \\ \mathcal{T}(q_1,q_2) = \omega}}
	\frac{1}{c_\omega(q_1,q_2)}
	\sum_{\substack{e\in E \;: \\e\in q_1\cup q_2}} 1.
\end{eqnarray*}
By floating the inner-most summation, we obtain
\begin{eqnarray*}
f_\omega
	&=
	\frac{1}{c_\omega}
	\sum_{e\in E}
	\sum_{q_1\in Q}
	\sum_{
		\substack{
			q_2\in Q \;: \\
			e\in q_1 \cup q_2 \\
			\mathcal{T}(q_1,q_2)=\omega
		}
	}
	1 \\
	&=
	\frac{1}{c_\omega}
	\sum_{e\in E}
	|\rho_\omega(G,e)|,
\end{eqnarray*}
hence equation \ref{eq:sum-cardinality}.

Whereas equation \ref{eq:sum-cardinality} is true for general graphs, in a cycle graph one has
\begin{eqnarray}
\label{eq:equality-cardinalities}
|\rho_\omega(\cycle,e_1)| = |\rho_\omega(\cycle,e_2)| = \cdots = |\rho_\omega(\cycle,e_n)|.
\end{eqnarray}
In words, the amount of occurrences of an edge of $\cycle$ in the elements of $Q(\cycle)\times Q(\cycle)$ is the same independently of the edge. This is true due to the graph's structure, i.e., the properties of one edge are the same as any other edges' properties. We use $|\rho_\omega(\cycle,e_*)|$ to denote any of the $|\rho_\omega(\cycle,e_j)|$, $1\le j\le n$, where $e_*$ denotes any edge. Thus, equation \ref{eq:sum-cardinality} becomes
\begin{eqnarray*}
\sum_{e\in E(\cycle)} |\rho_\omega(\cycle,e)|
	= n|\rho_\omega(\cycle,e_*)| = c_\omega f_\omega(G).
\end{eqnarray*}
Finally,
\begin{eqnarray}
\label{eq:rho-i-cycle-edges}
|\rho_\omega(\cycle,e_*)| = \frac{c_\omega}{n}f_\omega(\cycle).
\end{eqnarray}
Equation \ref{eq:equality-cardinalities} may also hold for other types of graphs, e.g., regular graphs, but such analysis is beyond the scope of the present article.

%	subsubsection
\subsection{Linear trees}
\label{sec:linear_trees}

In order to calculate the $f_\omega(\lintree)$ for every $\omega$ we apply the following strategy, whenever necessary:

\begin{enumerate}
	\item Convert $\lintree$ into a cycle graph $\cycle$ by joining the two leaves of $\lintree$. Let $e_n$ be the new edge joining the two leaves.

	\item Compute the value of $f_\omega(\cycle)$.
	
	\item Substract from $f_\omega(\cycle)$ the amount of elements of $Q(\cycle)\times Q(\cycle)$ classified as type $\omega$ in which $e_n$ appears, namely $|\rho_\omega(\cycle, e_n)|$. Thus, using equation \ref{eq:rho-i-cycle-edges}, the case of linear trees is simple to solve. Using that equation we get that
	\begin{eqnarray}
	\label{eq:linear-trees:all-types}
	f_\omega(\lintree)
		= f_\omega(\cycle) - |\rho_\omega(\cycle,e_n)|
		= f_\omega(\cycle) - \frac{c_\omega}{n} f_\omega(\cycle)
	\end{eqnarray}
	where the values of $c_\omega$ are given in equation \ref{eq:values-of-ci}.
\end{enumerate}

Each of the $f_\omega$'s below are derived using equation \ref{eq:linear-trees:all-types}.

\subsubsection{$\tau=0$, $\phi=0$}
\label{sec:linear_trees:00}

\begin{eqnarray*}
%\label{eq:linear-trees:00}
f_{00}(\lintree) = f_{00}(\cycle) - \frac{4}{n}f_{00}(\cycle) = 6{n - 4 \choose 4}.
\end{eqnarray*}

\subsubsection{$\tau=2$, $\phi=4$}
\label{sec:linear_trees:24}

This is trivially obtained via equation \ref{eq:general:24:final}
\begin{eqnarray*}
%\label{eq:linear-trees:24}
f_{24}(\lintree) = |Q(\lintree)| = {n - 2 \choose 2}.
\end{eqnarray*}

\subsubsection{$\tau=1$, $\phi=3$}
\label{sec:linear_trees:13}

\begin{eqnarray*}
%\label{eq:linear-trees:13}
f_{13}(\lintree) = f_{13}(\cycle) - \frac{3}{n}f_{13}(\cycle) = 4{n - 3 \choose 2}.
\end{eqnarray*}

\subsubsection{$\tau=1$, $\phi=2$}
\label{sec:linear_trees:12}

\begin{eqnarray*}
%\label{eq:linear-trees:12}
f_{12}(\lintree) = f_{12}(\cycle) - \frac{3}{n}f_{12}(\cycle) = 6{n - 3 \choose 3}.
\end{eqnarray*}

\subsubsection{$\tau=0$, $\phi=4$}
\label{sec:linear_trees:04}

\begin{eqnarray*}
%\label{eq:linear-trees:04}
f_{04}(\lintree) = 0
\end{eqnarray*}
because trees do not contain cycles.

\subsubsection{$\tau=0$, $\phi=3$}
\label{sec:linear_trees:03}

\begin{eqnarray*}
%\label{eq:linear-trees:03}
f_{03}(\lintree) = f_{03}(\cycle) - \frac{4}{n}f_{03}(\cycle) = 2n - 8.
\end{eqnarray*}

\subsubsection{$\tau=0$, $\phi=2$}
\label{sec:linear_trees:02}

\begin{eqnarray*}
%\label{eq:linear-trees:021}
f_{021}(\lintree) = f_{021}(\cycle) - \frac{4}{n}f_{021}(\cycle) = 4{n - 4 \choose 2}.
\end{eqnarray*}

Since $f_{021}(\cycle) = f_{022}(\cycle)$,
\begin{eqnarray*}
%\label{eq:linear-trees:022}
f_{022}(\lintree) = f_{021}(\lintree) = 4{n - 4 \choose 2}.
\end{eqnarray*}

\subsubsection{$\tau=0$, $\phi=1$}
\label{sec:linear_trees:01}

\begin{eqnarray*}
%\label{eq:linear-trees:01}
f_{01}(\lintree) = f_{01}(\cycle) - \frac{4}{n}f_{01}(\cycle) = 12{n - 4 \choose 3}.
\end{eqnarray*}

\subsubsection{Variance}
\label{sec:linear_trees:variance}

Applying the results above and equation \ref{eq:variance:freq-times-exp}, the variance in linear trees becomes
\begin{align*}
\gvar{C(\lintree)}
	&= \frac{1}{2}(4(n-3)(n-4)\gexpet{13} + (n-2)(n-3)\gexpet{24} + 4(n-4)\gexpet{03} \\
	& + (n-4)(n-5)( 2(n-3)\gexpet{12} + 4(\gexpet{021}+\gexpet{022}) )),
\end{align*}
which leads to (recall table \ref{table:types_of_combinations})

\begin{eqnarray*}
\lvar{C(\lintree)}
	= \frac{1}{45}n^3 - \frac{1}{18}n^2 - \frac{11}{45}n + \frac{2}{3}.
\end{eqnarray*}

%	subsubsection
\subsection{The scaling of $\lvar{C}$ as function of $n$. }

\begin{figure}
	\centering
	\includegraphics[scale = 0.8]{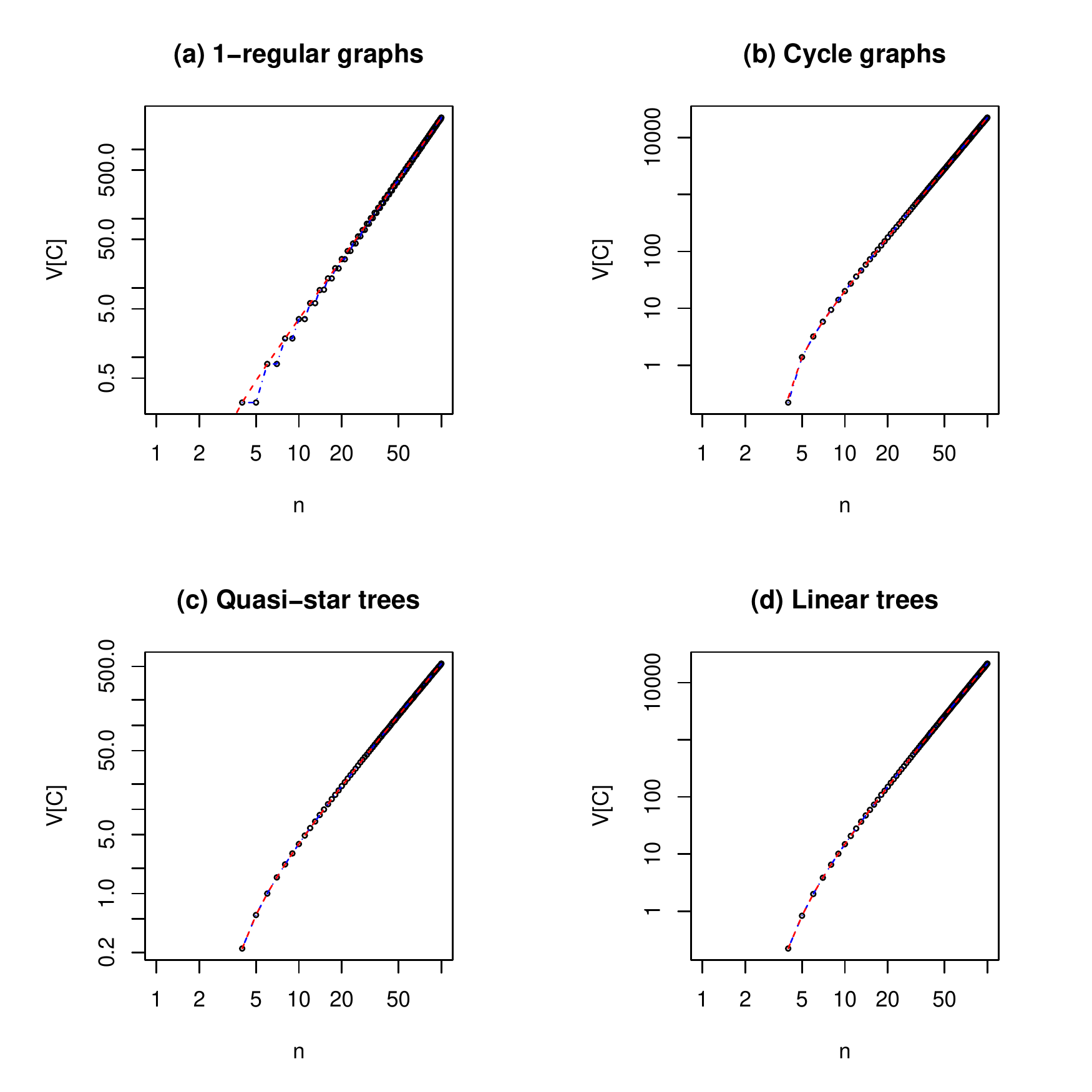}
	\caption{The variance of $C$, the number of crossings in random linear arrangements, as a function of $n$, the number of vertices of the graph, for different kinds of graphs.   For every $n$, the variance of $C$ is estimated over $T$ linear arrangements, i.e. all the $T=n!$ distinct arrangements for $n \le n^* = 10$ and $T=10^5$ uniformly random linear arrangements for $n > n^*$. Thus, the variance of $C$ matches $\lvar{C}$, the theoretical variance, when $n \leq n^*$, and estimates it when $n > n^*$ (circles). $\lvar{C}$ is calculated in two ways: table \ref{table:special_graphs_summary} (dashed red line) and the general formula for $\lvar{C}$ where the value of every $f_\omega$ is calculated with a brute force algorithm (dashed blue line). As the scale is log-log, values with $n<4$ are not shown because $\lvar{C}=0$.
	\label{fig:variance_of_number_of_crossings}
	}
\end{figure}

Figure \ref{fig:variance_of_number_of_crossings} shows $\lvar{C}$ as a function of $n$ for the special graphs where $\lvar{C}$ depends only on the number of vertices of the graph. According to table \ref{table:special_graphs_summary}, $\lvar{C}$ is expected to scale as $\sim n^{\gamma}$, with $\gamma = 2$ for quasi-star trees, $\gamma = 3$ for one-regular graphs and cycles, and $\gamma = 4$ for linear trees. Figure \ref{fig:variance_of_number_of_crossings} shows the agreement between the theoretical $\lvar{C}$ and numerical estimates. The testing protocol for $\lvar{C}$ is explained in \ref{testing_protocol_appendix}. 

%	subsubsection
\subsection{Summary}

Table \ref{table:var_C_freqs_all_graphs} summarizes the formulae for the number of products of each type as a function of $n$ that have been obtained in the preceding sections for particular graphs. See table \ref{table:types_of_combinations} for the valid range  of the parameter $n$ for each $f_\omega$. In particular,  $n \geq |\upsilon|$ is needed. All variances are $0$ for $n\le 3$ and all the expressions are valid for $n\ge 4$, with the exception of $\lvar{\cycle[4]}=2/9$. In the case of $\compbip$, the expression for the variance is valid for $n_1,n_2\ge 2$ (for $n_1,n_2< 2$ the variance is $0$).

\begin{landscape}
\fulltable{\label{table:var_C_freqs_all_graphs} Expressions for the $f_\omega$'s and variance on each type of graph as a function of $n$, the number of vertices. The expressions of the $f_\omega$'s are valid for values of $n\ge |\upsilon|$ (see table \ref{table:types_of_combinations} for details on the values of $|\upsilon|$). All variances are $0$ for $n \le 3$. In the case of $\compbip$ the variance is $0$ when $n_1,n_2<2$.
}
\cline{1-7}\noalign{\medskip}
$f_\omega$	& $\onereg$ 					& $\quasistar$ 				& $\complete$ 			& $\compbip$																							& $\cycle$								& $\lintree$ \\
\noalign{\medskip}\cline{1-7}\noalign{\medskip}
$f_{00}$	& $6{n/2 \choose 4}$				& $0$						& $630{n \choose 8}$		& $144{n_1 \choose 4}{n_2 \choose 4}$																		& $\frac{3}{2}n{n - 5 \choose 3}$		& $6{n-4 \choose 4}$ \\
$f_{24}$	& ${n/2 \choose 2}$				& $n - 3$					& $3{n \choose 4}$		& $2{n_1 \choose 2}{n_2 \choose 2}$																		& $\frac{1}{2}n(n - 3)$					& ${n-2 \choose 2}$ \\
$f_{13}$	& $0$							& $(n - 3)(n - 4)\;\;$		& $60{n \choose 5}$		& $12{n_1 \choose 3}{n_2 \choose 2} + 12{n_1 \choose 2}{n_2 \choose 3}$										& $2n(n - 4)$							& $4{n-3 \choose 2}$ \\
$f_{12}$	& $6{n/2 \choose 3}$				& $0$						& $90{n \choose 6}$		& $36{n_1 \choose 3}{n_2 \choose 3}$																		& $n(n-4)(n-5)$							& $6{n-3 \choose 3}$\\
$f_{04}$	& $0$							& $0$						& $6{n \choose 4}$		& $2{n_1 \choose 2}{n_2 \choose 2}$																		& $0$ (except $2$ for $n=4$)			& $0$ \\
$f_{03}$	& $0$							& $0$						& $120{n \choose 5}$		& $12{n_1 \choose 3}{n_2 \choose 2} + 12{n_1 \choose 2}{n_2 \choose 3}$										& $2n$									& $2n - 8$\\
$f_{021}$	& $0$							& $0$						& $360{n \choose 6}$		& $72{n_1 \choose 3}{n_2 \choose 3}$																		& $2n(n - 5)$							& $4{n-4 \choose 2}$	\\
$f_{022}$	& $0$							& $0$						& $360{n \choose 6}$		& $24{n_1 \choose 2}{n_2 \choose 4} + 24{n_1 \choose 4}{n_2 \choose 2} + 36{n_1 \choose 3}{n_2 \choose 3}\;\;$& $2n(n - 5)$							& $4{n-4 \choose 2}$	\\
$f_{01}$	& $0$							& $0$						& $1260{n \choose 7}\;\;$& $144{n_1 \choose 4}{n_2 \choose 3} + 144{n_1 \choose 3}{n_2 \choose 4}$									& $2n(n - 5)(n - 6)$					& $12{n-4 \choose 3}$ \\
\noalign{\medskip}\cline{1-7}\noalign{\medskip}
$\lvar{C}\;\;$
			& $\frac{1}{360}(n-2)n(n+6)\;\;$& $\frac{1}{18}n(n - 3)\;$	& $0$					& $\frac{1}{90}{n_1 \choose 2}{n_2 \choose 2}((n_1 + n_2)^2 + n_1 + n_2)$									&
\begin{tabular}{@{}ll}
$2/9$ (for $n=4$) \\
$\frac{1}{45}n^3 + \frac{1}{90}n^2 - \frac{1}{3}n$ (for $n>4$)
\end{tabular}
& $\frac{1}{45}n^3 - \frac{1}{18}n^2 - \frac{11}{45}n + \frac{2}{3}$ \\
\noalign{\medskip}\cline{1-7}
\endfulltable
\end{landscape}

% Section
\section{Discussion}
\label{sec:discussion}

In the preceding sections, we have investigated the expectation and the variance of $\gC$ in general graphs. We have also provided compact formulae for specific graphs (table \ref{table:special_graphs_summary}). The scaling of the expectation and the variance of $C$ as function of the number of vertices is asymptotically power-law-like (figures \ref{fig:expected_number_of_crossings} and \ref{fig:variance_of_number_of_crossings} and also tables \ref{table:special_graphs_summary} and \ref{table:var_C_freqs_all_graphs}).

$\gvar{C}$ turns out to be a weighted sum of the number of products of each type and their corresponding probabilities. As these probabilities are constant (they do not depend on the graph), $\gvar{C}$ is determined by the number of only seven types of products, and so is $\lvar{C}$. Furthermore, we have shown that the number of products of a given type is proportional to the number subgraphs of a certain kind (table \ref{table:summary_frequencies}). Then the calculation of $\gvar{C}$ reduces to counting the frequency of seven distinct types of graphettes, i.e. possibly disconnected subgraphs \cite{Hasan2017a}. Thus our work is related to research on meaningful substructures, e.g., motifs, in complex networks \cite{Milo2002a,Przulj2007a}. We have also provided simple exact formulae for these numbers in particular graphs (table \ref{table:var_C_freqs_all_graphs}). Our work has consisted of a first approximation to the calculation of $\lvar{C}$ that relies on counting the number of subgraphs of a certain kind (table \ref{table:summary_frequencies}). General formulae for the frequency of every type or $\gvar{C}$ as a whole should be investigated. These formulae may allow for a simpler calculation of $\gvar{C}$ from an algebraic or computational standpoint and may allow one to establish further connections with network theory \cite{Newman2010a}.

Our analysis on individual graphs has paved the way to study the distribution of crossings for classic ensembles of graphs, e.g., Erd\H{o}s-R\'enyi graphs \cite{Bollobas2002a} and uniformly random trees \cite{Aldous1990a,Broder1989a}, as well as other classes of random networks with more realistic characteristics \cite{Newman2010a,Ferrer2017a}.

In previous work on syntactic dependency trees, $C$ has been shown to be significantly low with respect to random linear arrangements with the help of Monte Carlo statistical tests \cite{Ferrer2017a}. Thanks to our novel knowledge about the expectation and the variance of $C$ in these arrangements, fast statistical tests of the number of crossings of graphs could be derived with the help of Chebychev-like inequalities to linguistic and biological networks were vertices form a 1-dimensional lattice \cite{Chen2009a, Ferrer2017a}. Such a procedure has already been outlined to check if $D$, the sum of edge lengths in a linear arrangement, is significantly low with respect to random linear arrangements \cite{Ferrer2018a}.

Our algebraic characterization of the types of products in $\gvar{C}$ (figure \ref{fig:bipartite_graphs} and table \ref{table:types_of_combinations}) was motivated by spatial networks in 1-dimensional lattices but it has been derived independently from that layout based on purely algebraic criteria. Therefore it is valid, as a first approximation, to study $\gvar{C}$ in other layouts or embeddings, e.g., lattices of two or three dimensions or the layout on the surface of a sphere in the pioneering research by Moon \cite{Moon1965a}. For this reason, we have established the foundations to revise Moon's work. In section \ref{sec:var_C}, we adapted Moon's derivation for the particular case of a complete graph whose vertices are arranged at random on the surface of a sphere. In the process, we discovered that some types of products that we have identified were omitted in the original derivation. Indeed, Moon’s derivation for the spherical case is inaccurate and a correction will be published somewhere else \cite{Alemany2018b}.

\section*{Acknowledgements}

We are grateful to Kosmas Palios for helpful discussions. We thank Antoni Lozano and Merc\`e Mora for helpful suggestions to improve the quality of the manuscript. This research was supported by the grant TIN2017-89244-R from MINECO (Ministerio de Econom{\'i}a y Competitividad) and the recognition 2017SGR-856 (MACDA) from AGAUR (Generalitat de Catalunya).

\appendix
\addtocontents{toc}{\fixappendix}
\setcounter{section}{0}

\section{Potential number of crossings of an edge}
\label{crossings_of_edge_appendix}

The potential number of crossings involving the edge formed by $u$ and $v$ can be defined as
\begin{eqnarray*}
q(u,v) = \frac{1}{2} \sum_{s\neq u,v} \sum_{t\neq u,v} a_{st}.
\end{eqnarray*}

Noting that 
\begin{eqnarray*}
\sum_{s\neq u,v} k_s = 2m - k_u - k_v \\
a_{uu}, a_{vv} = 0 \\
a_{uv} = 1
\end{eqnarray*}
and, applying iteratively
\begin{eqnarray*}
\sum_{s\neq u,v} a_{st} = k_t - a_{tu} - a_{tv},
\end{eqnarray*}
it is easy to see that $q(u,v)$ can be expressed as 
\begin{eqnarray*}
q(u,v) & = & \frac{1}{2} \sum_{s\neq u,v} (k_s - a_{su} - a_{sv})\\
           & = & \frac{1}{2} \left[ \sum_{s\neq u,v}  k_s - \sum_{s\neq u,v} a_{su} - \sum_{s\neq u,v} a_{sv}\right] \\
           & = & \frac{1}{2} \left[ 2m - k_u - k_v - (k_u - a_{uu} - a_{uv}) - (k_v - a_{vv} - a_{uv}) \right] \\
           & = & m - k_u - k_v + 1.
\end{eqnarray*}

%%%%%%%%%%%%%%%%%%%%%%%%%%%%%%%%%%%%%%%%%%%%%%%%%%%%%%%%%%%%%%%%%%%%%%%%%%%%%

\section{Toy examples}
\label{toy_examples_appendix}

\begin{figure}
	\centering
	\includegraphics{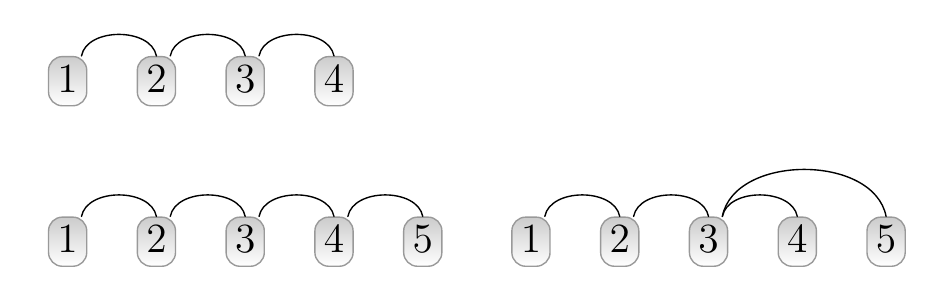}
	\caption{Some labeled trees. Top: $\lintree[4]$. Bottom: $\lintree[5]$ and $\quasistar[5]$. }
	\label{fig:abstract_trees}
\end{figure}

First we consider all trees up to $n=5$. Suppose that linear trees are labeled with integers starting at one in one leaf and increasing one unit per vertex till the other leaf (figure \ref{fig:abstract_trees}).  When $n < 4$, $\lvar{C}=0$ trivially because $|Q| = 0$ and then $\gC = 0$. The same happens to all star trees with $n>4$. When $n=4$, the only trees are $\startree[4]$ and $\lintree[4]$.

\subsection{4 vertices}

Here, $\mmtdeg{2} = 5/2$ and then $|Q| = 1$ (recall equation \ref{eq:potential_num_crossings:trees}).  For $\lintree[4]$ (figure \ref{fig:abstract_trees}), we have $Q= \{ \{12, 34\} \}$. As $Q \times Q = \{ \{\{12, 34\},\{12, 34\}\}\}$, $f_{24}=1$, and $f_\omega=0$ for the other types. Finally,
\begin{eqnarray*}
\lvar{C} = \sum_{\omega\in\Omega} \lexpet{\omega} = \lexpet{24} = \frac{2}{9}.
\end{eqnarray*}
$\lvar{C}$ could have been derived also as the variance of $C$, a Bernoulli variable  because $C \in \{0, 1\}$ due to $|Q| = 1$. As $\prob{C = 1} = 1/3$, namely the probability that two independent edges cross, 
\begin{eqnarray*}
\lvar{C} = \prob{C = 1} (1- \prob{C = 1}) = \frac{2}{9}.
\end{eqnarray*}
When $n=4$, we conclude that $\lvar{C} = 0$ if the tree is $\startree[4]$ or $\lvar{C} = 2/9$ if the tree is $\lintree[4]$. 

\begin{table}[H]
\caption{Types of products in $Q \times Q$
for $\lintree[5]$. }
\label{table:combinations_linear_tree}
\begin{indented}
\item[]
\begin{tabular}{@{}llll}
\br
          & $\{12,34\}$  & $\{12, 45\}$   & $\{23, 45\}$ \\ 
\mr
$\{12,34\}$ & 24         & 13           & 03 \\
$\{12,45\}$ & 13         & 24           & 13 \\ 
$\{23,45\}$ & 03         & 13           & 24 \\
\br
\end{tabular}
\end{indented}
\end{table}

\begin{table}[H]
\caption{Types of products in $Q \times Q$ for $\quasistar[5]$. }
\label{table:combinations_quasi_star_tree}
\begin{indented}
\item[]
\begin{tabular}{@{}lll}
\br
          & $\{12,34\}$ & $\{12, 35\}$ \\ 
\mr
$\{12,34\}$ & 24        & 13 \\
$\{12,35\}$ & 13        & 24 \\
\br
\end{tabular}
\end{indented}
\end{table}

\subsection{5 vertices}

When $n=5$, the only possible trees are $\lintree[5]$, $\quasistar[5]$ \cite{Ferrer2014f} and $\startree[5]$ (figure \ref{fig:abstract_trees}). As for the linear tree, applying $\mmtdeg{2} = 14/5$ (equation \ref{eq:2nd_mmt_deg:linear_tree}) to equation \ref{eq:potential_num_crossings:trees} gives $|Q| = 3$. Following the labeling in figure \ref{fig:abstract_trees}, it is easy to see that
\begin{eqnarray*}
Q = \{\{12,34\},\{12,45\},\{23,45\}\}.
\end{eqnarray*}
The types of products in $Q\times Q$ in table \ref{table:combinations_linear_tree} give $f_{24}=3$, $f_{13}=4$, $f_{03}=2$, $f_\omega=0$ for the other types, and then
\begin{eqnarray*}
\lvar{C} = 3\lexpet{24} + 4\lexpet{13} + 2\lexpet{03} =
           3 \frac{2}{9} + 4\frac{1}{18} - 2\frac{1}{36} = \frac{5}{6}.
\end{eqnarray*}

As for $\quasistar[n]$, we have \cite{Ferrer2014f}
\begin{eqnarray*}
\mmtdeg{2}= n - 3 + 6/n.
\end{eqnarray*}
Applying $\mmtdeg{2} = 16/5$ to equation \ref{eq:potential_num_crossings:trees} gives $|Q| = 2$. Following the labeling in figure \ref{fig:abstract_trees}, it is easy to see that
\begin{eqnarray*}
Q = \{\{12,34\},\{12,35\}\}.
\end{eqnarray*}
The summary of types of products in $Q\times Q$ in table \ref{table:combinations_quasi_star_tree}
gives $f_{24}=f_{13}=2$ and then
\begin{eqnarray*}
\lvar{C} = 2 \lexpet{24} + 2\lexpet{13}
		= 2 \left( \frac{2}{9} + \frac{1}{18} \right) = \frac{5}{9}.
\end{eqnarray*}

When $n=5$, we conclude that
\begin{itemize}
	\item $\lvar{C} = 0$ if the tree is $\startree[5]$.
	\item $\lvar{C} = 5/9$ if the tree is a $\quasistar[5]$.
	\item $\lvar{C} = 5/6$ if the tree is $\lintree[5]$.
\end{itemize}

\subsection{6 vertices}

Here we only consider $\lintree[6]$. The second moment degree of $\lintree[6]$ gives $\mmtdeg{2}=3$ (equation \ref{eq:2nd_mmt_deg:linear_tree}). Therefore $|Q| = 6$ (equation \ref{eq:potential_num_crossings:trees}). Indeed, by following the same labeling method, we see that
\begin{eqnarray*}
Q = \{\{12,34\},\{12,45\},\{12,56\},\{23,45\},\{23,56\},\{34,56\}\}.
\end{eqnarray*}

The summary of types of products in $Q \times Q$ is shown in table \ref{table:combinations_linear_tree_6}.

\begin{table}[H]
\caption{Types of products in $Q \times Q$ for a $\lintree[6]$.}
\label{table:combinations_linear_tree_6}
\begin{indented}
\item[]
\begin{tabular}{@{}lllllll}
\br
          & $\{12,34\}$ & $\{12,45\}$ & $\{12,56\}$ & $\{23,45\}$ & $\{23,56\}$ & $\{34,56\}$ \\
\mr
$\{12,34\}$ & 24        & 13        & 12        & 03        & 021       & 12    \\
$\{12,45\}$ & 13        & 24        & 13        & 13        & 022       & 022   \\
$\{12,56\}$ & 12        & 13        & 24        & 022       & 13        & 12    \\
$\{23,45\}$ & 03        & 13        & 022       & 24        & 13        & 03    \\
$\{23,56\}$ & 021       & 022       & 13        & 13        & 24        & 13    \\
$\{34,56\}$ & 12        & 021       & 12        & 03        & 13        & 24    \\
\br
\end{tabular}
\end{indented}
\end{table}

The summary of types of products in $Q\times Q$ in table \ref{table:combinations_linear_tree_6} gives $f_{24}=f_{12}=6$, $f_{13}=12$, $f_{03}=f_{021}=f_{022}=4$ and then
\begin{eqnarray*}
\lvar{C}
	&=& 6(\lexpet{24} + \lexpet{12}) + 12\lexpet{13} \\
	&& + 4(\lexpet{03} + \lexpet{021} + \lexpet{022}) \\
	&=& 6\left(\frac{2}{9} + \frac{1}{45} \right) + 12\frac{1}{18} + 4\left( -\frac{1}{36} - \frac{1}{90} + \frac{1}{180} \right) = 2.
\end{eqnarray*}

\subsection{7 vertices}

Here we only consider $\lintree[7]$. The second moment of degree of $\lintree[7]$ gives $\mmtdeg{2} = 22/7$. Therefore $|Q| = 10$ (see equation \ref{eq:potential_num_crossings:trees}). Indeed, by following the same labeling method, we see that
\begin{eqnarray*}
Q
	&=& \{\{12,34\},\{12,45\},\{12,56\},\{12,67\},\{23,45\},\\
	& & \{23,56\},\{23,67\},\{34,56\},\{34,67\},\{45,67\}\}.
\end{eqnarray*}

\fulltable{
\label{table:combinations_linear_tree_7} Types of products in $Q \times Q$ for $\lintree[7]$.}
\br
          &$\{12,34\}$&$\{12,45\}$&$\{12,56\}$&$\{12,67\}$&$\{23,45\}$&$\{23,56\}$&$\{23,67\}$&$\{34,56\}$&$\{34,67\}$&$\{45,67\}$ \\
\mr
$\{12,34\}$ & 24      & 13      & 12      & 12      & 03      & 021     & 021     & 12      & 12      & 01 \\
$\{12,45\}$ & 13      & 24      & 13      & 12      & 13      & 022     & 01      & 021     & 01      & 12 \\
$\{12,56\}$ & 12      & 13      & 24      & 13      & 022     & 13      & 022     & 12      & 01      & 021 \\
$\{12,67\}$ & 12      & 12      & 13      & 24      & 01      & 022     & 13      & 01      & 12      & 12 \\
$\{23,45\}$ & 03      & 13      & 022     & 01      & 24      & 13      & 12      & 03      & 021     & 12 \\
$\{23,56\}$ & 021     & 022     & 13      & 022     & 13      & 24      & 13      & 13      & 022     & 021 \\
$\{23,67\}$ & 021     & 01      & 022     & 13      & 12      & 13      & 24      & 022     & 13      & 12 \\
$\{34,56\}$ & 12      & 021     & 12      & 01      & 03      & 03      & 022     & 24      & 13      & 03 \\
$\{34,67\}$ & 12      & 01      & 01      & 12      & 021     & 022     & 13      & 13      & 24      & 13 \\
$\{45,67\}$ & 01      & 12      & 021     & 12      & 12      & 021     & 12      & 03      & 13      & 24 \\
\mr
\endfulltable

The summary of types of products in $Q \times Q$ is shown in table \ref{table:combinations_linear_tree_7}, and gives $f_{24}=10$, $f_{13}=f_{12}=24$, $f_{03}=6$, $f_{021}=f_{022}=f_{01}=12$, and $f_{00}=f_{24}=0$, and then
\begin{eqnarray*}
\lvar{C}
	&=& 10\lexpet{24} + 24(\lexpet{13} + \lexpet{12}) + 6\lexpet{03} \\
	&& + 12(\lexpet{021} + \lexpet{022} + \lexpet{01}) \\
	&=& 10\frac{2}{9} +24\left( \frac{1}{18} + \frac{1}{45}\right)
		+ 6 \left(- \frac{1}{36} \right)
		+ 12 \left( - \frac{1}{90} + \frac{1}{180} \right) = \frac{347}{90}.
\end{eqnarray*}

\subsection{Complete graphs}

Finally, we consider $\complete[4]$ and $\complete[5]$. We know that $\lvar{C}=0$. The case is not only interesting due to the cancellation of the products $f_\omega\lexpet{\omega}$ but also for showing products of type $04$ in $Q \times Q$, which cannot be found in trees. 

When $n=4$, we have $|Q| = 3$ (recall equation \ref{eq:potential_number_of_crossings_complete_graph}). In particular,
\begin{eqnarray*}
Q = \{ \{12, 34\}, \{13,24\}, \{14,23\} \}.
\end{eqnarray*} 
The types of products in table \ref{table:combinations_complete_graph_4} give
\begin{eqnarray*}
\lvar{C} = 3 \lexpet{24} + 6 \lexpet{04}
        = 3 \frac{2}{9} - 6 \frac{1}{9} = 0
\end{eqnarray*}
as expected.
 
\begin{table}[H]
\caption{Types of products in $Q \times Q$ for $\complete[4]$. }
\label{table:combinations_complete_graph_4}
\begin{indented}
\item[]
\begin{tabular}{@{}llll}
\br
          & $\{12,34\}$ & $\{13, 24\}$ & $\{14, 23\}$ \\
\mr
$\{12,34\}$ & 24        & 04        & 04 \\
$\{12,24\}$ & 04        & 24        & 04 \\ 
$\{14,23\}$ & 04        & 04        & 24 \\
\br
\end{tabular}
\end{indented}

\end{table}

When $n=5$, we have $|Q| = 15$ (recall equation \ref{eq:potential_number_of_crossings_complete_graph}) with
\begin{eqnarray*}
Q	&=& \{ \{12,34\}, \{12,35\}, \{12,45\}, \{13,24\}, \{13,25\}, \{13,45\}, \{14,23\}, \{14,25\}, \\
	& & \{14,35\}, \{15,23\}, \{15,24\}, \{15,34\}, \{23,45\}, \{24,35\}, \{25,34\} \}.
\end{eqnarray*}

%\begin{table}[H]
%\caption{Types of products in $Q \times Q$ for a complete graph with $n = 5$. }
%\label{combinations_complete_graph_5}
%\begin{indented}
%\item[]
%\tiny
%\begin{tabular}{@{}lllllllll}
\fulltable{
Types of products in $Q \times Q$ for $\complete[5]$.
\label{combinations_complete_graph_5}
}
\br
          &$\{12,34\}$&$\{12,35\}$&$\{12,45\}$&$\{13,24\}$&$\{13,25\}$&$\{13,45\}$&$\{14,23\}$&$\{14,25\}$ \\
\mr
$\{12,34\}$ & 24      & 13      & 13      & 04      & 03      & 03      & 04      & 03 \\
$\{12,35\}$ & 13      & 24      & 13      & 03      & 04      & 03      & 03      & 03 \\
$\{12,45\}$ & 13      & 13      & 24      & 03      & 03      & 13      & 03      & 04 \\
$\{13,24\}$ & 04      & 03      & 03      & 24      & 13      & 13      & 04      & 03 \\
$\{13,25\}$ & 03      & 04      & 03      & 13      & 24      & 13      & 03      & 13 \\
$\{13,45\}$ & 03      & 03      & 13      & 13      & 13      & 24      & 03      & 03 \\
$\{14,23\}$ & 04      & 03      & 03      & 04      & 03      & 03      & 24      & 13 \\
$\{14,25\}$ & 03      & 03      & 04      & 03      & 13      & 03      & 13      & 24 \\
$\{14,35\}$ & 03      & 13      & 03      & 03      & 03      & 04      & 13      & 13 \\
$\{15,23\}$ & 03      & 04      & 03      & 03      & 04      & 03      & 13      & 03 \\
$\{15,24\}$ & 03      & 03      & 04      & 13      & 03      & 03      & 03      & 04 \\
$\{15,34\}$ & 13      & 03      & 03      & 03      & 03      & 04      & 03      & 03 \\
$\{23,45\}$ & 03      & 03      & 13      & 03      & 03      & 13      & 13      & 03 \\
$\{24,35\}$ & 03      & 13      & 03      & 13      & 03      & 03      & 03      & 03 \\
$\{25,34\}$ & 13      & 03      & 03      & 03      & 13      & 03      & 03      & 13 \\
\mr
		  &$\{14,35\}$&$\{15,23\}$&$\{15,24\}$&$\{15,34\}$&$\{23,45\}$&$\{24,35\}$&$\{25,34\}$ \\
\mr
$\{12,34\}$ & 03  & 03      & 03      & 13      & 03      & 03      & 13  \\
$\{12,35\}$ & 13  & 04      & 03      & 03      & 03      & 13      & 03  \\
$\{12,45\}$ & 03  & 03      & 04      & 03      & 13      & 03      & 03  \\
$\{13,24\}$ & 03  & 03      & 13      & 03      & 03      & 13      & 03  \\
$\{13,25\}$ & 03  & 04      & 03      & 03      & 03      & 03      & 13  \\
$\{13,45\}$ & 04  & 03      & 03      & 04      & 13      & 03      & 03  \\
$\{14,23\}$ & 13  & 13      & 03      & 03      & 13      & 03      & 03  \\
$\{14,25\}$ & 13  & 03      & 04      & 03      & 03      & 03      & 13  \\
$\{14,35\}$ & 24  & 03      & 03      & 04      & 03      & 13      & 03  \\
$\{15,23\}$ & 03  & 24      & 13      & 13      & 13      & 03      & 03  \\
$\{15,24\}$ & 03  & 13      & 24      & 13      & 03      & 13      & 03  \\
$\{15,34\}$ & 04  & 13      & 13      & 24      & 03      & 03      & 13  \\
$\{23,45\}$ & 03  & 13      & 03      & 03      & 24      & 04      & 04  \\
$\{24,35\}$ & 13  & 03      & 13      & 03      & 04      & 24      & 04  \\
$\{25,34\}$ & 03  & 03      & 03      & 13      & 04      & 04      & 24  \\
\br
\endfulltable
%\end{tabular}
%\end{indented}
%\end{table}

The types of products in table \ref{combinations_complete_graph_5} give $f_{24}=15$, $f_{13}=60$, $f_{04}=30$, $f_{03}=120$, $f_{00}=f_{12}=f_{021}=f_{022}=f_{01}=0$, and then
\begin{eqnarray*}
\lvar{C}
	&=& 15 \lexpet{24} + 60 \lexpet{13} + 30 \lexpet{04} + 120 \lexpet{03} \\
	&=& 15 \frac{2}{9} + 60 \frac{1}{18} - 30 \frac{1}{9} - 120 \frac{1}{36} = 0
\end{eqnarray*}
as expected.

In all the examples above, we have obtained all the types of products that can appear when $n < 6$, namely types 1,2,4 and 5. $n=6$ is needed to obtain a combination of type 2. $n=8$ is needed to obtain a combination of type 0.

\section{Protocol for testing}
\label{testing_protocol_appendix}

The formulae for $\gexpe{C}$ are mathematically trivial and easy to compute. In contrast, the formulae and the algorithms for calculating $\lvar{C}$ (equation \ref{eq:variance_of_number_of_crossings}) are complex and require a validation protocol. That protocol is inspired by that of \cite{Esteban2016a} and consists of two types of tests: computational tests and manual mathematical tests. Computational tests take a certain class of graphs and, for each graph in the class, they calculate $\lvar{C}$ following two independent procedures. First, the $f_\omega$'s are calculated by brute force with a general algorithm and then plugged into equation \ref{eq:variance_of_number_of_crossings} to produce $\lvar{C}'$. Second,  $\lvar{C}$ is estimated over the space of $n!$ permutations of the vertices of the graph producing $\lvar{C}''$.

The computational tests were applied to the following classes of graphs taking every value of $n$ within the interval $[2, n_{max}]$
\begin{itemize} 
\item
All labeled undirected trees of $n$ vertices. The $n^{n-2}$ labeled undirected trees of $n$ vertices \cite{Cayley1889a} were generated using of Pr\"ufer codes \cite{Pruefer1918a}, with the algorithm described in \cite{Alonso1995a} (Chapter 3, section 3.3.4). $n_{max}=9$ was used.
% Notice that the cost of checking that $\lvar{C}' = \lvar{C}''$ for a given $n$ is about 
% \begin{equation}
% n^{n-2} n! {n - 1 \choose 2} 
% \end{equation}
% where ${n - 1 \choose 2}$ is the cost of calculating $C$ for a tree of $n$ vertices.
% Is it the same cost as that of the brute force algorithm above ???
\item Representatives of isomorphic classes of graphs \cite{McKay2013a}. An undirected graph of $n$ vertices without loops is defined simply by a triangle of the $n\times n$ adjacency matrix that has ${n \choose 2}$ cells. Therefore, the space of potential graphs of $n$ vertices is $2^{n \choose 2}$. To reduce the cost of testing, we consider a smaller space defined by representatives of isomorphic classes of \cite{McKay2013a}. These representatives were downloaded from a database \url{https://users.cecs.anu.edu.au/~bdm/data/graphs.html}. $n_{max}=9$ was used.

\item The specific kinds of graphs listed introduced in section \ref{sec:theoretical_graphs}: One-regular graphs, cycle graphs, quasi-star trees and linear trees. Their simplicity allows one to test larger values of $n$ compared to the preceding classes of graphs. $n_{max}=100$ was used.
\end{itemize}

For $n < n_{MC} = 10$, all the permutations were generated and therefore the estimate is the true value. A key point of this exhaustive exploration is that $\lvar{C}''$ has to be calculated via the biased estimator as no sampling bias is possible (the customary biased estimator yields wrong results). $\lvar{C}'$ and $\lvar{C}''$ are rational numbers that are simplified so that they can be compared easily. The test is successful if $\lvar{C}' = \lvar{C}''$. The rational numbers were represented, simplified and compared using the GMP Library (version 6.1.2, see \url{https://gmplib.org/}). Alternatively, real numbers for $\lvar{C}'$ and $\lvar{C}''$ could have been used but that would have required defining an error threshold to decide whether the two independent calculations yield the same result or not.

For $n\ge n_{MC}$, $\lvar{C}''$ was not calculated exactly, but estimated using a Monte Carlo procedure over $10^5$ random permutations.

For the special kinds of graphs, the computational testing procedure was extended. Figure \ref{fig:variance_of_number_of_crossings} shows the great accuracy of the Monte Carlo estimates for $n \geq n_{MC}$. Second, we calculated $\lvar{C}$ using the simple formulae in table \ref{table:special_graphs_summary} producing $\lvar{C}'''$. We checked that $\lvar{C}' =  \lvar{C}'' = \lvar{C}'''$ for $n_{min} \leq n \leq n_{MC}$. For simplicity, the Monte Carlo estimation was not applied to bipartite graphs, that were tested for $1 \leq n_1, n_2 \leq 32$, with $n_1 + n_2 \leq 9$.

In the tests where $\lvar{C}$ was computed either exactly or approximately, we made sure that the values of $C$ used to compute the variance were values such that $C\le |Q|$. When the variance was computed theoretically, namely via the amount of occurrences of the $9$ different types, we checked that
\begin{eqnarray*}
\sum_{\omega\in\Omega} f_\omega = |Q|^2
\end{eqnarray*}
and that $f_\omega$ is even for any $\omega\in\Omega$ excluding $\omega=24$.

The manual mathematical tests consisted of checking that $\lvar{C}=0$ in complete graphs (section \ref{sec:complete_graphs:variance}) and star trees, as it is expected by definition. The case of star tree is trivial because $|Q| = 0$. The values of $\lvar{C}$ and the $f_\omega$'s for the toy graphs in \ref{toy_examples_appendix} were calculated by hand. These independent results provide test cases for  all the computer algorithms on small graphs. 

Finally, independently of the other tests, we also checked the expressions of the $f_\omega$ derived in section \ref{sec:general_formulas} (summarized in table \ref{table:summary_frequencies}) using the ensemble of Erd\"os-R\'enyi random graphs \cite{Bollobas2002a}, denoted as $G_{n,p}$. A graph $G_{n,p}$ is a graph of $n$ vertices where each edge is taken from $\complete$ with probability $p$. In these tests, we calculated the $f_\omega$'s by listing all the elements of $Q\times Q$ and classifying them accordingly. Then we compared these results against the corresponding counts via $a_\omega n_G(F_\omega)$ in table \ref{table:summary_frequencies} using a custom algorithm. The graphs used were of various sizes but always with $n\le 50$. The values of $p$ were usually $p=0.1, 0.2, 0.5$. High values of $p$ were combined with low values of $n$.

%%%%%%%%%%%%%%%%%%%%%%%%%%%%%%%%%%%%%%%%%%%%%%%%%%%%%%%%%%%%%%%%%%%%%%%%%%

\section{Alternative proof for $f_{021} = 2n_G(\lintree[4] \oplus \lintree[2])$. }
\label{alternative_proof_f021_appendix}

We can see that type 021 counts over the $\lintree[4]\oplus\lintree[2]$. This can be readily seen by noting that, for a fixed $\{st,uv\}\in Q$, the graphs counted by $\varphi_{st}$, $\varphi_{uv}$ and the $\varepsilon_{..}$ are of the form
\begin{align*}
\varphi_{st}&:     \{(z_s,s,t,z_t), (u,v)\}, &\varphi_{uv}&:     \{(s,t), (z_u,u,v,z_v)\}, \\
\varepsilon_{su}^*&: \{(t,s,u,v), (z_1,z_2)\}, &\varepsilon_{sv}&: \{(t,s,v,u), (z_1,z_2)\}, \\
\varepsilon_{tu}&: \{(s,t,u,v), (z_1,z_2)\}, &\varepsilon_{tv}&: \{(s,t,v,u), (z_1,z_2)\}.
\end{align*}
We marked $\varepsilon_{su}$ with $^*$ for later reference. We now show that each $\lintree[4]\oplus\lintree[2]$ is counted twice in equation \ref{eq:general:021}. Consider a fixed $\lintree[4]\oplus\lintree[2]$ of the form $H=\{(s,t,u,v), (w,x)\}$. Notice that $Q(H)$ has four elements, i.e.
\begin{equation*}
q_1=\{st,uv\}, \; q_2=\{st,wx\}, \; q_3=\{tu,wx\}, \; q_4=\{uv,wx\},
\end{equation*}
only two of which make equation \ref{eq:general:021} count over $H$, once for each of the two. This can be seen by noting that the graphs counted by the equation have to preserve the order of the vertices of the $\lintree[4]$, achieved only by $\varepsilon_{su}$ when the equation is evaluated with $q_1$ (see the description above), and by $\varphi_{tu}$ when evaluated with $q_3$ (both marked with $^*$ above and below). For the sake of brevity we only show the form of the graphs counted by equation \ref{eq:general:021} for $q_2$ and $q_3$. For $q_4$ are derived similarly. Notice that for $q_2$ none of the graphs is isomorphic to $H$ and that the same happens with $q_4$.
\begin{align*}
q_2=\{st,wx\}
	&&\varphi_{st}&:     \{(z_s,s,t,z_t), (w,x)\}, &\varphi_{wx}&:     \{(s,t), (z_w,w,x,z_x)\}, \\
	&&\varepsilon_{sw}&: \{(t,s,w,x), (z_1,z_2)\}, &\varepsilon_{sx}&: \{(t,s,x,w), (z_1,z_2)\}, \\
	&&\varepsilon_{tw}&: \{(s,t,w,x), (z_1,z_2)\}, &\varepsilon_{tx}&: \{(s,t,x,w), (z_1,z_2)\}, \\
q_3=\{tu,wx\}
	&&\varphi_{tu}^*&:     \{(z_t,t,u,z_u), (w,x)\}, &\varphi_{wx}&:     \{(t,u), (z_w,w,x,z_x)\}, \\
	&&\varepsilon_{tw}&: \{(u,t,w,x), (z_1,z_2)\}, &\varepsilon_{tx}&: \{(u,t,w,x), (z_1,z_2)\}, \\
	&&\varepsilon_{uw}&: \{(t,u,w,x), (z_1,z_2)\}, &\varepsilon_{ux}&: \{(t,u,x,w), (z_1,z_2)\}.
\end{align*}
Therefore, each $\lintree[4]\oplus\lintree[2]$ is only counted once by each of the only two elements of $Q(H)$ that can form it, namely
\begin{eqnarray*}
f_{021} = 2n_G(\lintree[4] \oplus \lintree[2]).
\end{eqnarray*}
Interestingly, the analysis above shows that a $\lintree[4]\oplus\lintree[2]$ is only counted in exactly one of the two $\varphi_{..}$ and in exactly one of the four $\varepsilon_{..}$, hence equations \ref{eq:021:sum1} and \ref{eq:021:sum2}.

\section*{References}

\bibliographystyle{unsrt}

\begin{thebibliography}{10}

\bibitem{Barthelemy2011a}
Marc Barth\'elemy.
\newblock Spatial networks.
\newblock {\em Physics Reports}, 499(1):1 -- 101, 2011.

\bibitem{Eppstein2017a}
David Eppstein and Siddharth Gupta.
\newblock Crossing patterns in nonplanar road networks.
\newblock In {\em Proceedings of the 25th ACM SIGSPATIAL International
  Conference on Advances in Geographic Information Systems}, SIGSPATIAL'17,
  pages 40:1--40:9, New York, NY, USA, 2017. ACM.

\bibitem{Ferrer2017a}
R.~{Ferrer-i-Cancho}, C.~{G\'omez-Rodr{\'i}guez}, and J.~L. Esteban.
\newblock Are crossing dependencies really scarce?
\newblock {\em Physica A: Statistical Mechanics and its Applications},
  493:311--329, 2018.

\bibitem{Ferrer2016a}
R.~{Ferrer-i-Cancho} and C.~G\'omez-Rodr\'iguez.
\newblock Liberating language research from dogmas of the 20th century.
\newblock {\em Glottometrics}, 33:33--34, 2016.

\bibitem{Chen2009a}
W.~Y.~C. Chen, H.~S.~W. Han, and C.~M. Reidys.
\newblock Random $k$-noncrossing {RNA} structures.
\newblock {\em Proceedings of the National Academy of Sciences},
  106(52):22061--22066, 2009.

\bibitem{Barthelemy2018a}
M.~Barth\'elemy.
\newblock {\em Morphogenesis of Spatial Networks}.
\newblock Springer, Cham, Switzerland, 2018.

\bibitem{Moon1965a}
J.~W. Moon.
\newblock On the distribution of crossings in random complete graphs.
\newblock {\em Journal of the Society for Industrial and Applied Mathematics},
  13:506--510, 1965.

\bibitem{Gomez2016a}
C.~G\'omez-Rodr\'iguez and R.~{Ferrer-i-Cancho}.
\newblock Scarcity of crossing dependencies: {A} direct outcome of a specific
  constraint?
\newblock {\em Physical Review E}, 96:062304, 2017.

\bibitem{Ferrer2013b}
R.~{Ferrer-i-Cancho}.
\newblock Hubiness, length, crossings and their relationships in dependency
  trees.
\newblock {\em Glottometrics}, 25:1--21, 2013.

\bibitem{Ferrer2013d}
R.~{Ferrer-i-Cancho}.
\newblock Random crossings in dependency trees.
\newblock {\em Glottometrics}, 37:1--12, 2017.

\bibitem{Milo2002a}
R.~S. Milo, S.~Shen-Orr, S.Itzkovitz, N.~Kashtan, D.~Chklovskii, and U.~Alon.
\newblock Network motifs: simple building blocks of complex networks.
\newblock {\em Science}, 298:824–827, 2002.

\bibitem{Ferrer2018a}
R.~{Ferrer-i-Cancho}.
\newblock The sum of edge lengths in random linear arrangements.
\newblock {\em Journal of Statistical Mechanics}, page 053401, 2019.

\bibitem{Ferrer2004b}
R.~{Ferrer-i-Cancho}.
\newblock {Euclidean} distance between syntactically linked words.
\newblock {\em Physical Review E}, 70:056135, 2004.

\bibitem{Esteban2016a}
J.~L. Esteban, R.~{Ferrer-i-Cancho}, and C.~G\'omez-Rodr\'iguez.
\newblock The scaling of the minimum sum of edge lengths in uniformly random
  trees.
\newblock {\em Journal of Statistical Mechanics}, page 063401, 2016.

\bibitem{Ferrer2006d}
Ramon {Ferrer-i-Cancho}.
\newblock Why do syntactic links not cross?
\newblock {\em Europhysics Letters}, 76(6):1228--1235, 2006.

\bibitem{Ferrer2014c}
Ramon {Ferrer-i-Cancho}.
\newblock A stronger null hypothesis for crossing dependencies.
\newblock {\em Europhysics Letters}, 108:58003, 2014.

\bibitem{Ferrer2015c}
Ramon {Ferrer-i-Cancho} and Carlos G\'omez-Rodr\'iguez.
\newblock Crossings as a side effect of dependency lengths.
\newblock {\em Complexity}, 21:320--328, 2016.

\bibitem{Ferrer2014f}
R.~{Ferrer-i-Cancho}.
\newblock Non-crossing dependencies: least effort, not grammar.
\newblock In A.~Mehler, A.~L{\"u}cking, S.~Banisch, P.~Blanchard, and B.~Job,
  editors, {\em Towards a theoretical framework for analyzing complex
  linguistic networks}, pages 203--234. Springer, Berlin, 2016.

\bibitem{Hasan2017a}
Adib Hasan, Po-Chien Chung, and Wayne Hayes.
\newblock Graphettes: Constant-time determination of graphlet and orbit
  identity including (possibly disconnected) graphlets up to size 8.
\newblock {\em PLOS ONE}, 12(8):1--12, 08 2017.

\bibitem{Chimani2018a}
M.~Chimani, S.~Felsner, S.~Kobourov, T.~Ueckerdt, P.~Valtr, and A.~Wolff.
\newblock On the maximum crossing number.
\newblock {\em Journal of Graph Algorithms and Applications}, 22:67--87, 2018.

\bibitem{Piazza1991a}
Barry~L. Piazza, Richard~D. Ringeisen, and Sam~K. Stueckle.
\newblock Properties of nonminimum crossings for some classes of graphs.
\newblock In Yousef~Alavi et~al., editor, {\em Proc. 6th Quadrennial Int. 1988
  Kalamazoo Conf. Graph Theory Combin. Appl.}, volume~2, pages 975--989, New
  York, 1991. Wiley.

\bibitem{Bollobas1998a}
B.~Bollob\'as.
\newblock {\em Modern graph theory}.
\newblock Springer-Verlag, 1998.

\bibitem{Wilson1996a}
R.~J. Wilson.
\newblock {\em Introduction to graph theory}.
\newblock Longman, Harlow, England, 4th edition, 1996.

\bibitem{Rosen2017a}
K.H. Rosen, D.~R. Shier, and W.~Goddard.
\newblock {\em Hanbook of discrete and combinatorial mathematics}.
\newblock CRC Press, Boca Raton, FL, 2017.

\bibitem{Bernhart1979a}
F.~Bernhart and P.~C. Kainen.
\newblock The book thickness of a graph.
\newblock {\em Journal of Combinatorial Theory, Series B}, 27(3):320 -- 331,
  1979.

\bibitem{Harary1969a}
F.~Harary.
\newblock {\em Graph Theory}.
\newblock Addison-Wesley, Reading, MA, 1969.

\bibitem{isgci}
H.~N. de~Ridder et~al.
\newblock
  {{I}}nformation~{S}ystem~on~{G}raph~{C}lasses~and~their~{I}nclusions~({I}{S}{G}{C}{I}).
\newblock \url{http://www.graphclasses.org}.

\bibitem{Beren2015a}
D.~Berend, S.~Dolev, and A.~Hanemann.
\newblock Graph degree sequence solely determines the expected {Hopfield}
  network pattern stability.
\newblock {\em Neural Computation}, 27(1):202--210, 2015.

\bibitem{Catellano2009a}
C.~Castellano, S.~Fortunato, and V.~Loreto.
\newblock Statistical physics of social dynamics.
\newblock {\em Rev. Mod. Phys.}, 81:591--646, 2009.

\bibitem{Liu2017a}
H.~Liu, C.~Xu, and J.~Liang.
\newblock Dependency distance: {A} new perspective on syntactic patterns in
  natural languages.
\newblock {\em Physics of Life Reviews}, 21:171--193, 2017.

\bibitem{Smit2002a}
J.~Smit.
\newblock {\em Introduction to Quantum Fields on a Lattice}.
\newblock Cambridge University Press, Cambridge, 2002.

\bibitem{Przulj2007a}
Nata\v{s}a Pr\v{z}ulj.
\newblock Biological network comparison using graphlet degree distribution.
\newblock {\em Bioinformatics}, 23(2):e177--e183, 2007.

\bibitem{Deutsch2002a}
E.~Deutsch and M.~Noy.
\newblock Statistics on non-crossing trees.
\newblock {\em Discrete Mathematics}, 254(1):75 -- 87, 2002.

\bibitem{Newman2010a}
M.~E.~J. Newman.
\newblock {\em Networks. An introduction}.
\newblock Oxford University Press, Oxford, 2010.

\bibitem{Bollobas2002a}
B.~Bollob\'as and O.~Riordan.
\newblock Mathematical results on scale-free random graphs.
\newblock In S.~Bornholdt and H.~Schuster, editors, {\em Handbook of graphs and
  networks: from the genome to the Internet}, pages 1--34. Wiley-VCH, Berlin,
  2003.

\bibitem{Aldous1990a}
D.~Aldous.
\newblock The random walk construction of uniform spanning trees and uniform
  labelled trees.
\newblock {\em SIAM J. Disc. Math.}, 3:450--465, 1990.

\bibitem{Broder1989a}
A.~Broder.
\newblock Generating random spanning trees.
\newblock In {\em Symp. Foundations of Computer Sci., IEEE}, pages 442--447,
  New York, 1989.

\bibitem{Alemany2018b}
L.~Alemany-Puig, M.~Mora, and R.~Ferrer i~Cancho.
\newblock Reappraising the distribution of the number of edge crossings of
  graphs on a sphere.
\newblock {\em in preparation}, 2019.

\bibitem{Cayley1889a}
A.~Cayley.
\newblock A theorem on trees.
\newblock {\em Quart. J. Math}, 23:376--378, 1889.

\bibitem{Pruefer1918a}
H.~Pr\"ufer.
\newblock {Neuer Beweis eines Satzes \"uber Permutationen}.
\newblock {\em Arch. Math. Phys}, 27:742--744, 1918.

\bibitem{Alonso1995a}
L.~Alonso and R.~Schott.
\newblock {\em Random generation of trees. Random generators in computer
  science}.
\newblock Springer, Dordrecht, 1995.

\bibitem{McKay2013a}
B.~D. McKay and A.~Piperno.
\newblock Practical graph isomorphism, {II}.
\newblock {\em Journal of Symbolic Computation}, 60:94--112, 2013.

\end{thebibliography}

\newcommand{\beeksort}[1]{}

\end{document}